\documentclass[fleqn,usenatbib]{mnras}

\usepackage{txfonts}

\usepackage[T1]{fontenc}
\usepackage{ae,aecompl}

\usepackage[normalem]{ulem}

\usepackage{graphicx}
\usepackage{tabularx}
\usepackage{comment}

\usepackage{amssymb}	
\usepackage{epsfig}
\usepackage{upgreek}
\usepackage{mathptmx}
\usepackage[utf8]{inputenc}
\usepackage{dcolumn}
\usepackage{natbib}
\usepackage{enumerate}
\usepackage{longtable,lscape}
\usepackage{grffile}
\usepackage{array,float}
\usepackage{multirow}
\usepackage{lscape}
\usepackage{pdflscape}
\usepackage{afterpage}
\usepackage{siunitx}
\usepackage{capt-of}
\usepackage{xargs}

\usepackage{float}

\usepackage[dvipsnames]{xcolor}
\usepackage{adjustbox}
\usepackage{booktabs}

\usepackage{tikz}
\usetikzlibrary{shapes.geometric, arrows,calc}

\tikzstyle{startstop} = [rectangle, rounded corners, minimum width=2cm, minimum height=1cm,text centered, draw=black, fill=red!30]

\tikzstyle{io} = [trapezium, trapezium left angle=70, trapezium right angle=110, minimum width=2cm, minimum height=0.8cm, text centered, draw=black, fill=blue!30]

\tikzstyle{process} = [rectangle, minimum width=3cm, minimum height=0.8cm, text centered, text width=2cm, draw=black, fill=orange!30]
\tikzstyle{decision} = [diamond, minimum width=2cm, minimum height=0.8cm, text centered, draw=black,text width=2cm, fill=green!30]

\tikzstyle{arrow} = [thick,->,>=stealth]

\newcolumntype{d}[1]{D{.}{\cdot}{#1}}

\newcolumntype{.}{D{.}{.}{-1}}

\newcommand{\lsun}{L$_\odot$}
\newcommand{\msun}{M$_\odot$}

\newcommand{\mclump}{\emph{M}$_{\rm{fwhm}}$}

\newcommand{\lm}{$L_{\rm{bol}}/M_{\rm{fwhm}}$}

\newcommand{\vlsr}{V$_{\rm{LSR}}$}

\newcommand{\mum}{$\mu$m}

\newcommand{\kms}{\ensuremath{\textrm{km\,s}^{-1}}}

\newcommand{\x}{$\times$}

\newcommand{\hi}{H{\sc i}}
\newcommand{\hii}{H{\sc ii}}

\newcommand{\uchii}{UC\,H{\sc ii}}

\newcommand{\poi}{Poisson}

\newcommand{\new}[1]{ \textcolor{black}{ {#1}}}

\title[Evolutionary trends in HMSF]{ATLASGAL -- Evolutionary trends in high-mass star formation\thanks{The full version of Tables\, 2, 3, 4 and 6 are only available in electronic form at the CDS via anonymous ftp to cdsarc.u-strasbg.fr (130.79.125.5) or via http://cdsweb.u-strasbg.fr/cgi-bin/qcat?J/MNRAS/.}}
\author[J.\,S.\,Urquhart et al.]{J.\,S.\,Urquhart$^{1}$\thanks{E-mail: j.s.urquhart@gmail.com}, M.\,R.\,A.\,Wells$^{1}$, T.\,Pillai$^{2}$, S.\,Leurini$^{3}$, A.\,Giannetti$^4$, T.\,J.\,T.\,Moore$^{5}$, \newauthor M.\,A.\,Thompson$^{6}$, C.\,Figura$^{7}$, D.\,Colombo$^{8}$, A.\,Y.\,Yang$^{8}$, C.\,K\"onig$^{8}$, F.\,Wyrowski$^{8}$, \newauthor K.\,M.\,Menten$^{8}$, A.\,J.\,Rigby$^9$, D.\,J.\,Eden$^{5}$, S.\,E.\,Ragan$^9$  \\
\\
$^{1}$ Centre for Astrophysics and Planetary Science, University of Kent, Canterbury, CT2\,7NH, UK \\
$^{2}$ Institute for Astrophysical Research, Boston University, Boston, MA 02215, USA\\
$^{3}$ INAF–Osservatorio Astronomico di Cagliari, Via della Scienza 5, 09047 Selargius (CA), Italy\\
$^{4}$ INAF-Istituto di Radioastronomia, Via P. Gobetti 101, I-40129 Bologna, Italy\\
$^{5}$ Astrophysics Research Institute, Liverpool John Moores University, Liverpool Science Park, 146 Brownlow Hill, Liverpool, L3\,5RF, UK\\
$^{6}$School of Physics and Astronomy, University of Leeds, Leeds LS2 9JT, UK\\
$^{7}$ Wartburg College, Waverly, IA, 50677, USA\\
$^{8}$ Max-Planck-Institut für Radioastronomie, Auf dem Hügel 69, 53121 Bonn, Germany\\
$^{9}$ School of Physics and Astronomy, Cardiff University, Queen’s Buildings, The Parade, Cardiff, CF24 3AA, UK\\
}

\date{Accepted XXX. Received YYY; in original form ZZZ}

\pubyear{2021}

\begin{document}
\label{firstpage}
\pagerange{\pageref{firstpage}--\pageref{lastpage}}
\maketitle

\begin{abstract}

ATLASGAL is a 870-\mum\ dust survey of 420 square degrees of the inner Galactic plane and has been used to identify $\sim$10\,000 dense molecular clumps. Dedicated follow-up observations and complementary surveys are used to characterise the physical properties of these clumps, map their Galactic distribution and investigate the evolutionary sequence for high-mass star formation. The analysis of the ATLASGAL data is ongoing: we present an up-to-date version of the catalogue. We have classiﬁed 5007 clumps into four evolutionary stages (quiescent, protostellar, young stellar objects and \hii\ regions) and ﬁnd similar numbers of clumps in each stage, suggesting a similar lifetime. The luminosity-to-mass (\lm) ratio curve shows a smooth distribution with no signiﬁcant kinks or discontinuities when compared to the mean values for evolutionary stages indicating that the star-formation process is continuous and that the observational stages do not represent fundamentally different stages or changes in the physical mechanisms involved. We compare the evolutionary sample with other star-formation tracers (methanol and water masers, extended green objects and molecular outﬂows) and ﬁnd that the association rates with these increases as a function of evolutionary stage, conﬁrming that our classiﬁcation is reliable. This also reveals a high association rate between quiescent sources and molecular outﬂows, revealing that outﬂows are the earliest indication that star formation has begun and that star formation is already ongoing in many of the clumps that are dark even at 70\,\mum.

\end{abstract}



\section{Introduction}

High-mass stars are a key component of many areas of astrophysics. They dominate the energy budget of galaxies, driving their evolution and ultimately that of the Universe as a whole (\citealt{kennicutt2012}). They are also responsible for the production of all heavy elements, which are returned to the interstellar medium through their strong stellar winds and in supernovae. This action changes the composition and chemistry of their local environments and provides the raw material required for new generations of star and planet formation \citep{2007ARA&A..45..481Z}. It is therefore crucial to understand the high-mass star-formation process, not only as a stand-alone theory, but also as a tool to understand both the history and future of massive star clusters  and to allow us to construct complex models of the evolution of the Universe via numerical simulations \citep{Motte_2018}.

High-mass stars are rare and only contribute a few per\,cent of the stellar population, normally characterised by the initial mass function (IMF; \citealt{kroupa2001}). A consequence of their rarity is that regions of high-mass star formation are statistically less likely to be found nearby, with most having distances $> 2$\,kpc. They evolve very rapidly, reaching the main sequence while still deeply embedded, and so the earliest stages take place behind many 100s of magnitudes of visual extinction and thus remain hidden at visual and near- and even mid-infrared wavelengths. High-mass stars nearly always form in clusters, which significantly complicates the identification and analysis of individual protostellar objects. All of these complications have hindered the development of a comprehensive evolutionary framework for high-mass star formation. 

A first step to improve our understanding of high-mass star formation is the availability of high-resolution images at submillimetre wavelengths over large areas \citep{2007ARA&A..45..481Z}. These are now readily available for the whole Galactic mid-plane (HiGAL; \citealt{Molinari2010}) and the inner Galactic plane (APEX Telescope Large Area Survey of the Galaxy (ATLASGAL); \citealt{schuller2009_full}) with resolutions between $\sim$10-40\arcsec. These provide unbiased samples of dense clumps identified by their dust emission that are a fundamental starting point for star formation as well as being intimately associated with all stages in the process.  

The ATLASGAL survey is an unbiased 870-$\mu$m submillimetre continuum survey of the inner Galactic plane ($300\degr < \ell < 60\degr$ and $|b| < 1.5\degr$, with a spatial resolution of $19^{\prime\prime}$) conducted with the 12-m Atacama Pathfinder Experiment (APEX) telescope (\citealt{gusten2006}). This survey was later extended to include the Carina tangent ($280\degr < \ell < 300\degr$ and $-2\degr < b  < 1\degr$). It has provided a large inventory of dense molecular clumps ($\sim$10\,000 clumps; \citealt{contreras2013,urquhart2014_csc,csengeri2014}) and includes samples of sources in all of the early evolutionary stages associated with high-mass star formation, from very young starless objects to evolved \hii\ regions that are starting to break out of their natal clumps (\citealt{konig2017, urquhart2014_atlas, urquhart2018}). 

The ATLASGAL compact source catalogue (CSC; \citealt{contreras2013, urquhart2014_csc}) consists of $\sim$10\,000 dense clumps. Accurate categorisation of the ATLASGAL catalogue has the potential not only to provide large samples of sources in all of the main evolutionary stages and allow them to be robustly compared to each other, but will also provide a more complete picture of star formation activity across the inner Galactic disc.
The ATLASGAL CSC has therefore been the focus of an intensive campaign to characterise the properties and evolutionary state of the clumps. These have included  dedicated molecular-line studies to produce the kinematics, temperatures, chemistry and determine kinematic distances (\citealt{wienen2012,giannetti2014,wienen2015, csengeri2016_sio, kim2017, kim2018, wienen2018,tang2018, navarete2019,urquhart2019}) and detailed analysis of clumps associated with star-formation tracers such as \hii\ regions (\citealt{urquhart2013_cornish}), methanol masers (\citealt{urquhart2013_methanol, urquhart2015_methanol,billington2019_meth, billington2020_timescale}), massive young stellar objects (MYSO; \citealt{urquhart2014_csc, konig2017, urquhart2018}) and filamentary structures (\citealt{li2016,mattern2018, lin2019}). This programme of follow-up observations and use of complementary multi-wavelength studies has resulted in the ATLASGAL sample being the most well characterised sample of high-mass star-forming clumps currently available. 

This characterisation of the sample is ongoing as new data become available and our analysis improves. In this paper, we provide an updated version of the catalogue presented in \citet{urquhart2018}. We have used new radial velocity measurements taken from the SEDIGISM  (Structure, Excitation and Dynamics of the Inner Galactic Interstellar Medium) survey (\citealt{schuller2017, schuller2021, cabral2021}) to derive distances to $\sim900$ clumps ($\sim600$ new distances and $\sim300$ distances have been updated). We also describe a robust set of criteria that has been applied to multi-wavelength mid- and far-infrared images to determine accurately the evolutionary stages and properties of the clumps. 
The main aim of this work is to produce a high-reliability sample of star-forming clumps that are well characterised in terms of their physical properties and their evolutionary stage, and will serve as a starting point for more detailed investigations. 

The structure of the paper is as follows: we describe how the velocities and distances have been determined in Sect.\,\ref{sect:distances}.  In Sect.\,\ref{sect:physical_properties} we describe how the physical parameters are derived.  We outline our evolutionary classification criteria in Sect.\,\ref{sect:evolutionary_sequence} and compare the physical properties characterising various stages identified. In Sect.\,\ref{sect:discussion} we present the correlation between various parameters and investigate the evolutionary sequence for high-mass star formation. We summarise the work presented and outline future work in Sect.\,\ref{sect:summary}.

\section{Radial Velocities and Distances}
\label{sect:distances}

\subsection{Radial velocities}

\citet{urquhart2018} combined the results from a number of molecular-line surveys and targeted observations towards $\sim$1000 ATLASGAL clumps to provide velocities and distances to nearly 8000 clumps. This study focused on clumps located more than 5\degr\ away from the Galactic centre as few high-resolution molecular-line surveys were available for this region. This situation has improved significantly in recent years with data now becoming available from the SEDIGISM and CHIMPS2 surveys (\citealt{schuller2021} and \citealt{eden2021} respectively). \new{These new surveys allow us to extend our previous work into the Galactic Centre (GC) region.}

The SEDIGISM survey is a $^{13}$CO and C$^{18}$O (2-1) survey conducted with the APEX telescope.\footnote{Other transitions have also been covered. These are much weaker in general and only detected towards the brightest ATLASGAL sources and are less useful; for more details see \citealt{schuller2021}).} In a previous paper, (\citealt{urquhart2021}) we combined the ATLASGAL sample of clumps with the reduced data cubes provided by the SEDIGISM catalogue to extract spectra towards clumps where the two surveys overlapped (i.e. $300\degr < \ell < 18\degr$ and $|b| < 0.5\degr$). This comparison resulted in 5148 spectra being extracted and analysed, allowing us to assign velocities to an additional 1108 ATLASGAL clumps and extend our velocity coverage into the Galactic centre region (i.e., $|\ell| < 5$\degr). In addition, we also include the 31 clumps for which radial velocities have been taken from the literature.

A comparison of the velocities assigned from the analysis of the SEDIGISM data to those previously assigned found agreement within 3\,\kms\ for 3608 of the 3936 clumps ($\sim$92\,per\,cent; see \citealt{urquhart2021} for details) in the SEDIGISM region. The disagreements are due to multiple line-emission components being seen towards the clumps, which reduce the reliability of identifying the correct velocity component. In these cases, we were able to take advantage of the SEDIGISM data to create maps of the different velocity components, which were then compared to the dust emission to identify the component with the best morphological correlation. Lower-resolution CO~(1-0) data had been used to make the initial velocity assignments to many of the clumps with disagreements and the higher-resolution SEDIGISM observations were better able to resolve the line-of-sight confusion (see \citealt{urquhart2021} for more details). The high-resolution and higher critical density of the $^{13}$CO (2-1) line and the ability to match the morphology of gas and dust means velocities based on SEDIGISM data are more reliable than values from some of the other surveys used previously. We therefore adopted the SEDIGISM velocities for 269 of the 318 clumps where the velocities do not agree. The allocation of a velocity component was uncertain for the remaining 49, and so the previous velocities assigned to these clumps were discarded. 

In total, there are 9817 clumps within the main ATLASGAL region (i.e., $300\degr < \ell < 60\degr$ and $|b| < 1.5$\degr) and we have now managed to assign a velocity to 8899 of these ($\sim$90\,per\,cent of the sample). Many of the remaining clumps are located within the Galactic Centre region where the spectra have either poor baselines or are very broad ($> 30$\,\kms) and often blended components making any velocity allocation difficult (\citealt{urquhart2021}). We are now, therefore, as complete with regards to velocities for the ATLASGAL catalogue as we are likely to get.

\subsection{Kinematic distances}

\setlength{\tabcolsep}{6pt}

\begin{table}
\begin{center}
\caption{Summary of the kinematic distance solutions for the 8417 ATLASGAL sources located in the Galactic disk ($\ell = 300\degr-357\degr$ and $\ell = 3\degr-60\degr$). The Roman numerals (i) to (viii) given in Column\,1 refer to the various steps described in \citet{urquhart2018} (see text for brief summary), while the (ix) and (x) are used to identify sources where a distance could not be determined and clumps for which no velocity is available, respectively.}
\label{tbl:distance_solutions}
\begin{minipage}{\linewidth}
\begin{tabular}{clc}
\hline 
\hline
Step & Description   & Total number distances  \\
 & of method   & assigned  \\
\hline
 (i) & Parallax/Spectroscopic  & 114 \\
(ii) & Outer Galaxy  & 136  \\
(iii)&  Tangent  & 766  \\
(iv) & Z distance   & 1280   \\
(v) & \hi EA   & 76   \\
(vi) &  \hi SA Near (?) & 2725 (174) \\
(vi) & \hi SA Far  (?)& 904  (182) \\
(vii) & IRDC Associations  &  39 \\
(viii) & Literature  & 1707  \\
\hline
(ix) & Ambiguous (Solar Circle) & 325 (61) \\
(x) & No \vlsr\ available  & 345  \\
\hline
\end{tabular}
\end{minipage}
\end{center}

\end{table}
\setlength{\tabcolsep}{6pt}

\setlength{\tabcolsep}{3pt}

\begin{table*}

\begin{center}
\caption{Summary of the kinematic distance analysis.}
\label{tbl:source_vlsr}
\begin{minipage}{\linewidth}
\small
\begin{tabular}{l.l.....cll.}
\hline \hline
\multicolumn{1}{c}{}&  \multicolumn{1}{c}{}&\multicolumn{1}{c}{}&	\multicolumn{3}{c}{Reid Distances}  && \multicolumn{3}{c}{Adopted Kinematic Solution}  & \multicolumn{1}{c}{} & \multicolumn{1}{c}{} \\
\cline{4-6}\cline{8-10}
\multicolumn{1}{c}{ATLASGAL}&  \multicolumn{1}{c}{\vlsr}&\multicolumn{1}{c}{\vlsr\ Ref.}&	\multicolumn{1}{c}{Bayesian}  & \multicolumn{1}{c}{Near}  & \multicolumn{1}{c}{Far} && \multicolumn{1}{c}{Distance} & \multicolumn{1}{c}{Solution} & \multicolumn{1}{c}{Dist.} & \multicolumn{1}{c}{Association} & \multicolumn{1}{c}{ Distance} \\

\multicolumn{1}{c}{CSC name}&  \multicolumn{1}{c}{(\kms)}&\multicolumn{1}{c}{}&	\multicolumn{1}{c}{(kpc)}  & \multicolumn{1}{c}{(kpc)}  & \multicolumn{1}{c}{(kpc)} && \multicolumn{1}{c}{(kpc)} & \multicolumn{1}{c}{Flag$^a$} & \multicolumn{1}{c}{ Ref.$^b$} & \multicolumn{1}{c}{Name} & \multicolumn{1}{c}{(kpc)} \\

\hline
AGAL003.613$-$00.104	&	36.2	&	\citet{urquhart2021}	&	10.9	&	6.7	&	9.9	&&	10.9	&	(vi)	&	\multicolumn{1}{c}{$\cdots$}	&	\multicolumn{1}{c}{$\cdots$}	&	\multicolumn{1}{c}{$\cdots$}	\\
AGAL003.651$-$00.127	&	7.9	&	\citet{urquhart2021}	&	2.9	&	2.5	&	14.6	&&	2.9	&	(vi)	&	\multicolumn{1}{c}{$\cdots$}	&	G003.705$-$00.068	&	2.9	\\
AGAL003.661$-$00.114	&	3.6	&	\citet{urquhart2021}	&	2.9	&	1.5	&	16.7	&&	2.9	&	(viii)	&	\multicolumn{1}{c}{$\cdots$}	&	G003.705$-$00.068	&	2.9	\\
AGAL003.683$-$00.096	&	7.5	&	\citet{urquhart2021}	&	2.9	&	2.4	&	14.7	&&	2.9	&	(vi)	&	\multicolumn{1}{c}{$\cdots$}	&	G003.705$-$00.068	&	2.9	\\
AGAL003.688$-$00.147	&	4.3	&	\citet{urquhart2021}	&	2.9	&	1.6	&	16.3	&&	2.9	&	(vi)	&	\multicolumn{1}{c}{$\cdots$}	&	G003.705$-$00.068	&	2.9	\\
AGAL003.689$-$00.274	&	-30.9	&	\citet{urquhart2021}	&	18.8	&	\multicolumn{1}{c}{$\cdots$}	&	\multicolumn{1}{c}{$\cdots$}	&&	18.8	&	(ii)	&	\multicolumn{1}{c}{$\cdots$}	&	\multicolumn{1}{c}{$\cdots$}	&	\multicolumn{1}{c}{$\cdots$}	\\
AGAL003.698$-$00.102	&	7.8	&	\citet{urquhart2021}	&	2.9	&	2.4	&	14.6	&&	2.9	&	(vi)	&	\multicolumn{1}{c}{$\cdots$}	&	G003.705$-$00.068	&	2.9	\\
AGAL003.719$+$00.016	&	9.1	&	\citet{urquhart2021}	&	2.9	&	2.7	&	14.1	&&	2.9	&	(viii)	&	\multicolumn{1}{c}{$\cdots$}	&	G003.705$-$00.068	&	2.9	\\
AGAL003.744$+$00.019	&	7.8	&	\citet{urquhart2021}	&	2.9	&	2.4	&	14.6	&&	2.9	&	(vi)	&	\multicolumn{1}{c}{$\cdots$}	&	G003.705$-$00.068	&	2.9	\\
AGAL003.746$-$00.071	&	4.0	&	\citet{urquhart2021}	&	2.9	&	1.5	&	16.4	&&	2.9	&	(vi)	&	\multicolumn{1}{c}{$\cdots$}	&	G003.705$-$00.068	&	2.9	\\
\hline\\
\end{tabular}\\

Notes: Only a small portion of the data is provided here. The full table is available in electronic form at the CDS via anonymous ftp to cdsarc.u-strasbg.fr (130.79.125.5) or via http://cdsweb.u-strasbg.fr/cgi-bin/qcat?J/MNRAS/.\\ 
$^a$ The distance solution flags refer to the different steps described in Sect.\,\ref{sect:distances} and Table\,\ref{tbl:distance_solutions}.\\
$^b$ References for distance solutions: (1) \citet{anderson2009a}: (2) \citet{araya2002}; (3) \citet{battisti2014}; (4) \citet{urquhart2013_cornish}; (5) \citet{dunham2011_bgps_vii}; (6) \citet{fish2003}, (7) \citealt{reid2014}, (8) \citet{bessel_sanna2014}, (9) \citet{bessel_wu2014}, (10) \citet{bessel_xu2009}, (11) \citet{fish2003}, (12) \citealt{downes1980}, (13) \citet{giannetti2015}, (14) \citet{green2011b}, (15) \citet{immer2012}, (16) \citet{Kolpak2003}; (17) \citet{pandian2009}; (18) \citet{sewilo2003}; (19) \citet{moises2011}; (20) \citet{pandian2008}; (21) \citet{roman2009}; (22) \citet{stead2010}; (23) \citet{bessel_sanna2009}; (24) \citet{bessel_sato2010_w51}; (25) \citet{urquhart2012_hiea}; (26) \citet{watson2003}; (27) \citet{xu2011}; (28) \citet{bessel_zhang2013}; (29) (\citet{nagayama2011}.

\end{minipage}

\end{center}
\end{table*}
\setlength{\tabcolsep}{6pt}

We have new or updated velocities for  1408 clumps taken from a combination of velocities obtained from the literature and from SEDIGISM (\citealt{urquhart2021}). This sample has allowed us to determine radial velocities for many sources toward the Galactic centre ($355\degr < \ell < 5\degr$) that were excluded from our previous paper. We will use these radial velocities to determine kinematic distances, which will subsequently be used to calculate their physical properties and investigate their Galactic distribution. We still, however, need to exclude sources within a few degrees of the Galactic centre  (i.e., $357\degr < \ell < 3\degr$) because kinematic distances for sources in this region are extremely unreliable. This exclusion reduces the number of sources for which we need to determine distances to 581.

\setlength{\tabcolsep}{6pt}

\begin{table*}

\begin{center}\caption{Derived Association parameters. }
\label{tbl:cluster_velocities_distances}
\begin{minipage}{\linewidth}
\small
\begin{tabular}{llcccc.c.l}
\hline \hline
  \multicolumn{1}{c}{Association}& \multicolumn{1}{c}{Literature} & \multicolumn{1}{c}{\# of}  &\multicolumn{1}{c}{$\ell$}  &\multicolumn{1}{c}{$b$}  &\multicolumn{1}{c}{\vlsr} &\multicolumn{1}{c}{$\Delta$\vlsr}  &\multicolumn{1}{c}{Distance}   &\multicolumn{1}{c}{$\Delta$Distance}   &	\multicolumn{1}{c}{Ref.}\\ 
  
  \multicolumn{1}{c}{name}& \multicolumn{1}{c}{name} &\multicolumn{1}{c}{members}  &\multicolumn{1}{c}{(\degr)}  &\multicolumn{1}{c}{(\degr)}  &\multicolumn{1}{c}{(\kms)} &\multicolumn{1}{c}{(\kms)}  &\multicolumn{1}{c}{(kpc)}  &\multicolumn{1}{c}{(kpc)}  \\
\hline

G005.950$-$01.183	&	M8 - Lagoon Nebula	&	24	&	5.950	&	-1.183	&	12.4	&	2.0	&	0.95	&	\multicolumn{1}{c}{$\cdots$}	&	\citet{moises2011}	\\
G006.183$-$00.358	&	W28	&	78	&	6.183	&	-0.358	&	12.6	&	4.9	&	3	&	0.2	&	\citet{reid2019}	\\
G008.581$-$00.329	&	W30 (SNR?)	&	38	&	8.581	&	-0.329	&	37.1	&	1.3	&	4.45	&	0.23	&	\multicolumn{1}{c}{$\cdots$}	\\
G010.147$-$00.237	&	W31-south	&	65	&	10.147	&	-0.237	&	16.9	&	11.6	&	2.9	&	0.3	&	\citet{moises2011}	\\
G010.474$-$00.012	&	G010.47+00.02	&	10	&	10.474	&	-0.012	&	68.4	&	2.8	&	8.5	&	0.6	&	\citet{reid2019}	\\
G010.639$-$00.400	&	W31C (North)	&	17	&	10.639	&	-0.400	&	-2.9	&	1.2	&	3.9	&	0.5	&	\citet{reid2019}	\\
G010.901$-$00.070	&	Snake	&	34	&	10.901	&	-0.070	&	25.8	&	4.9	&	4.1	&	0.2	&	\citet{reid2019}	\\
G011.916+00.726	&	N4 Bubble	&	8	&	11.916	&	0.726	&	26.9	&	3.8	&	2.81	&	0.29	&	\multicolumn{1}{c}{$\cdots$}	\\
G012.910+00.460	&	IRAS-18089-1732	&	8	&	12.910	&	0.460	&	32.8	&	1.3	&	2.5	&	0.3	&	\citet{reid2019}	\\
G012.960$-$00.168	&	W33 (G012.80-00.20)	&	85	&	12.960	&	-0.168	&	37.8	&	7.6	&	2.6	&	0.2	&	\citet{reid2019}	\\
G015.079$-$00.625	&	M17	&	55	&	15.079	&	-0.625	&	19.5	&	1.9	&	2	&	0.1	&	\citet{reid2019}	\\
G016.996+00.719	&	M16	&	62	&	16.996	&	0.719	&	22.1	&	2.8	&	1.86	&	0.13	&	\multicolumn{1}{c}{$\cdots$}	\\
G018.936$-$00.350	&	W39	&	58	&	18.936	&	-0.350	&	63.8	&	3.4	&	4	&	1.1	&	\multicolumn{1}{c}{$\cdots$}	\\
\hline\\
\end{tabular}\\
Notes: A more complete table that includes parameters for all 877 clusters is available in electronic form at the CDS via anonymous ftp to cdsarc.u-strasbg.fr (130.79.125.5) or via http://cdsweb.u-strasbg.fr/cgi-bin/qcat?J/MNRAS/.
\end{minipage}

\end{center}
\end{table*}
\setlength{\tabcolsep}{6pt}

We have given a detailed description of the method used to determine distances to the ATLASGAL clumps in \citet{urquhart2018}, and so we only provide a brief outline of the main steps here. We combine the radial velocity measurements with a model of the Galactic rotation to obtain a kinematic distance. We have used the Galactic rotation curve of \citet{reid2016}, which has utilised the results of maser parallax distances measured by the Bar and Spiral Structure Legacy (BeSSeL) Survey (\citealt{reid2019}).\footnote{http://bessel.vlbi-astrometry.org/} An unavoidable consequence of using a rotation model is that there are two possible solutions for any given radial velocity for sources located inside the solar circle: these are located at equal distances on either side of the tangent distance, and are commonly referred to the near and far distances. 

In Table\,\ref{tbl:distance_solutions}, we provide a summary of the different methods used and the number of sources  to which these have been applied. Here we give a brief description of the methods: (i) For sources with a reliable spectroscopic (\citealt{moises2011}) or maser parallax distance (\citealt{reid2019}), these have been adopted. (ii) Sources located outside the Solar circle (i.e., Galactocentric radius (R$_{\rm gc}$) $>8.35$\,kpc) are not affected by distance ambiguity. (iii) Sources located within 10 \kms\ of the tangent velocity are placed at the tangent distance. (iv) Sources where a far distance would place them more than 120\,pc from the Galactic mid-plane are placed at the near distance as a far distance is very unlikely. (v) Distance ambiguity resolved using the \hi\ Emission-absorption method (\hi\,EA; e.g. \citealt{Kolpak2003,araya2002,urquhart2011a}). (vi) Distance ambiguity resolved using the \hi\ self-absorption method (\hi\,SA; e.g. \citealt{jackson2003,roman2009, wienen2015}). Question marks indicate the number of sources where the resolution is less certain. We use archival data from the Southern and VLA Galactic Plane surveys (SGPS; \citealt{mcclure2005} and VGPS; \citealt{stil2006}) for the \hi\ distance resolutions. (vii) Sources associated with an infrared dark cloud (IRDC), which are seen in silhouette against the bright mid-infrared Galactic background, are considered to be at the near distance.  (viii) Distance solutions taken from the literature (see Table\,\ref{tbl:source_vlsr} for references).

We have conducted a comprehensive literature review of similar studies to confirm our distance analysis and to allocate distances where our methods have been unsuccessful. Comparisons with the distances reported by other studies were made in \citet{urquhart2018} and distances were found to agree in $\sim$70-80\,per\,cent of cases. As an additional check, we compare our distances to those reported by \citet{elia2021} based on anaylsis the HiGAL survey (\citealt{Molinari2010}). Restricting the matches to those within 20\arcsec\ and radial velocities within 5\,\kms, we find agreement in the assigned velocities to be $\sim$72\,per\,cent (i.e., distances agree within 2\,kpc to allow for slight differences in the assigned radial velocities and the ways the distances are determined).

\begin{figure}
	\centering

	\includegraphics[width=0.49\textwidth, trim= 20 0 20 0]{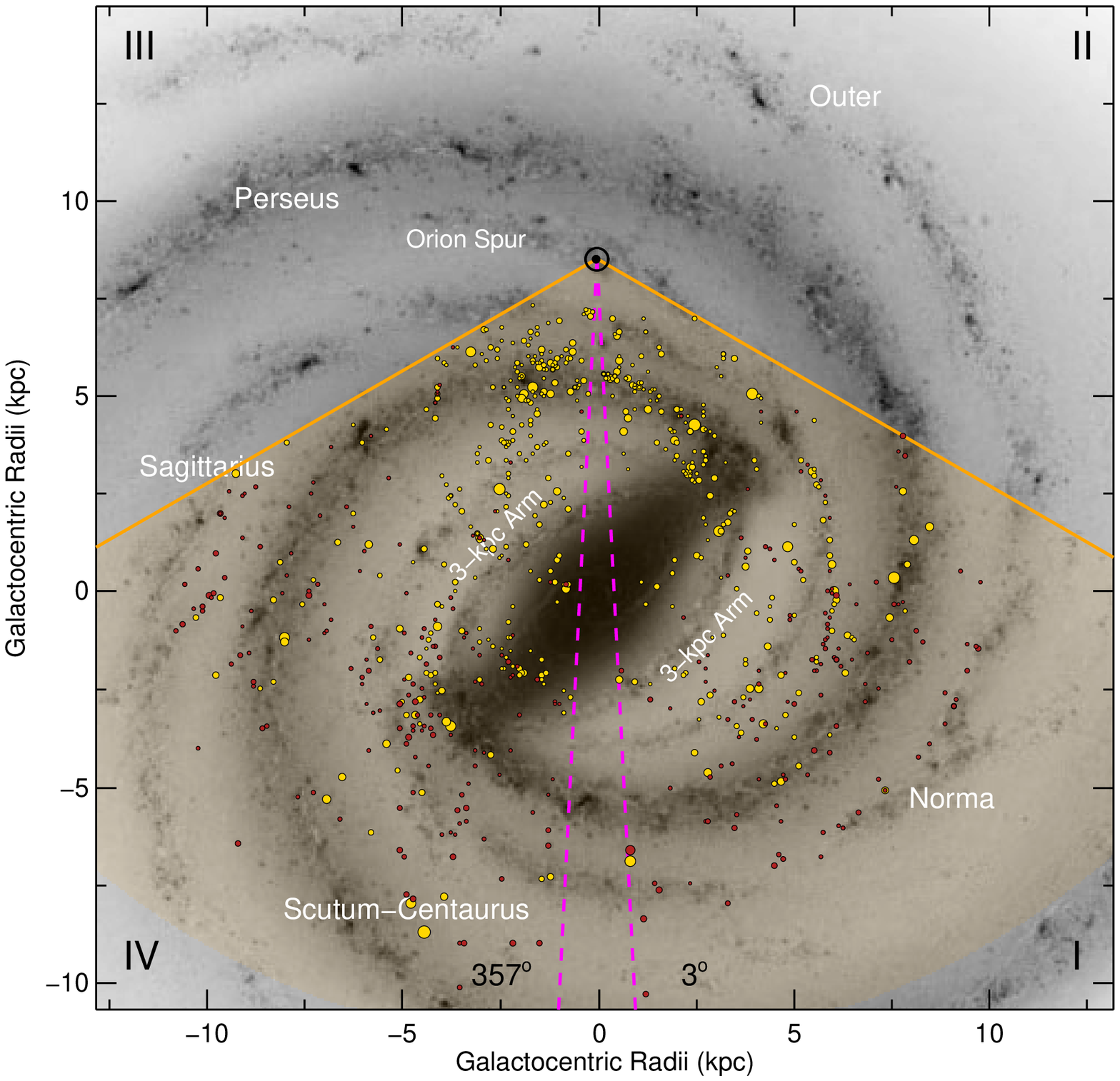}

	\caption{2-D images showing expected large-scale features of the Milky Way as viewed from the Galactic pole. The positions of ATLASGAL sources have been overlaid to facilitate comparison of their distribution to the large-scale structure of the Galaxy. The yellow circles show positions of the clusters while the red circles show the positions of individual clumps. The sizes of the circles give an indication of the masses. The background image is a schematic of the Galactic disc as viewed from the Northern Galactic Pole (courtesy of NASA/JPL-Caltech/R. Hurt (SSC/Caltech)). The Sun is located at the apex of the wedge and is indicated by the $\odot$ symbol.  The spiral arms are labelled in white and Galactic quadrants are given by the Roman numerals in the corners of the image. The magenta line shows the innermost region toward the Galactic centre where distances are not reliable.  }

	\label{fig:galactic_cluster_distribution}

\end{figure}

\subsection{Identification of complexes}

We have grouped the individual clumps together into complexes. This grouping is done by using friends-of-friends clustering analysis to associate clumps that are localised in longitude, latitude, and velocity ($\ell b v$-space). The initial groups identified by the algorithm are subsequently inspected to ensure reliability and that the distance solutions are self-consistent within the uncertainties associated with the different methods used. For example, the \hi SA method is approximately 80\,per\,cent reliable (\citealt{busfield2006}) and so the individual distances within a grouping should have a similar level of agreement. For complexes where the distances are not consistent, we have repeated the clustering analysis using increasingly tighter angular and velocity offsets between clumps until structures that are coincident in position and distance are identified. Where a reliable distance can be assigned to a complex, it is also assigned to the clumps associated with it. This process is more fully described in \citet{urquhart2018} and we refer the reader to that paper for more details.

We have re-run the analysis of the 776 associations identified in \citet{urquhart2018}, updating velocities for $\sim300$ clumps that were included in the previous analysis, and providing new velocities and distances for a similar number of clumps. This new analysis has identified 877 associations, 684 of which were previously identified, with 47 located in the $\ell = 355\degr-357\degr$ and $\ell = 3\degr-5\degr$ ranges and are therefore new identifications. The modified and new velocities have resulted in the modification of 146 associations, in most cases only very slightly. Of the 684 associations that were previously identified, the distances agree for all but 49, and so distances have been amended for $\sim$7\,per\,cent of the associations.  We provide a list of the associations identities and their positions, velocities, and derived distances in Table\,\ref{tbl:cluster_velocities_distances}.

\subsection{Summary of distances}

There are a total of 8417 clumps in the $3\degr < |\ell| <60\degr$ region. 6673 of these clumps have been associated with one of the 877 associations identified by the $\ell b v$ friends-of-friends  analysis. The remaining 1744 are considered to be isolated. We have allocated distances through the methods described above to 8130 clumps in this region ($\sim$97\,per\,cent). This re-analysis represents a modest increase ($\sim10$\,per\,cent) in the number of distances presented in \citet{urquhart2018} and so the overall Galactic distribution shown in Fig.\,\ref{fig:galactic_cluster_distribution} is broadly unchanged. \new{The overall reliability and completeness of the ATLASGAL catalogue, however, have been improved.}

\section{Physical Properties}
\label{sect:physical_properties}

In this section, we describe the methods used to determine physical properties for the 8417 clumps located within the $3\degr < |\ell| <60\degr$ region for which a distance has been determined. We have previously derived the clump masses, peak column densities, bolometric luminosity, sizes and virial parameter in \citet{urquhart2018}, and we refer the reader to that paper for a detailed description. One issue we noted in that paper was a trend for higher masses and sizes as a function of evolution (i.e., larger for \hii\ regions compared to quiescent sources, suggesting an evolutionary trend). This trend is caused by the heating of the clump by the embedded young star(s), which results in more of the clump's envelope being raised above the detection threshold. This effect produces an observational bias in which the clumps appear to increase in size and their volume densities decrease with evolution (see figure\,33 in \citealt{urquhart2018}). 

\citet{billington2019_meth} demonstrated that the observational bias can be completely eliminated by using the FWHM source sizes (determined from the pixels above the half power level) and masses (see Figure 9 of that paper). These FWHM parameters are therefore much more reliable when comparing properties of different evolutionary subsamples. Following the method described by \citet{billington2019_meth}, we have recalculated the FWHM sizes, masses, H$_2$ volume density ($n$), and free-fall times for the clumps using the updated distances discussed in the previous section. We provide a brief description of how these source properties have been determined below.

\setlength{\tabcolsep}{3pt}

\begin{table*}

\begin{center}\caption{Derived clump parameters.}
\label{tbl:derived_clump_para}
\begin{minipage}{\linewidth}
\small
\begin{tabular}{ll..........}
\hline \hline
  \multicolumn{1}{c}{CSC}&  
  \multicolumn{1}{c}{Evolution}&
  \multicolumn{1}{c}{T$_{\rm{dust}}$} & 
   \multicolumn{1}{c}{$V_{\rm{LSR}}$}   &
  \multicolumn{1}{c}{Distance} &
  \multicolumn{1}{c}{R$_{\rm{GC}}$}&
  \multicolumn{1}{c}{R$_{\rm{fwhm}}$} & 
  \multicolumn{1}{c}{Log[$L_{\rm{bol}}$]}	&
  \multicolumn{1}{c}{Log[$M_{\rm{fwhm}}$]} & 
  \multicolumn{1}{c}{Log[$n_{\rm fwhm}$(H$_2$)]} & 
   \multicolumn{1}{c}{$\tau_{\rm ff}$} & 
    \multicolumn{1}{c}{\lm-ratio} \\

    \multicolumn{1}{c}{name }&  
    \multicolumn{1}{c}{type} & 
    \multicolumn{1}{c}{(K)}  &
     \multicolumn{1}{c}{(km\,s$^{-1}$)}&	
    \multicolumn{1}{c}{(kpc)} &
    \multicolumn{1}{c}{(kpc)}&
    \multicolumn{1}{c}{(pc)}&
    \multicolumn{1}{c}{(\lsun)} &
    \multicolumn{1}{c}{(\msun)} &
    \multicolumn{1}{c}{(cm$^{-3}$)} & 
     \multicolumn{1}{c}{(Myr)} & 
      \multicolumn{1}{c}{(\lsun/\msun)} \\
    
\hline
AGAL003.008+00.111	&	Ambiguous	&	15.2	&	152.4	&	15.4	&	7.03	&	0.75	&	3.298	&	4.000	&	4.907	&	0.115	&	0.20	\\
AGAL003.016+00.141	&	Ambiguous	&	12.7	&	151.1	&	15.4	&	7.03	&	1.78	&	2.831	&	4.734	&	4.522	&	0.179	&	0.01	\\
AGAL003.016-00.179	&	Ambiguous	&	17.8	&	-4.8	&	21.3	&	12.97	&	4.31	&	4.595	&	5.066	&	3.703	&	0.834	&	0.34	\\
AGAL003.021-00.067	&	Ambiguous	&	12.2	&	-0.9	&	2.9	&	5.44	&	0.35	&	1.852	&	2.955	&	4.856	&	0.104	&	0.08	\\
AGAL003.028-00.094	&	Ambiguous	&	11.9	&	9.3	&	2.9	&	5.42	&	0.51	&	2.314	&	2.820	&	4.248	&	0.209	&	0.31	\\
AGAL003.034+00.402	&	Quiescent	&	13.9	&	13.0	&	2.9	&	5.43	&	0.34	&	2.318	&	2.422	&	4.356	&	0.185	&	0.79	\\
AGAL003.049+00.392	&	Ambiguous	&	15.7	&	11.6	&	2.9	&	5.43	&	0.27	&	1.656	&	2.502	&	4.740	&	0.119	&	0.14	\\
AGAL003.049-00.047	&	Ambiguous	&	19.1	&	9.0	&	2.9	&	5.42	&	0.18	&	2.379	&	2.033	&	4.805	&	0.110	&	2.22	\\
AGAL003.056+00.151	&	Ambiguous	&	13.5	&	153.4	&	15.4	&	7.03	&	1.18	&	3.047	&	4.201	&	4.525	&	0.179	&	0.07	\\
AGAL003.093+00.422	&	Quiescent	&	9.7	&	22.7	&	10.7	&	2.40	&	1.87	&	3.035	&	4.127	&	3.855	&	0.243	&	0.08	\\
\hline\\
\end{tabular}\\
Notes: Only a small portion of the data is provided here. The full table is available in electronic form at the CDS via anonymous ftp to cdsarc.u-strasbg.fr (130.79.125.5) or via http://cdsweb.u-strasbg.fr/cgi-bin/qcat?J/MNRAS/.
\end{minipage}

\end{center}
\end{table*}
\setlength{\tabcolsep}{6pt}

The sizes of clumps are determined from the number of pixels within the FWHM contour, i.e., above 50\,per\,cent of the peak of the ATLASGAL dust continuum emission (c.f. \citealt{shirley2003}):

\begin{equation}
R_{\rm{fwhm}} = \sqrt{\frac{A}{\pi}}
\end{equation}

\noindent where $A$ is the number of pixels within the 50-per-cent flux contour multiplied by the square of the pixel size in arcsec (i.e., 36\,sq.\,arcsec). The 50-per-cent flux level is below the detection threshold for sources with a SNR below 6$\sigma$ and cannot be reliably determined and so these sources are excluded. We also exclude clumps with fewer pixels than the beam integral ($\Theta_{\rm{fwhm}}^2\times 1.133 = 11.3$\,pixels, where $\Theta_{\rm{fwhm}}$ is 19.2\,arcsec) as the flux measurements will be overestimated. Of the 8417 clumps considered here, 5866 satisfy these two conditions and are deconvolved from the ATLASGAL beam using:

\begin{equation}
R_{\rm{fwhm,decon}} = \sqrt{R_{\rm{fwhm}}^2-\left(\frac{\Theta_{\rm{fwhm}}}{2}\right)^2}
\end{equation}

The distances determined in Sect.\,\ref{sect:distances} are used to convert the deconvolved radius into a physical radius, $R_{\rm{fwhm, pc}}$. The FWHM clump mass, $M_{\rm{fwhm}}$, is determined using:

\begin{equation}
\label{eqn:mass}
M_{\rm{fwhm}} \, = \, \frac{D^2 \, S_{\rm \nu, fwhm} \, \gamma}{B_\nu(T_{\rm{dust}}) \, \kappa_\nu},
\end{equation}

\noindent where $S_{\rm \nu, fwhm}$ is the integrated 870-\mum\ flux density within the 50-per-cent contour, $D$ is the distance to the source, $\gamma$ is the gas-to-dust mass ratio, $B_\nu$ is the Planck function for a dust temperature $T_{\rm{dust}}$, and $\kappa_\nu$ is the dust absorption coefficient taken as 1.85\,cm$^2$\,g$^{-1}$ (\citealt{schuller2009_full} and references therein).  We use the dust temperatures derived from fits to the spectral-energy distributions of the clumps (SEDs; \citealt{urquhart2018}) and assume that these values are reasonable estimates of the average temperatures of the inner parts of the clumps (see also \citealt{konig2017}). The gas-to-dust ratio is calculated using the following relationship empirically determined by \citet{giannetti2017}:

\begin{equation}
    {\rm log}(\gamma) = \left(0.087\left[^{+0.045}_{-0.025}\right]\pm 0.007 \right)R_{\rm gc} +  \left(1.44\left[^{-0.45}_{+0.21}\right]\pm 0.03 \right)
\end{equation}

\noindent where $R_{\rm gc}$ is the Galactocentric distance expressed in kpc. The systematic uncertainties are given in the square brackets. This prescription gives a value of $\gamma$ between 130 and 145 at the distance of the Sun, which is in very good agreement with the local value of 136 (see \citealt{giannetti2017} for a detailed discussion).

We calculate the mean FWHM volume density by dividing the $M_{\rm{fwhm}}$ by the volume:

\begin{equation}
n\left({\rm H_2}\right) = \frac{3 }{4 \pi  }\frac{M_{\rm{fwhm}}}{\mu m_{\rm p} R_{\rm{fwhm}}^3} 
\end{equation}

\noindent where $n\left({\rm H_2}\right) $ is the hydrogen particle density per cm$^{-3}$, $\mu$ is the mean molecular weight per hydrogen atom (taken as 2.8; \citealt{Kauffmann2008}), $m_{\rm p}$ is the mean proton mass, and $M_{\rm{fwhm}}$ and $R_{\rm{fwhm}}$ are as previously defined. We make the assumption that each clump is generally spherical and is not extended along the line of sight. 

The clump free-fall times are useful in providing a lower limit to the star-formation timescales and can be readily determined from the mean densities using:

\begin{equation}
    \tau_{\rm ff} = \sqrt{\frac{3\pi}{32 G \bar\rho}}
\end{equation}

\noindent where $\bar\rho  = \frac{3M_{\rm fwhm}}{4\pi R_{\rm fwhm}^3}$ is the mean density of the clump. The free-fall times are typically between 1-3$\times 10^5$\,yr, which compare well with the \hii\ region phase lifetimes derived empirically by \citet{mottram2011b} and from a numerical model by \citet{davies2011} ($\sim$2-$4\times 10^5$\,yr). 

Finally, we define two additional distance-independent parameters, these being the luminosity-to-mass ratio ($L/M_{\rm fwhm}$) and the mass surface density ($M_{\rm fwhm}/\pi R_{\rm fwhm}^2$). The luminosities have been taken directly from \citet{urquhart2018}; these have been rescaled to the distances presented here. The first of these two ratios is considered to be a good diagnostic for evolution and measure of the instantaneous star formation efficiency (\citealt{molinari2008, urquhart2014_atlas, urquhart2018}) and has been used and discussed extensively in previous ATLASGAL papers. 

\begin{figure*}
    \centering
  
    \includegraphics[width=.45\textwidth, trim= 0 0 0 0]{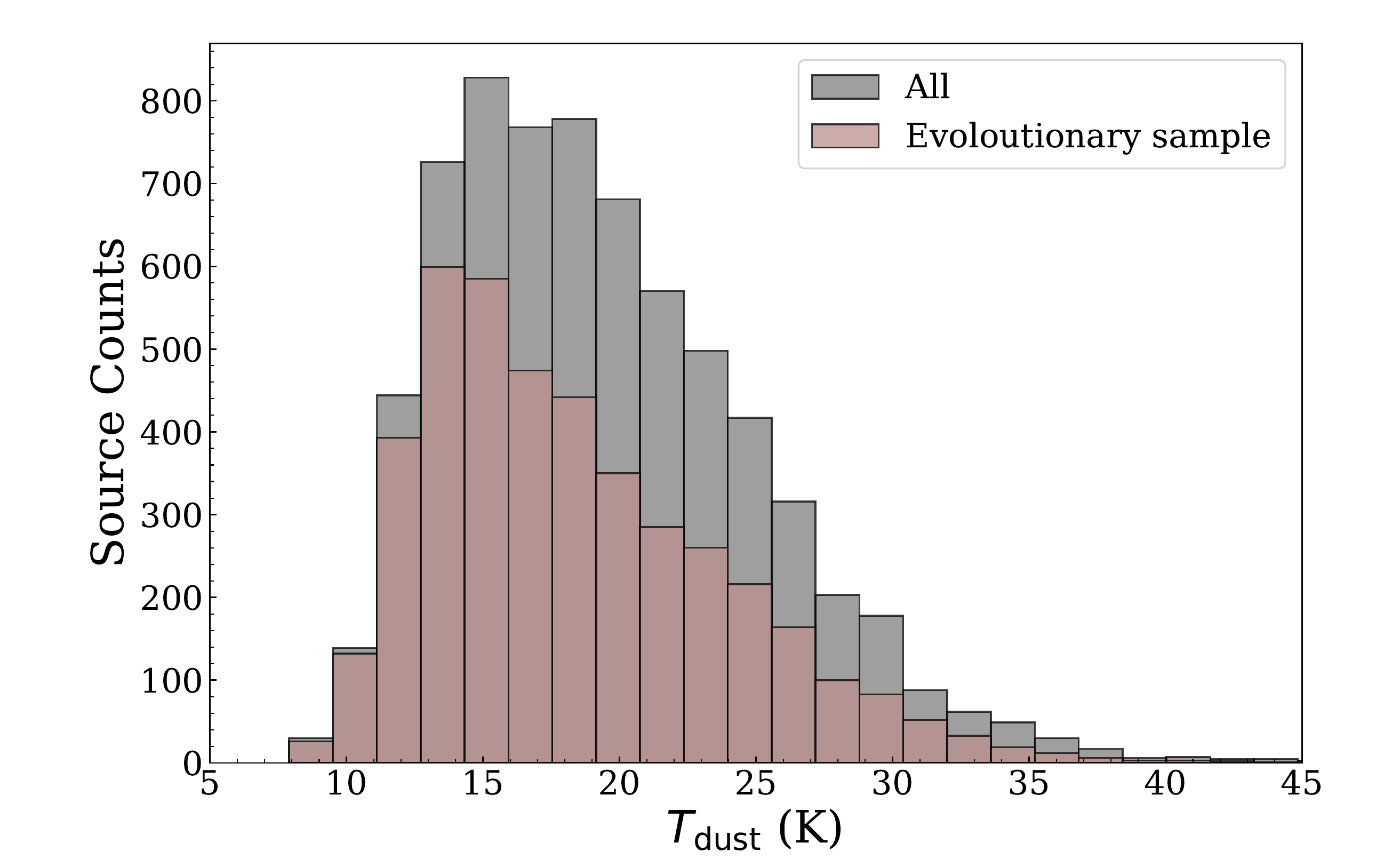}
     \includegraphics[width=.45\textwidth, trim= 0 0 0 0]{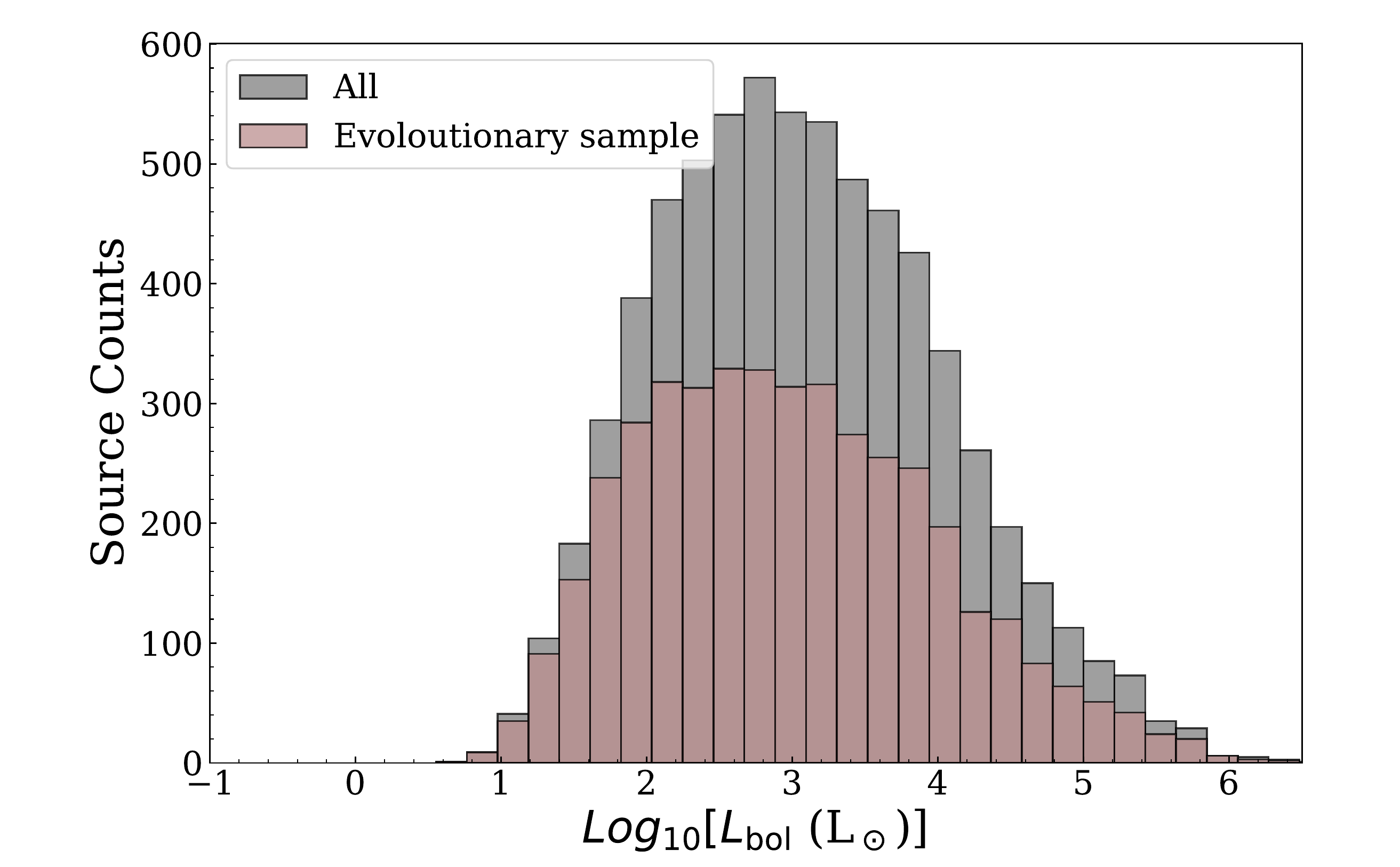}\\
        \includegraphics[width=.45\textwidth, trim= 0 0 0 0]{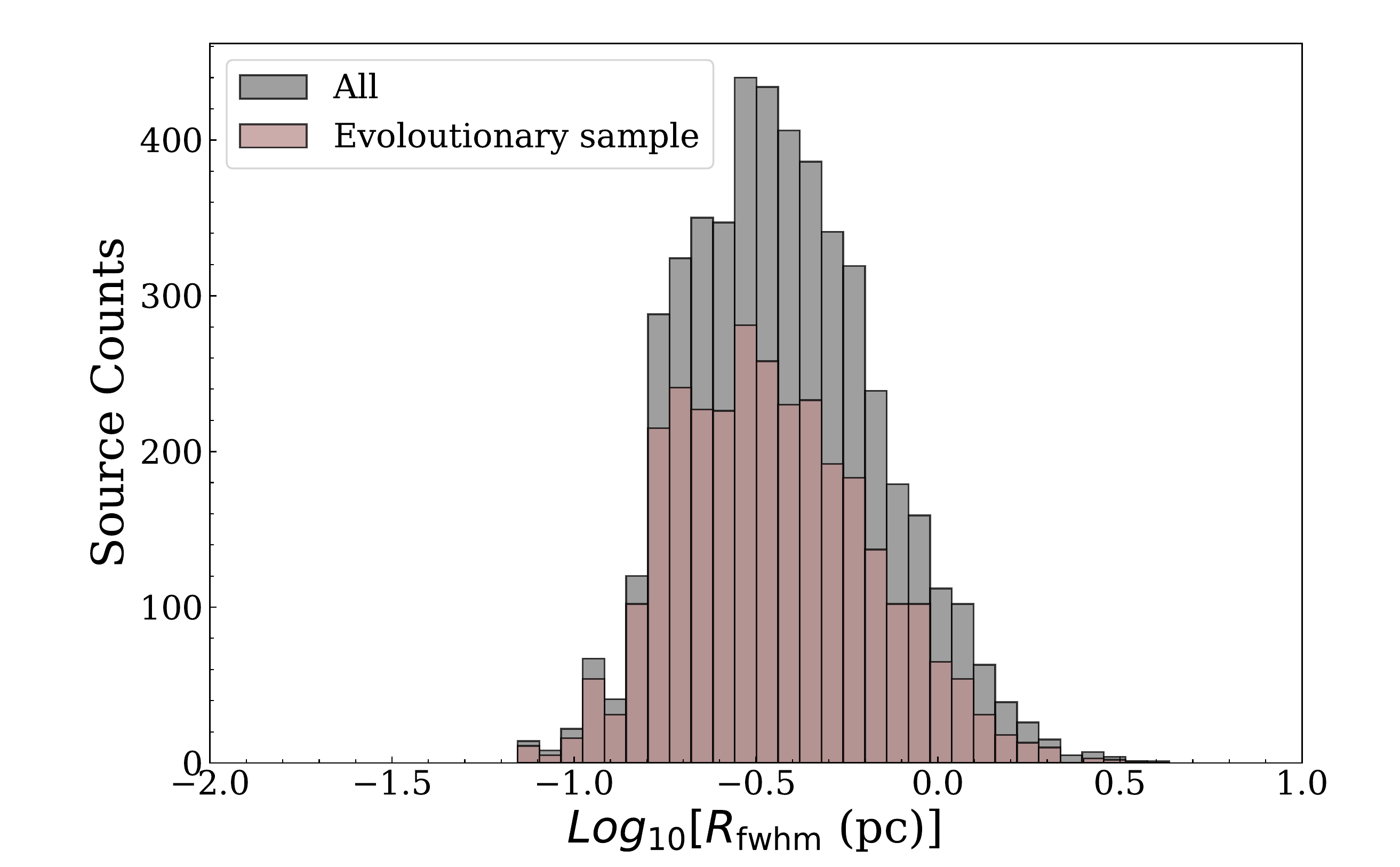}
   \includegraphics[width=.45\textwidth, trim= 0 0 0 0]{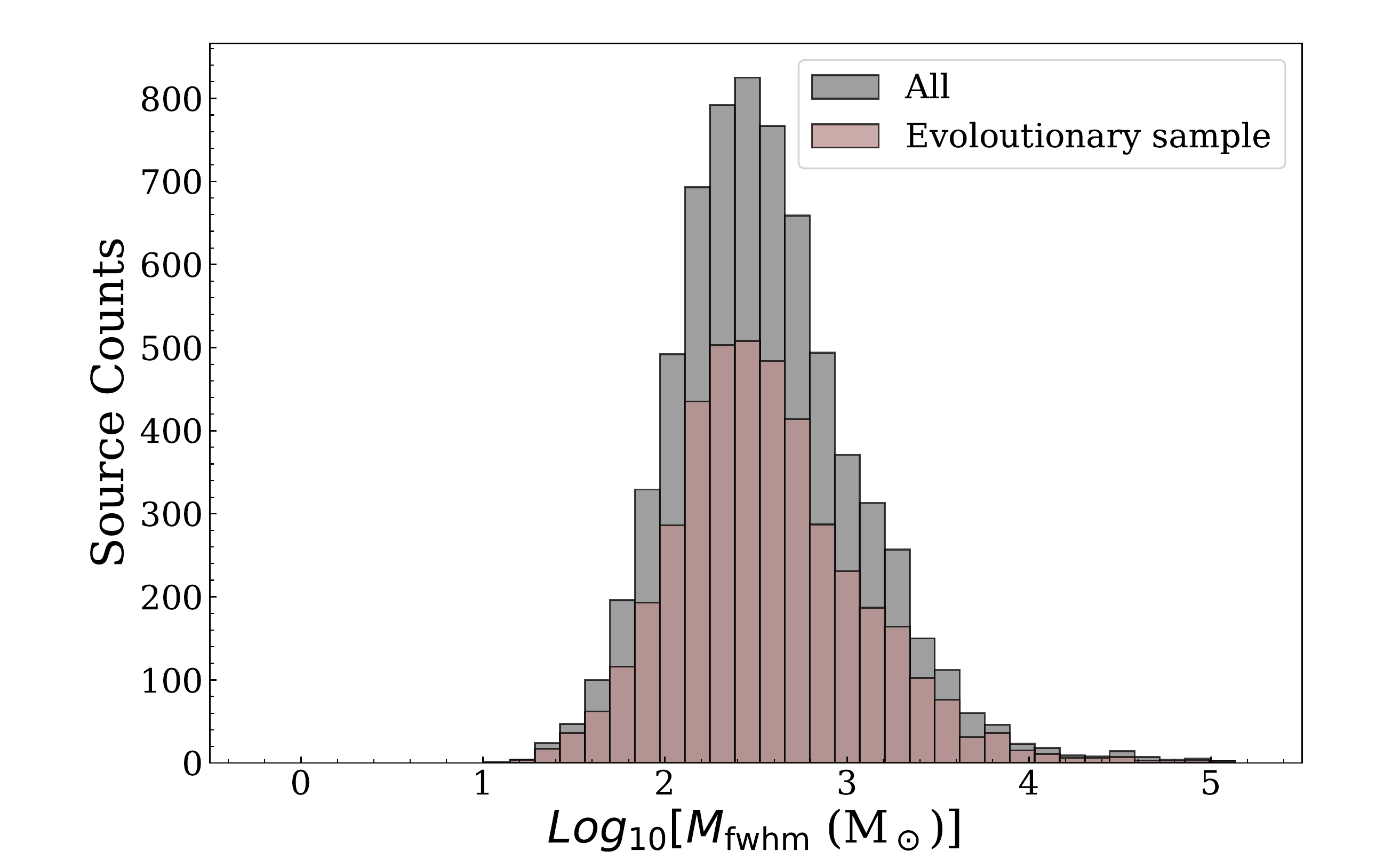}\\
     \includegraphics[width=.45\textwidth, trim= 0 0 0 0]{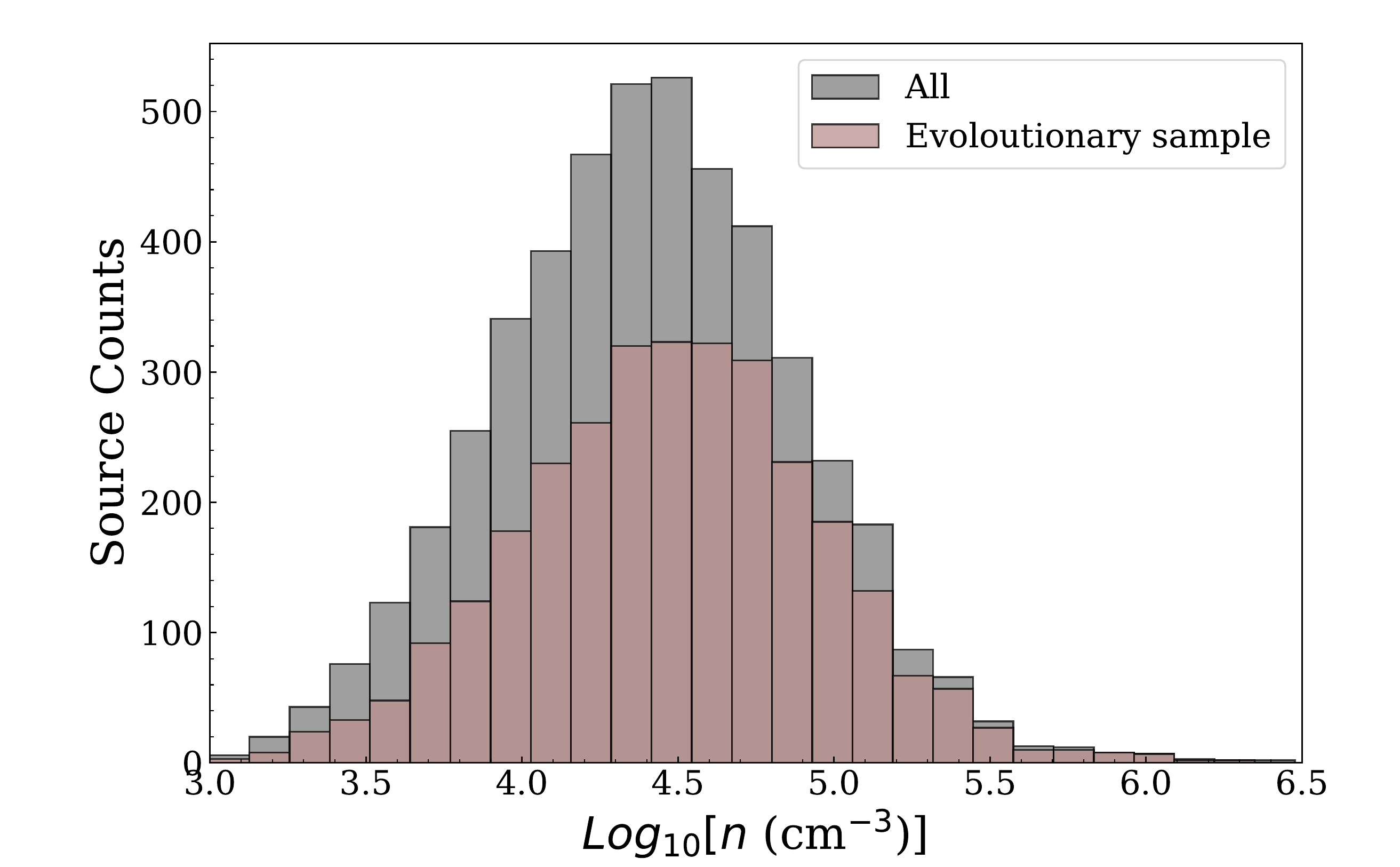}
   \includegraphics[width=.45\textwidth, trim= 0 0 0 0]{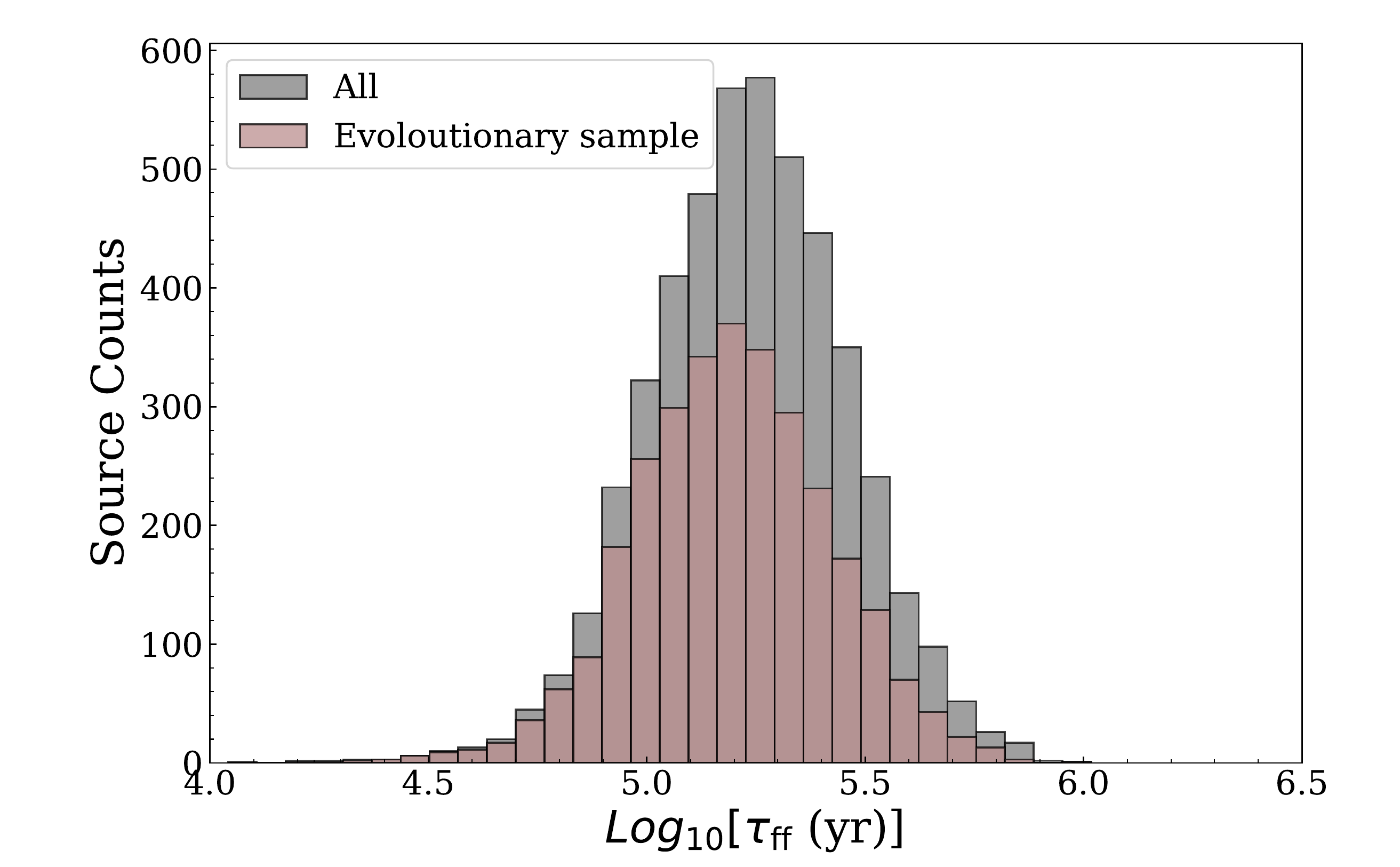}
\caption{Distributions of clump properties discussed in Sect.\,\ref{sect:physical_properties}, namely dust temperature and bolometric luminosity (top panels), FWHM size and mass  (middle panels), average density and free-fall time (bottom panels). The grey histograms show the distribution for the whole sample while the light-red shaded  histograms show the distribution for the clumps identified in one of four evolutionary stages that are described in Sect.\,\ref{sect:evolutionary_sequence}. We have excluded clumps within 2\,kpc of the Sun to avoid biasing the distribution to nearby low-mass and low-luminosity clumps. }

\label{fig:hist_properties}
\end{figure*}

We provide a full set of the physical properties described in this section in Table\,\ref{tbl:derived_clump_para}. Distances are available for 8130 clumps in the sample (96.7\,per\,cent) and masses, sizes and volume densities have been determined for 7992 (95\,per\,cent) clumps, 5696 (67.8\,per\,cent) clumps, and 5576 (66.2\,per\,cent) clumps, respectively. We show the distributions of the parameters described in the paragraphs above in Figure\,\ref{fig:hist_properties} and, for completeness, we also include the dust temperature and luminosity derived in \citealt{urquhart2018}. The grey histograms show the distribution of all clumps for which a value is available, while the light-red shaded histograms show the distribution of the clumps classified as being either quiescent, protostellar, YSO, or \hii\ regions. These groupings constitute our high-reliability ``evolutionary'' sample of clumps that are described in the next section. We present a summary of the properties of the clumps in Table\,\ref{tbl:summary_derived_para_all}.  We note that the clumps have sizes of $\sim$0.5\,pc and masses of $\sim$650\,\msun, and so are capable of forming a cluster of stars with at least one high-mass star (\citealt{csengeri2014}).

\setlength{\tabcolsep}{6pt}

\begin{table*}

\begin{center}\caption{Summary of physical properties. In Col.\,(2) we give the number of clumps in each subsample, in Cols.\,(3-5) we give the mean values, the error in the mean and the standard deviation, respectively, and in Cols.\,(6-7) we give the minimum and maximum values of the samples, respectively. }
\label{tbl:summary_derived_para_all}
\begin{minipage}{\linewidth}
\small
\begin{tabular}{lc......}
\hline \hline
  \multicolumn{1}{l}{Parameter}&  \multicolumn{1}{c}{\#}&	\multicolumn{1}{c}{$\bar{x}$}  &	\multicolumn{1}{c}{$\frac{\sigma}{\sqrt(N)}$} &\multicolumn{1}{c}{$\sigma$} &	
  \multicolumn{1}{c}{$x_{\rm{min}}$}& \multicolumn{1}{c}{$x_{\rm{max}}$}\\
\hline
T$_{\rm dust}$\,(K)	&	6853	&	19.416	&	0.069	&	5.678	&	7.9	&	56.1	\\
Log$_{10}$[$M_{\rm fwhm}$\,(M$_\odot$)]	&	6853	&	2.823	&	0.007	&	0.539	&	1.159	&	5.036	\\
Log$_{10}$[$L$\,(L$_\odot$)]	&	6853	&	3.077	&	0.012	&	0.971	&	0.552	&	6.906	\\
$R_{\rm fwhm}$\,(pc)	&	4859	&	0.459	&	0.005	&	0.338	&	0.07	&	4.31	\\
Log$_{10}$[$n$\,(cm$^{-3}$)]	&	4779	&	4.512	&	0.007	&	0.453	&	2.942	&	6.895	\\
$\tau_{\rm ff}$\,(Myr)	&	4779	&	0.195	&	0.002	&	0.105	&	0.011	&	1.041	\\
Log$_{10}$[$\Sigma$\,(M$_\odot$/pc$^2$)]	&	4708	&	1.837	&	0.006	&	0.441	&	0.219	&	4.072	\\
\hline\\
\end{tabular}\\

\end{minipage}

\end{center}
\end{table*}

\setlength{\tabcolsep}{6pt}

We present an updated version of the 30 most massive complexes located within the Solar circle in Table\,\ref{tbl:derived_cluster_para}. The content of this table was originally presented in Table\,8 of \citet{urquhart2018} and has been modified to reflect the changes in association membership, distances and mass measurements. The majority of the most prominent complexes are unchanged and this table is primarily included for completeness.

\setlength{\tabcolsep}{1.25pt}

\begin{table*}

\begin{center}\caption{Physical parameters for the thirty most massive complexes identified, ordered by their total dense-gas mass. In the first column we give the complex name, which is constructed from the mean longitude and latitude position of member clumps. In Column 2 we give the more common name found in the literature. In Columns 3-6 we give the number of clumps in the association, and the mean position and velocity of the associated members and their standard deviation. In Columns 7 and 8 we give the distance to the complex and the corresponding distance from the Galactic centre. In Column 8 we give the mean and standard deviation for the temperature determined by averaging over the associated clumps. In Columns 10 and 12 we give the total luminosity and mass of the clumps in each complex, while in Columns 11 and 13 we give the luminosity and mass as a percentage of total luminosity and mass in the disk (as determined in this paper). In the final column we give the luminosity to mass ratio determined by summing up the luminosity of all clumps and dividing it by the total mass of the associated clumps (i.e., dividing column 10 by column 12).}
\label{tbl:derived_cluster_para}
\begin{minipage}{\linewidth}
\small
\begin{tabular}{lcccccccccccccc}
\hline \hline
  \multicolumn{1}{c}{Complex}& \multicolumn{1}{c}{Literature} & \multicolumn{1}{c}{\# of}  &\multicolumn{1}{c}{$\ell$}  &\multicolumn{1}{c}{$b$}  &\multicolumn{1}{c}{\vlsr}  &\multicolumn{1}{c}{Distance}   
  &\multicolumn{1}{c}{$R_{\rm gc}$} 
  &	\multicolumn{1}{c}{T} &\multicolumn{1}{c}{Log($L_{\rm{bol}}$)}&\multicolumn{1}{c}{$L_{\rm{bol}}$} & \multicolumn{1}{c}{Log($M_{\rm{fwhm}}$)} &	\multicolumn{1}{c}{$M_{\rm{fwhm}}$} &\multicolumn{1}{c}{\lm} \\
  
  \multicolumn{1}{c}{name}& \multicolumn{1}{c}{name} &\multicolumn{1}{c}{members}  &\multicolumn{1}{c}{(\degr)}  &\multicolumn{1}{c}{(\degr)}  &\multicolumn{1}{c}{(\kms)}  &\multicolumn{1}{c}{(kpc)}   
  &\multicolumn{1}{c}{(kpc)}
  &	\multicolumn{1}{c}{(K)} &\multicolumn{1}{c}{(\lsun)}&\multicolumn{1}{c}{(\%)} & \multicolumn{1}{c}{(\msun)} &	\multicolumn{1}{c}{(\%)} &\multicolumn{1}{c}{(\lsun\,/\msun)} \\
  
 \multicolumn{1}{c}{(1)}& \multicolumn{1}{c}{(2)} &\multicolumn{1}{c}{(3)}  &\multicolumn{1}{c}{(4)}  &\multicolumn{1}{c}{(5)}  &\multicolumn{1}{c}{(6)}  &\multicolumn{1}{c}{(7)}   
  &\multicolumn{1}{c}{(8)}
  &	\multicolumn{1}{c}{(9)} &\multicolumn{1}{c}{(10)}&\multicolumn{1}{c}{(11)} & \multicolumn{1}{c}{(12)} &	\multicolumn{1}{c}{(13)} &\multicolumn{1}{c}{(14)} \\
\hline
G030.653-00.017	&	W43	&	250	&	30.65$\pm$0.47	&	-0.02$\pm$0.17	&	97.5$\pm$9.0	&	4.90	&	4.83	&	19.9$\pm$6.1	&	6.56	&	2.91	&	5.10	&	1.83	&	29.12	\\
G043.141-00.018	&	W49	&	14	&	43.14$\pm$0.05	&	-0.02$\pm$0.03	&	9.5$\pm$4.0	&	11.10	&	7.60	&	25.6$\pm$4.6	&	7.18	&	12.02	&	5.04	&	1.60	&	137.95	\\
G049.261-00.318	&	W51	&	93	&	49.26$\pm$0.22	&	-0.32$\pm$0.10	&	60.9$\pm$6.9	&	5.20	&	6.33	&	22.3$\pm$5.1	&	7.04	&	8.72	&	5.01	&	1.50	&	106.82	\\
G336.996-00.002	&		&	107	&	337.00$\pm$0.19	&	-0.00$\pm$0.14	&	-73.9$\pm$4.6	&	6.36	&	4.51	&	21.6$\pm$6.1	&	6.51	&	2.60	&	4.99	&	1.42	&	33.63	\\
G024.322+00.187	&		&	61	&	24.32$\pm$0.16	&	0.19$\pm$0.09	&	113.3$\pm$3.9	&	7.60	&	3.44	&	17.9$\pm$5.4	&	5.73	&	0.43	&	4.89	&	1.13	&	6.90	\\
G333.121-00.353	&	G333	&	245	&	333.12$\pm$0.44	&	-0.35$\pm$0.23	&	-50.8$\pm$4.3	&	3.60	&	5.41	&	21.5$\pm$5.7	&	6.52	&	2.64	&	4.82	&	0.96	&	50.41	\\
G033.527-00.011	&		&	26	&	33.53$\pm$0.19	&	-0.01$\pm$0.03	&	103.3$\pm$2.2	&	8.80	&	4.96	&	18.1$\pm$5.5	&	5.47	&	0.24	&	4.79	&	0.91	&	4.78	\\
G305.453+00.065	&	G305	&	102	&	305.45$\pm$0.27	&	0.06$\pm$0.20	&	-37.5$\pm$3.9	&	4.00	&	6.85	&	22.9$\pm$6.4	&	6.34	&	1.74	&	4.76	&	0.84	&	38.33	\\
G035.569+00.015	&	G35.58-0.03	&	15	&	35.57$\pm$0.05	&	0.01$\pm$0.08	&	52.4$\pm$3.2	&	10.40	&	6.06	&	21.2$\pm$4.8	&	5.69	&	0.38	&	4.57	&	0.54	&	13.18	\\
G331.389-00.106	&		&	87	&	331.39$\pm$0.28	&	-0.11$\pm$0.12	&	-89.0$\pm$5.0	&	3.90	&	5.27	&	21.6$\pm$6.1	&	6.24	&	1.37	&	4.50	&	0.46	&	54.70	\\
G336.469-00.166	&		&	29	&	336.47$\pm$0.08	&	-0.17$\pm$0.09	&	-84.5$\pm$5.2	&	10.20	&	4.21	&	23.6$\pm$2.6	&	6.20	&	1.25	&	4.47	&	0.43	&	53.42	\\
Bania Clump 1	&		&	17	&	354.75$\pm$0.08	&	0.32$\pm$0.06	&	89.9$\pm$9.8	&	8.40	&	0.76	&	16.6$\pm$5.4	&	5.40	&	0.20	&	4.46	&	0.42	&	8.81	\\
G023.392-00.166	&	W41	&	35	&	23.39$\pm$0.07	&	-0.17$\pm$0.08	&	99.3$\pm$3.4	&	5.90	&	3.76	&	20.6$\pm$6.3	&	5.99	&	0.78	&	4.45	&	0.42	&	34.44	\\
G020.737-00.089	&		&	17	&	20.74$\pm$0.04	&	-0.09$\pm$0.06	&	56.6$\pm$1.8	&	11.70	&	4.86	&	23.1$\pm$3.7	&	5.76	&	0.46	&	4.44	&	0.40	&	21.12	\\
G023.072-00.349	&	W41	&	41	&	23.07$\pm$0.23	&	-0.35$\pm$0.10	&	73.8$\pm$7.3	&	5.00	&	4.23	&	16.9$\pm$4.1	&	5.10	&	0.10	&	4.42	&	0.38	&	4.83	\\
G331.094-00.414	&		&	53	&	331.09$\pm$0.21	&	-0.41$\pm$0.06	&	-65.8$\pm$2.2	&	4.00	&	5.23	&	20.0$\pm$6.2	&	5.82	&	0.52	&	4.39	&	0.36	&	26.94	\\
G006.183-00.358	&	W28	&	78	&	6.18$\pm$0.29	&	-0.36$\pm$0.17	&	12.6$\pm$4.9	&	3.00	&	5.38	&	18.4$\pm$4.6	&	5.67	&	0.37	&	4.37	&	0.34	&	20.18	\\
G336.951+00.000	&		&	30	&	336.95$\pm$0.16	&	0.00$\pm$0.08	&	-118.3$\pm$3.4	&	7.70	&	3.27	&	21.6$\pm$4.9	&	5.86	&	0.58	&	4.36	&	0.33	&	31.74	\\
G037.684-00.209	&	W47	&	18	&	37.68$\pm$0.18	&	-0.21$\pm$0.13	&	58.0$\pm$6.5	&	9.80	&	6.03	&	23.9$\pm$4.2	&	5.99	&	0.77	&	4.34	&	0.32	&	44.10	\\
G045.486+00.071	&		&	13	&	45.49$\pm$0.05	&	0.07$\pm$0.06	&	58.6$\pm$2.2	&	7.70	&	6.23	&	23.1$\pm$7.0	&	6.17	&	1.17	&	4.32	&	0.31	&	69.99	\\
G019.627-00.110	&		&	14	&	19.63$\pm$0.06	&	-0.11$\pm$0.03	&	59.4$\pm$2.8	&	11.60	&	4.63	&	21.6$\pm$5.1	&	5.53	&	0.27	&	4.32	&	0.31	&	16.29	\\
G008.581-00.329	&	W30 	&	38	&	8.58$\pm$0.17	&	-0.33$\pm$0.05	&	37.1$\pm$1.3	&	4.40	&	4.00	&	17.4$\pm$5.8	&	5.36	&	0.18	&	4.32	&	0.31	&	10.94	\\
G018.936-00.350	&	W39	&	58	&	18.94$\pm$0.12	&	-0.35$\pm$0.20	&	63.8$\pm$3.4	&	4.00	&	4.75	&	19.3$\pm$4.8	&	5.41	&	0.20	&	4.30	&	0.29	&	12.84	\\
G321.082-00.516	&		&	13	&	321.08$\pm$0.06	&	-0.52$\pm$0.03	&	-60.8$\pm$0.7	&	9.30	&	5.93	&	24.0$\pm$4.0	&	5.73	&	0.42	&	4.30	&	0.29	&	26.88	\\
G024.049+00.495	&		&	31	&	24.05$\pm$0.27	&	0.50$\pm$0.06	&	96.5$\pm$3.7	&	5.80	&	3.85	&	16.9$\pm$4.5	&	5.72	&	0.41	&	4.27	&	0.27	&	27.96	\\
G010.903-00.068	&	Snake	&	35	&	10.90$\pm$0.19	&	-0.07$\pm$0.11	&	25.7$\pm$4.9	&	4.10	&	4.39	&	15.8$\pm$6.2	&	4.72	&	0.04	&	4.26	&	0.27	&	2.88	\\
G340.248-00.273	&		&	61	&	340.25$\pm$0.13	&	-0.27$\pm$0.13	&	-49.0$\pm$3.5	&	3.60	&	5.14	&	17.7$\pm$4.7	&	5.46	&	0.23	&	4.26	&	0.26	&	16.01	\\
G332.124-00.070	&		&	67	&	332.12$\pm$0.30	&	-0.07$\pm$0.12	&	-49.2$\pm$1.8	&	3.10	&	5.79	&	17.7$\pm$5.7	&	5.00	&	0.08	&	4.25	&	0.26	&	5.71	\\
G350.216+00.082	&		&	33	&	350.22$\pm$0.08	&	0.08$\pm$0.05	&	-65.5$\pm$5.2	&	6.00	&	2.67	&	20.4$\pm$4.5	&	5.99	&	0.77	&	4.24	&	0.25	&	55.53	\\
G012.960-00.168	&	W33	&	85	&	12.96$\pm$0.26	&	-0.17$\pm$0.15	&	37.8$\pm$7.6	&	2.60	&	5.85	&	16.3$\pm$4.2	&	4.98	&	0.08	&	4.24	&	0.25	&	5.50	\\

\hline\\
\end{tabular}\\
Notes: A more complete table that includes parameters for all 877 is available in electronic form at the CDS via anonymous ftp to cdsarc.u-strasbg.fr (130.79.125.5) or via http://cdsweb.u-strasbg.fr/cgi-bin/qcat?J/MNRAS/.
\end{minipage}

\end{center}
\end{table*}
\setlength{\tabcolsep}{6pt}

\section{Evolutionary Sequence}
\label{sect:evolutionary_sequence}

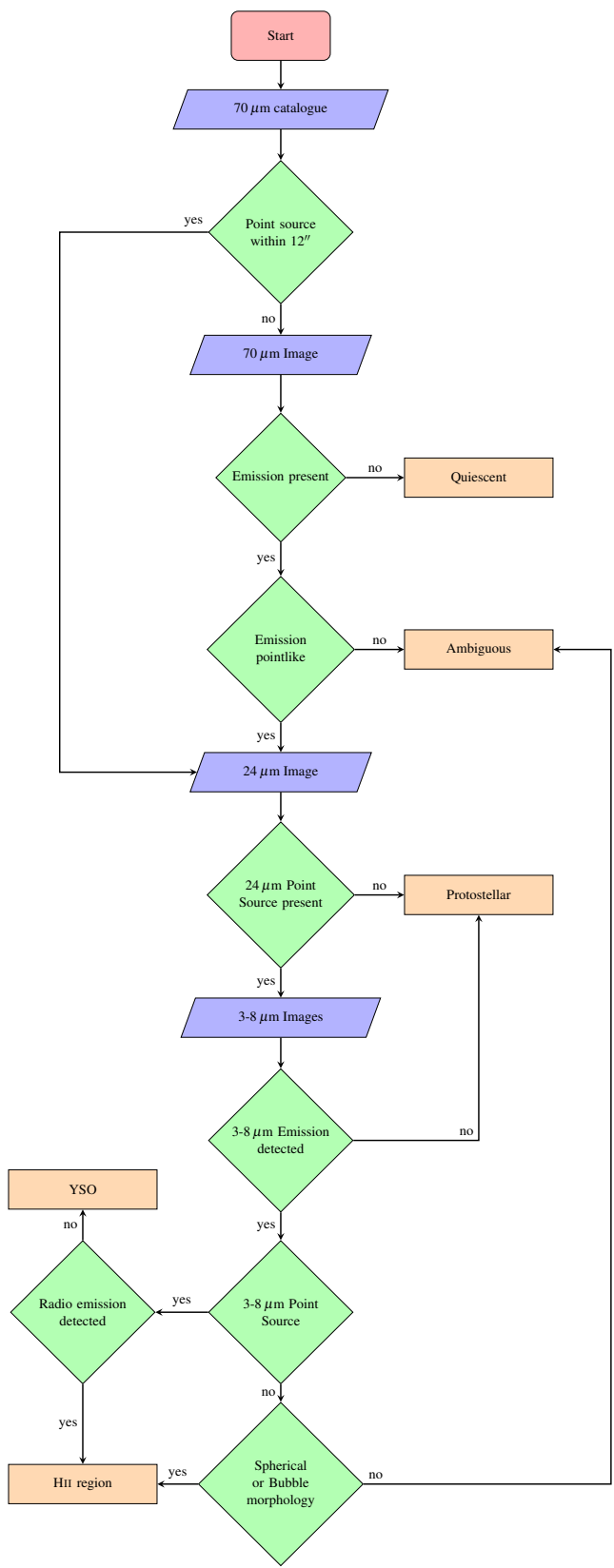
\begin{figure}
\centering
\resizebox{0.35\textheight}{!}{%
\begin{tikzpicture}[node distance=2cm]
\node (start) [startstop] {Start};

\node (in0) [io, below of=start, yshift=0.5cm] {70\,\mum\ catalogue};
\node (dec0) [decision, below of=in0, yshift=-0.5cm] {Point source within 12\arcsec};

\node (in1) [io, below of=dec0, yshift=-0.5cm] {70\,\mum\ Image};
\node (dec1) [decision, below of=in1, yshift=-0.5cm] {Emission present};

\node (pro1) [process, right of=dec1 , xshift=2.cm] {Quiescent};

\node (dec2) [decision, below of=dec1, yshift=-1.5cm] {Emission pointlike};

\node (pro2) [process, right of=dec2 , xshift=2.cm] {Ambiguous};

\node (in2) [io, below of=dec2, yshift=-0.5cm] {24\,\mum\ Image};
\node (dec3) [decision, below of=in2, yshift=-0.5cm] {24\,\mum\ Point Source present};

\node (pro3) [process, right of=dec3 , xshift=2.cm] {Protostellar};
\node (in3) [io, below of=dec3, yshift=-0.5cm] {3-8\,\mum\ Images};
\node (dec4) [decision, below of=in3, yshift=-0.5cm] {3-8\,\mum\ Emission detected};
\node (dec8) [decision, below of=dec4, yshift=-1.5cm] {3-8\,\mum\ Point Source};
\node (dec6) [decision, left of=dec8,xshift=-2.cm] {Radio emission detected};

\node (pro4) [process, above of=dec6 , yshift=0.5cm] {YSO};
\node (dec5) [decision, below of=dec8, yshift=-1.5cm] {Spherical or Bubble morphology};

\node (pro5) [process, left of=dec5 , xshift=-2.cm] {\hii\ region};

\draw [arrow] (start) --  (in0);
\draw [arrow] (in0) --  (dec0);
\draw [arrow] (in1) --  (dec1);
\draw [arrow] (dec0.west)  node[above left] {yes} -- + (-3.0,0) |- (in2);
\draw [arrow] (dec0) -- node[anchor=east] {no} (in1);
\draw [arrow] (dec1) -- node[anchor=east] {yes} (dec2);
\draw [arrow] (dec1) -- node[anchor=south] {no} (pro1);
\draw [arrow] (dec2) -- node[anchor=south] {no} (pro2);
\draw [arrow] (dec2) -- node[anchor=east] {yes} (in2);
\draw [arrow] (in2) --  (dec3);
\draw [arrow] (dec3) -- node[anchor=south] {no} (pro3);
\draw [arrow] (dec3) -- node[anchor=east] {yes} (in3);
\draw [arrow] (in3) --  (dec4);
\draw [arrow] (dec8) -- node[anchor=south] {yes} (dec6);
\draw [arrow] (dec4) -- node[anchor=east] {yes} (dec8);
\draw [arrow] (dec8) -- node[anchor=east] {no} (dec5);
\draw [arrow] (dec5) -- node[anchor=south] {yes} (pro5);
\draw [arrow] (dec6) -- node[anchor=east] {no} (pro4);
\draw [arrow] (dec6) -- node[anchor=east] {yes} (pro5);
\draw [arrow] (dec4) -| node[above left] {no} (pro3);

\draw [arrow] (dec5.east) node[above right] {no}  -- +  (5.0,0) |- (pro2);


\end{tikzpicture}

}
\caption{Flowchart showing the sequence of steps taken and the images used to classify the ATLASGAL clumps into one of five categories. }
\label{fig:flowchart}

\end{figure}

\begin{figure*}
    \centering
     \includegraphics[width=.95\textwidth]{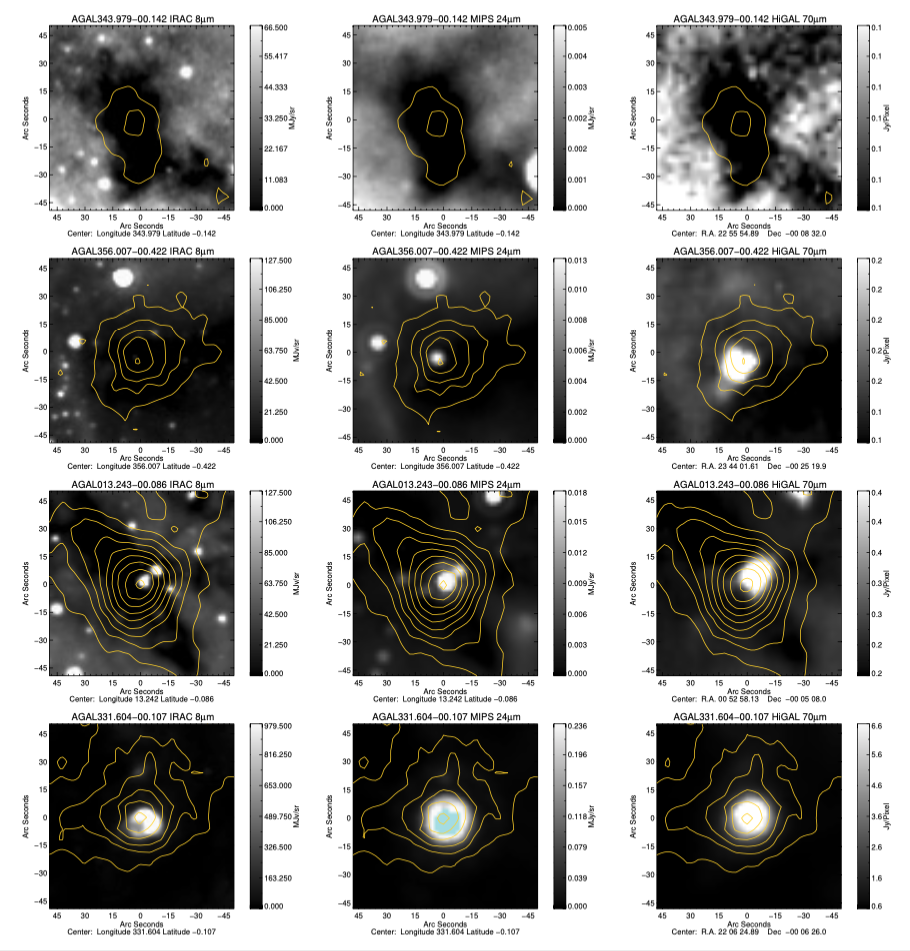}

	\caption{Examples of the 8-\mum\ IRAC, 24-\mum\ MIPS and 70-\mum\ HiGAL images used in the classification process. From top-down we show an example of a quiescent, protostellar, YSO, and \hii\ region clump. The contours show the distribution of the 870-\mum\ dust emission traced by ATLASGAL. These start at 3$\sigma$ and increase in steps determined using the dynamic range power-law fitting scheme where $D = N^i +3$ where $D$ is the dynamic range, $S_{\rm peak}/\sigma$, $N$ is the number of contours, in this case 8, and $i$ is the power-law that determines the separation between consecutive contours. This approach is a modified version of a scheme developed by \citealt{thompson2006}. The light-blue coloured region seen in the lower middle panel indicates the presence of saturated pixels. }

	\label{fig:classification_images}
\end{figure*}

\subsection{Classification scheme}
\label{sect:classification_scheme}

The large number of sources in the catalogue previously inhibited our ability to visually inspect the multi-wavelength data to classify every clump reliably. To date, only a relatively modest fraction of the clumps has been reliably classified through visual inspection ($\sim$100; \citealt{konig2017}) or by cross-matching with reliable \hii\ region and YSO catalogues ($\sim$1000; \citealt{urquhart2013_cornish, urquhart2014_atlas}) such as the Red MSX survey (RMS; \citealt{lumsden2013})\footnote{The RMS catalogue has produced the largest and most reliable catalogue of \hii\ regions and massive-YSOs to date.}, and the CORNISH survey (\citealt{hoare2012,purcell2013}). The majority of the rest of the catalogue, therefore, has been automatically classified into one of three categories (quiescent, protostellar or YSO) depending on their spectral energy distributions (\citealt{urquhart2018}). YSOs and \hii\ regions have very similar SEDs and so, where available, radio emission was used to distinguish between these two evolutionary stages (e.g., \citealt{becker1994, urquhart2007_atca, urquhart2009_vla,purcell2013,medina2019}).

Although these automatic classifications are useful for deriving statistical properties, they are not always reliable indicators of the various evolutionary stages, particularly towards intense star-forming complexes where clumps in different evolutionary stages are often found in close proximity to each other. Furthermore, as the brightness of the dust emission traced by ATLASGAL is a function of both the column density and the temperature, a detection can indicate a combination of high column density and low temperatures or vice versa. The catalogue is therefore also likely to contain sources with significant emission from warm diffuse gas and photon-dominated regions (PDRs) that are often found in star forming regions. These kinds of sources are not part of the evolutionary sequence and need to be removed to avoid unduly biasing the derived properties of any particular stage, and the only way this exclusion can be reliably achieved is through visual inspection of multi-wavelength images.

The MALT90 team (\citealt{jackson2013}) used mid-infrared images (3-24\,\mum) from the GLIMPSE (\citealt{churchwell2009}) and MIPSGAL (\citealt{carey2009}) surveys to classify $\sim$3000 clumps into four categories (i.e., PDR, \hii\ region, protostellar, and quiescent; \citealt{rathborne2016}). Our classification scheme follows a similar set of criteria but also takes advantage of the availability of the 70-\mum\ images and catalogue that are now available from HiGAL (\citealt{Molinari2010, molinari2016}). These data are indispensable in revealing embedded protostars at longer wavelengths in clumps that may be missed, even at 24\,\mum. This scheme has been applied to all 8417 clumps located outside the Galactic centre region, regardless of whether they satisfy the FWHM criterion discussed in the previous section.

The SEDs of quiescent and star-forming clumps can be described as a modified version of blackbody radiation, which is referred to as a greybody. The temperatures of quiescent clumps are $\sim$10-15\,K and, at these temperatures and with no internal heating source, these clumps are dark at wavelengths of 70\,\mum\ or less. The formation of a protostellar object inside a clump results in an increase in the global temperature of the clump due to accretion onto the embedded source. This luminosity warms the envelope and therefore shifts the peak of the SED to shorter wavelengths.  As a result, the protostar becomes detectable at 70\,\mum, but not at mid-infrared wavelengths (i.e., $\leq$24\,\mum). As the protostar continues to accrete material, it will increase the global temperature of the clump and will start to appear first at mid-infrared and then near-infrared wavelengths. At this stage, we refer to the embedded protostar as a young stellar object (YSO). For a low- or intermediate-mass star, accretion will end and the YSO will continue to contract and disperse its circumstellar envelope and eventually join the main sequence. High-mass stars, however, produce a significant amount of their flux at UV wavelengths, which ionizes their local environments.  This ionization creates a rapidly expanding bubble of hot, ionized gas, an \hii\ region that can be easily distinguished by its compact radio emission and mid-infrared morphology.

Given the gradual changes in the SED as the embedded object first appears at 70\,\mum\ and then evolves towards the main sequence, it is possible to use a carefully selected set of far- and mid-infrared wavelength images (3-8\,\mum, 24\,\mum\ and 70\,\mum) to initially distinguish between quiescent and star-forming clumps and, for the latter, to classify the embedded objects (e.g., protostellar, YSO, or \hii\ region). This method has been successfully applied by \citet{konig2017} to a small subsample of ATLASGAL clumps that had been specially selected to include examples of the four evolutionary stages discussed in this paragraph.

Figure\,\ref{fig:flowchart} illustrates the sequence of steps used to classify the clumps into one of five categories. This scheme is similar to the steps described by \citet{konig2017} with one small modification. We previously distinguished between protostars and YSOs using a $\sim$22\,\mum\ flux threshold of 2.7\,Jy\footnote{This was the sensitivity limit of the MSX survey (\citealt{price2001}) at 21\,\mum\ and was used to identify massive YSOs by the RMS survey team (\citealt{lumsden2013}).}, but this is somewhat arbitrary given the improved sensitivities of WISE (\citealt{Wright2010}) and MIPSGAL (\citealt{carey2009}) and so we have relaxed this criterion and simply use the presence or absence of an 8\,\mum\ point source to distinguish between a YSO and protostar. 

We have applied this set of criteria to GLIMPSE 3--8-\mum, MIPSGAL 24-\mum, and Hi-GAL 70-\mum\ postage-stamp images of all of the ATLASGAL sources for which the data are available (i.e., $|b| < 1\degr$) to produce reliable samples of the different evolutionary stages. The assignment of evolutionary stages described above works well when the regions are relatively isolated and are dominated by a single infrared object. Given the sizes and masses of these clumps ($\sim$0.5\,pc and $\sim$500\,\msun), however, it is reasonable to assume that these will all form clusters and so we should expect to find different evolutionary stages present in the same clump. In these cases, we classify the clump evolutionary stage according to the most evolved embedded object, assuming that it will dominate the clump's physical properties. 

Below, we provide a brief description of the six of the most common categories identified using the 8-\mum\, 24-\mum\ and 70-\mum. The first four of these represent an evolutionary sequence (see  Fig.\,\ref{fig:classification_images} for examples).

\begin{itemize}

    \item {\bf Quiescent:} These clumps are cold (10-15\,K), devoid of any embedded objects and dark at 70\,\mum\ (see upper panel of Fig.\,\ref{fig:classification_images} for an example). A first step was to cross-match the ATLASGAL clumps with the Hi-GAL 70\,\mum\ point source catalogue (\citealt{elia2017}) and exclude any clumps where a match was found within 12\arcsec\ of the submm emission peak.\footnote{We have used a radius of 12\arcsec\ for determining associations with other catalogues as this is three times the positional uncertainty of the ATLASGAL survey \citep{schuller2009_full}.} The rationale for this criterion is that these clumps are already undergoing star formation. This step reduced the sample that needed to be inspected by a factor of two. The next step was to inspect the 70-\mum\ maps to ensure that the central part of the clumps (as bounded by the 50-per-cent contour of the 870-\mum\ emission) is clear of any large-scale extended 70-\mum\ emission that might hide an embedded point source.    \\
    
    \item {\bf Protostellar:} This stage is the earliest in the star-formation process when the protostar begins to warm its natal clump and becomes visible at 70\,\mum. A comparison with the HiGAL 70-\mum\ catalogue revealed the presence of a 70-\mum\ point source located within 12\arcsec\ of the centre of the clump, causing them to be immediately flagged as being associated with star formation. If these sources have no counterpart at 3--8\,\mum\ within 12\arcsec\ of the centre of the clump, they are classified as protostellar (see upper-middle panel of Fig.\,\ref{fig:classification_images} for an example). We initially made a distinction between 70-\mum\ sources that had a 24-\mum\ counterpart and those that did not.  We found no significant difference in the physical properties of these two groups  (i.e., dust temperature, luminosity and luminosity-to-mass ratio, all of which are considered to be good evolutionary diagnostics) and so decided that the distinction was not useful and merged them.\\

     \item {\bf YSO:} At this point in the protostar's evolution, it has warmed its natal clump sufficiently enough to be visible at both 3--8\,\mum\ and 24\,\mum\ (see lower-middle panel of Fig.\,\ref{fig:classification_images} for an example). The protostar and YSOs are much smaller than the resolution of the GLIMPSE image ($\sim$2\arcsec) and so the emission at all mid-infrared wavelengths will be unresolved (\citealt{mottram2007, de-buizer2005}). \\
    
    \item {\bf \hii\ region:} These are the most evolved sources in  embedded stages investigated and as such will be the warmest and most clearly observable at all of the wavelengths inspected. Due to the fact that they are surrounded by a rapidly expanding ionized nebula, we would expect all but the very youngest of these regions to be extended in the 3--8\,\mum\ images and many will be clearly associated with bubbles (see lower panel of Fig.\,\ref{fig:classification_images} for an example). The only reliable way of identifying the youngest (hypercompact and ultracompact) \hii\ regions is to search for a correlation between the peak of the submillimetre emission and compact radio emission at frequencies of $\sim$5\,GHz. Coincidence of radio emission with the extended  3--8\,\mum\ is also useful in confirming that they are indeed \hii\ regions. Our analysis has identified two populations of \hii\ regions, which we refer to as radio loud and radio quiet. Comparing the properties of these two types of \hii\ regions, we noticed that the former are warmer and more luminous and consequently have  higher luminosity-to-mass ratios. Looking at the luminosity-mass plot shown in Fig.\,\ref{fig:lum_mass_dist_HII}, however,  we see that the two samples form a continuous distribution in this parameter space. We note that the radio-loud \hii\ regions are almost an order of magnitude more luminous than the radio-quiet \hii\ regions. This difference suggests that, while all of these \hii\ regions are associated with high-mass stars, the radio-quiet \hii\ regions are primarily associated with early B-type stars, based on their luminosities seen in Fig.\,\ref{fig:lum_mass_dist_HII} ($M_\star < 20$\,\msun). For lower mass stars,  while the intensity of the radio emission quickly drops below the detection limit of the radio surveys. \\ 
    
\begin{figure}
    \centering
  
    	\includegraphics[width=.45\textwidth, trim= 0 0 0 0]{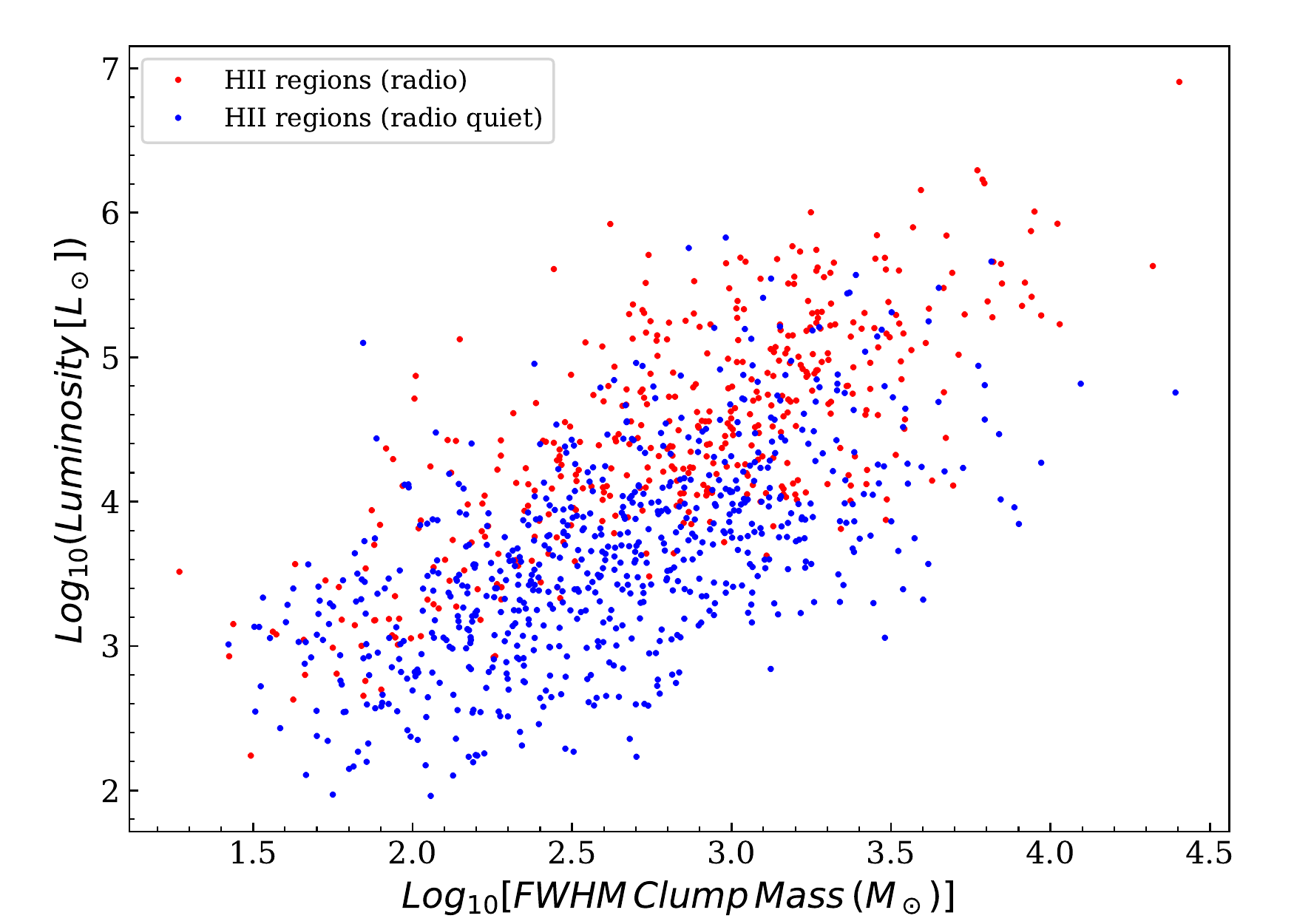}

		\caption{Correlation between luminosity and mass for all \hii\ regions identified in the sample. The dashed green line shows the results of a linear least-squares fit to the distance limited sample of \hii\ regions (i.e., distances between 2 and 4\,kpc).}

		\label{fig:lum_mass_dist_HII}
\end{figure}

    \item {\bf Ambiguous:} In many cases, the ATLASGAL clump is coincident with significant amounts of extended mid-infrared emission across most of one or more images, and no distinguishable embedded point source is seen.  These sources are then classified as `ambiguous.' This appearance is indicative of regions of warm diffuse gas that may obscure the presence of any protostellar objects at a similar or cooler temperature. There a few different kinds of sources that fall into this category such as clumps associated with PDRs found around the edges of evolved \hii\ regions, and star formation complexes where many sub-regions at different stages of evolution are located in close proximity to each other. This ambiguity  is not to say that many of these regions are not associated with star formation, but that it is impossible to estimate the fraction that are so associated, or their evolutionary state. It is, therefore, not possible to make a reliable determination about the current level of star formation in these clumps and so they are excluded from the statistical analysis that follows. In total, 3395 clumps have been classified as ambiguous.\\

  
    \item {\bf Planetary nebulae (PNe):} These form during the latest stages in the life cycle of lower-mass evolved stars when they are shedding their outer layers. These objects are therefore enshrouded in an envelope of hot dust that can be detected in the ATLASGAL survey. PNe are the only unresolved sources detected in the ATLASGAL catalogue and appear to be isolated and featureless in mid-infrared images (star-forming regions are nearly always associated with dust lanes and/or nebulosity) and their SEDs peak at mid-infrared wavelengths. Only four of these have been detected in ATLASGAL. Given their rarity, these are not shown explicitly in the flowchart presented in Fig.\,\ref{fig:flowchart}.  \\

    \end{itemize}

\setlength{\tabcolsep}{6pt}

\begin{table*}

\begin{center}\caption{Summary of mean and standard deviations for the clump properties by evolutionary stage.}
\label{tbl:properties_by_source_type}
\begin{minipage}{\linewidth}
\small
\begin{tabular}{lcccccccc}
\hline \hline
  \multicolumn{1}{l}{Classification}&  
  \multicolumn{1}{c}{Number}&
  \multicolumn{1}{c}{$T_{\rm dust}$}   &
  \multicolumn{1}{c}{Distance}   &
  \multicolumn{1}{c}{$R_{\rm fwhm}$}   &	
  \multicolumn{1}{c}{Log[$M_{\rm fwhm}$]} &
  \multicolumn{1}{c}{Log[$n_{\rm H_2}$]} &
   \multicolumn{1}{c}{Log[$L_{\rm bol}$]} &
  \multicolumn{1}{c}{Log[$L_{\rm{bol}}$/$M_{\rm{fwhm}}$]}\\

    \multicolumn{1}{c}{type }&  
      \multicolumn{1}{c}{}&  
\multicolumn{1}{c}{(K)} &
    \multicolumn{1}{c}{(kpc)}&
    \multicolumn{1}{c}{(pc)}&
    \multicolumn{1}{c}{(\msun)} &
    \multicolumn{1}{c}{(cm$^{-3}$)} &
    \multicolumn{1}{c}{(\lsun)} &
    \multicolumn{1}{c}{(\lsun\,\msun$^{-1}$)} \\
\hline
\multicolumn{9}{c}{Full sample}\\
\hline
Quiescent 	&	1218	&	13.8$\pm$2.56	&	4.4$\pm$2.78	&	0.43$\pm$0.35	&	2.511	$\pm$	0.515	&	4.537	$\pm$	0.449	&	2.09$\pm$0.76	&	-0.42	$\pm$	0.59	\\
Protostellar	&	1010	&	15.7$\pm$3.46	&	4.5$\pm$3.13	&	0.34$\pm$0.25	&	2.408	$\pm$	0.558	&	4.702	$\pm$	0.492	&	2.42$\pm$0.88	&	0.02	$\pm$	0.70	\\
YSO	&	1543	&	18.4$\pm$3.89	&	5.1$\pm$3.64	&	0.35$\pm$0.28	&	2.350	$\pm$	0.627	&	4.614	$\pm$	0.509	&	2.86$\pm$0.85	&	0.51	$\pm$	0.65	\\
HII Region	&	1236	&	24.4$\pm$4.87	&	6.5$\pm$4.33	&	0.46$\pm$0.34	&	2.517	$\pm$	0.687	&	4.445	$\pm$	0.574	&	3.82$\pm$0.95	&	1.30	$\pm$	0.64	\\
\hline
\multicolumn{9}{c}{Distance limited sample (2-4\,kpc)}\\
\hline
Quiescent		&	572	&	13.7$\pm$2.74	&	3.1$\pm$0.49	&	0.30$\pm$0.11	&	2.384	$\pm$	0.291	&	4.660	$\pm$	0.336	&	1.92$\pm$0.60	&	-0.47$\pm$	0.60	\\
Protostellar		&	457	&	15.5$\pm$3.55	&	3.1$\pm$0.50	&	0.25$\pm$0.09	&	2.301	$\pm$	0.334	&	4.790	$\pm$	0.337	&	2.26$\pm$0.71	&	-0.04$\pm$	0.71	\\
YSO		&	626	&	17.9$\pm$3.84	&	3.1$\pm$0.52	&	0.23$\pm$0.09	&	2.189	$\pm$	0.348	&	4.780	$\pm$	0.367	&	2.63$\pm$0.67	&	0.44$\pm$	0.66	\\
HII Region		&	373	&	24.5$\pm$5.50	&	3.2$\pm$0.52	&	0.24$\pm$0.09	&	2.166	$\pm$	0.373	&	4.698	$\pm$	0.463	&	3.49$\pm$0.80	&	1.33$\pm$	0.66	\\
\hline

\hline\\
\end{tabular}\\

\end{minipage}

\end{center}
\end{table*}
\setlength{\tabcolsep}{6pt}

\subsection{Classification Statistics for Embedded Objects}
\label{sect:class_stats}

\begin{figure}
    \centering
    \includegraphics[width=.45\textwidth, trim= 50 30 50 10,clip]{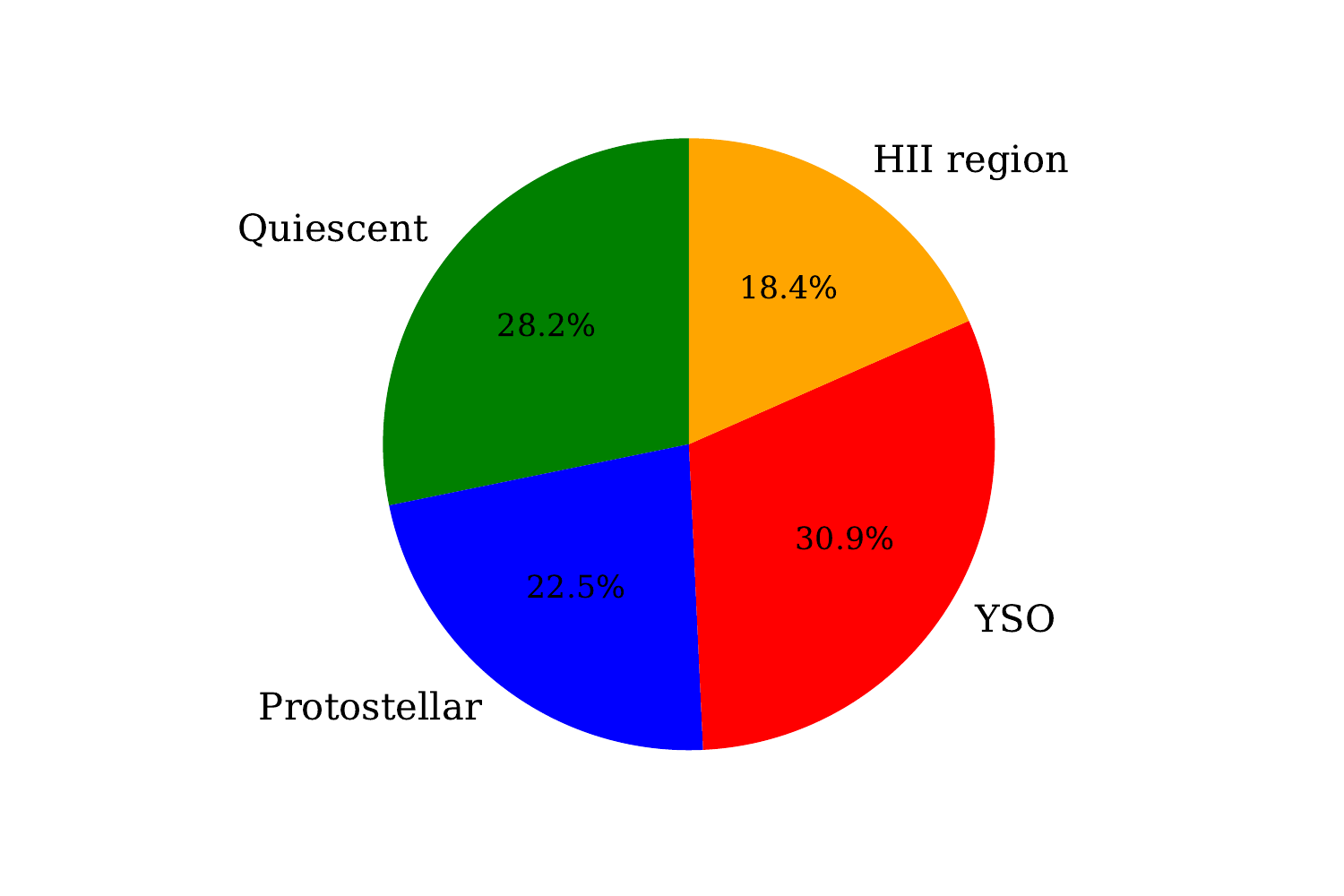}
\caption{Breakdown of the proportions of the clumps from the distance-limited sample in the four evolutionary stages. The \poi\ uncertainties on these values  are of order one per\,cent and so the differences are significant.}

\label{fig:evolution_pie_chart}
\end{figure}

We have four reliable evolutionary subsamples, each comprising a similar number of clumps. If we look at a distance-limited sample (i.e., see Fig.\,\ref{fig:evolution_pie_chart}), however,  we find that Quiescent and YSO-associated clumps are the most numerous, each with approximately 30\,per\,cent, with the protostellar and \hii\ regions having $\sim$20\,per\,cent each. Together, these make up approximately 60\,per\,cent of the classified sample, with the remaining 40\,per\,cent being classified as Ambiguous. In Figure\,\ref{fig:hist_properties}, we show the distributions of physical properties for the whole ATLASGAL sample and for those sources associated with one of the four evolutionary stages described in the previous section. It is interesting to note the similarities between the distributions of the two samples and this suggests that many of the clumps excluded from one of the evolutionary samples may also be associated with star formation but that the images are complicated or hard to interpret. We present a summary of results of the classification process giving the statistics and mean and standard deviation for the physical properties in Table\,\ref{tbl:properties_by_source_type}. 

If we assume that the fraction of sources in each stage is proportional to a similar fraction of the star-formation timescale, the statistics suggest that the lifetimes of the quiescent and YSO stages are similar to each other but 50\,per\,cent longer than the protostellar and \hii\ region stages. Although the fraction of quiescent sources is approximately a factor of two higher than reported in \citealt{urquhart2018}, the number of quiescent sources is actually similar. The reason for the increase in this fraction is that a larger proportion of the sources previously assigned to  one of the other three source types is now classified as being Ambiguous, as these are more often associated with diffuse or complicated far- and mid-infrared emission. The impact of this reclassification can be seen in the temperature distribution plot shown in the upper-left panel of Fig.\,\ref{fig:hist_properties}, which shows that the fraction of clumps classified as one of the four star-forming types decreases with increasing temperature. The fraction of quiescent sources given here is therefore likely to be an upper limit and the true fraction is likely to be somewhere between between 12 and 28\,per\,cent.

\begin{figure*}
    \centering
  
    \includegraphics[width=.45\textwidth, trim= 0 0 0 0]{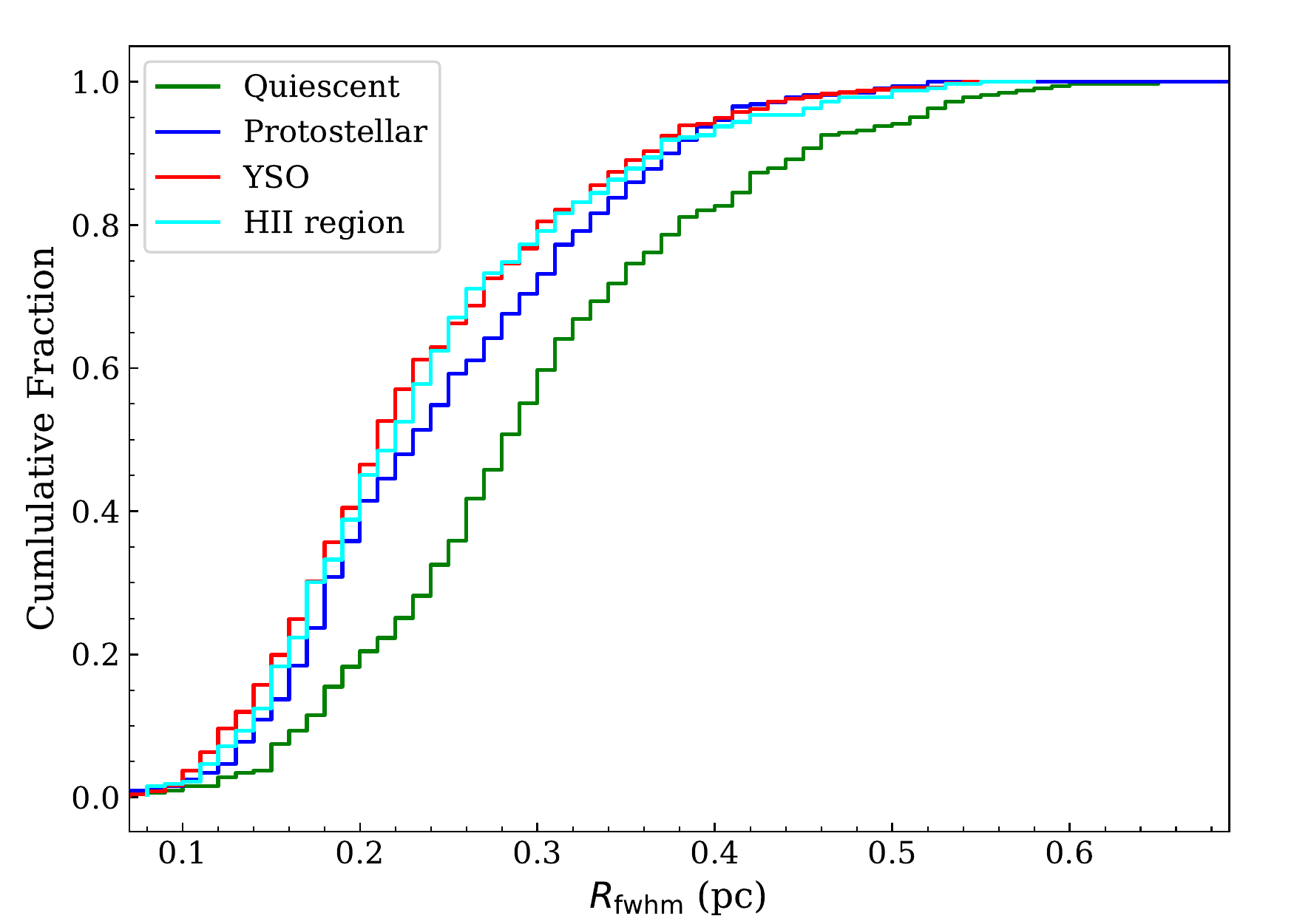}
    \includegraphics[width=.45\textwidth, trim= 0 0 0 0]{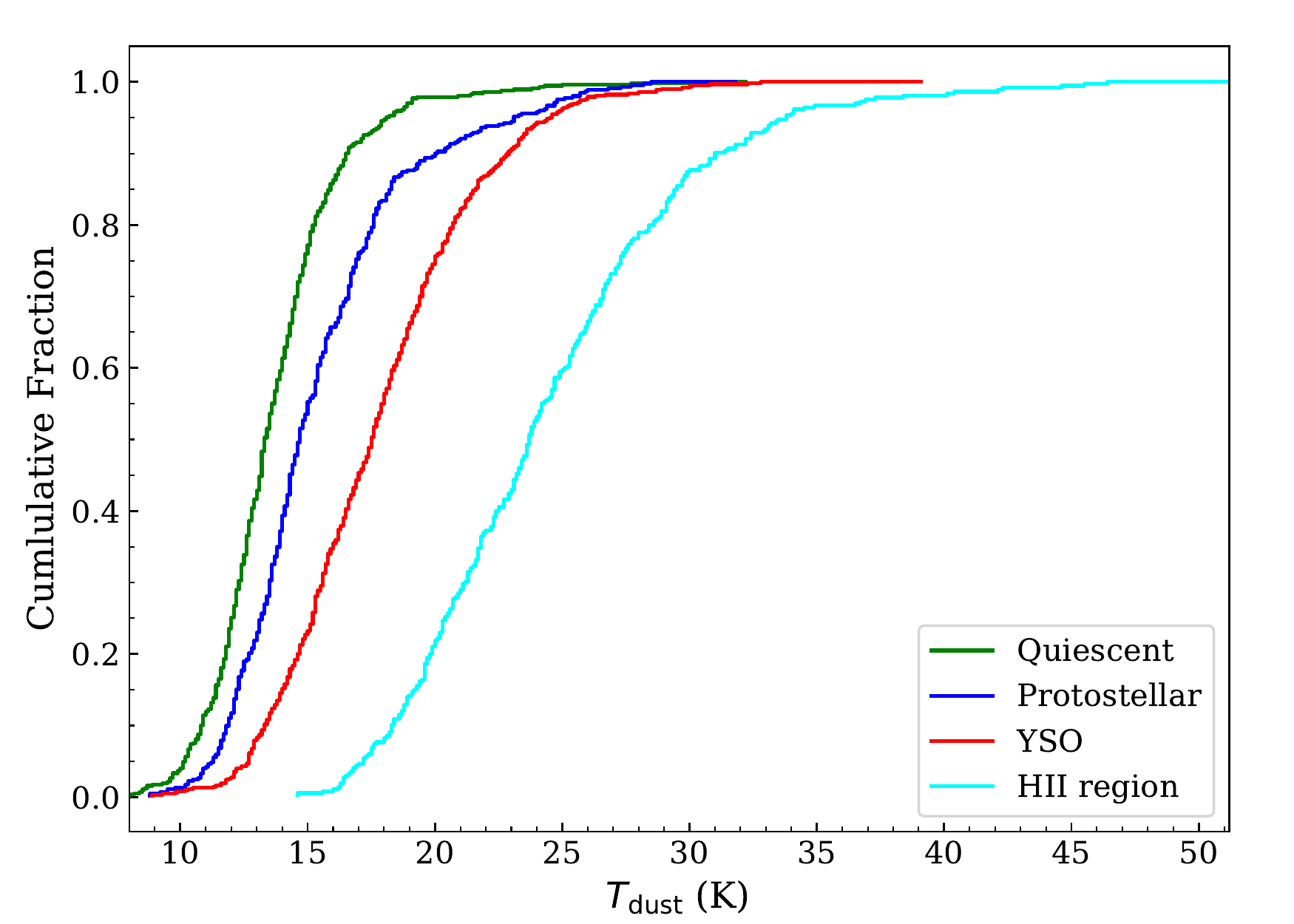}\\
     \includegraphics[width=.45\textwidth, trim= 0 0 0 0]{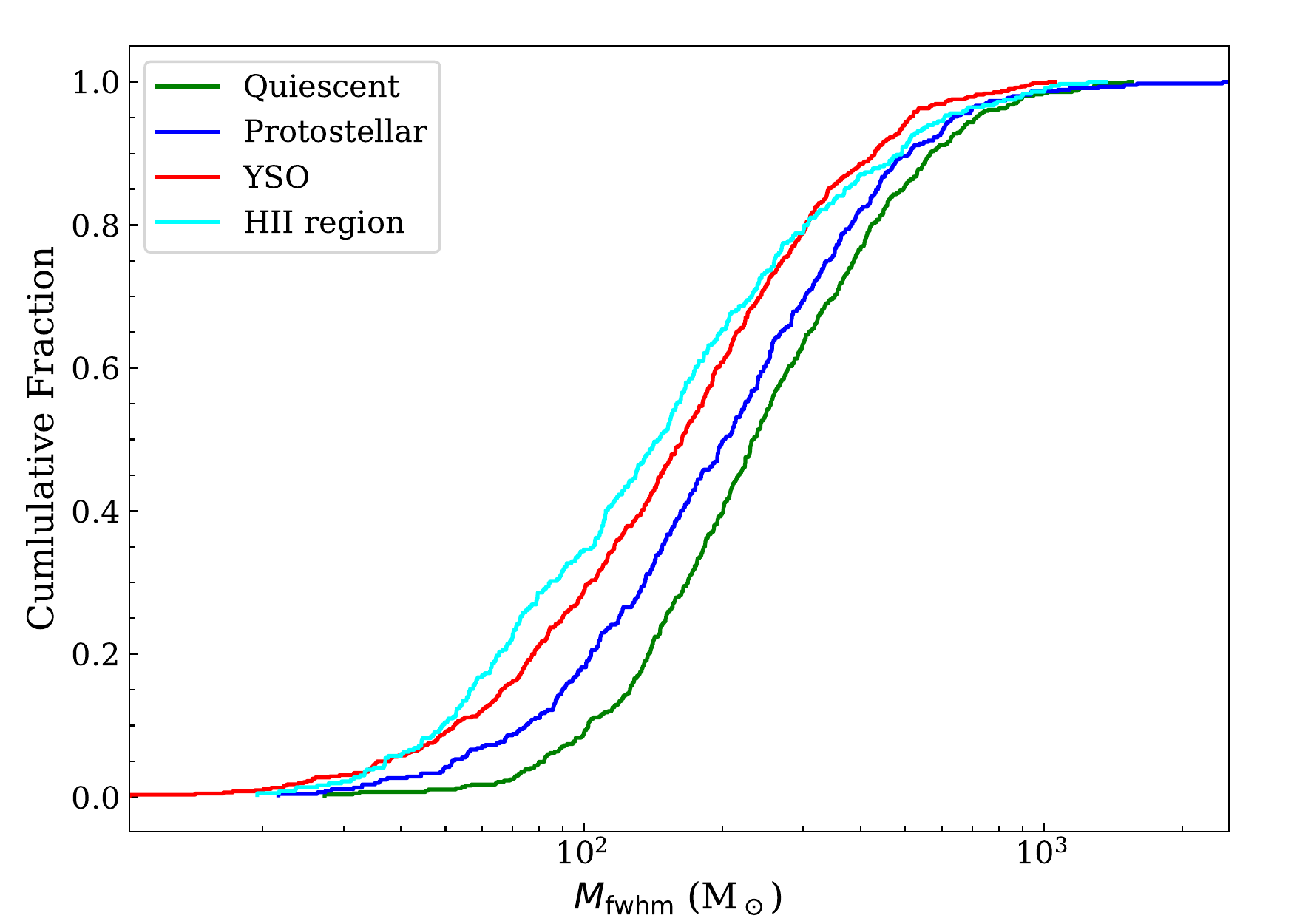}
     \includegraphics[width=.45\textwidth, trim= 0 0 0 0]{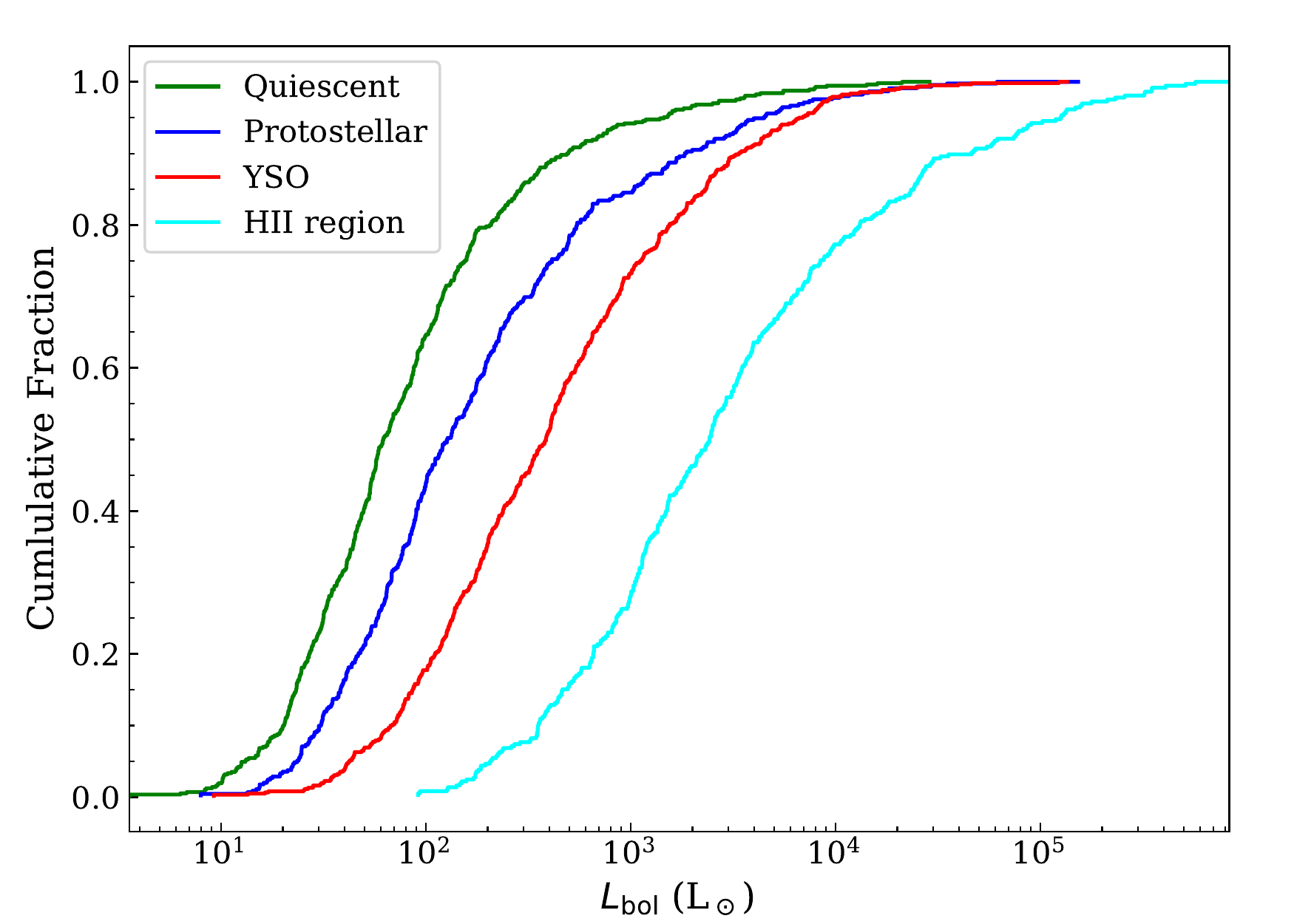}\\
   \includegraphics[width=.45\textwidth, trim= 0 0 0 0]{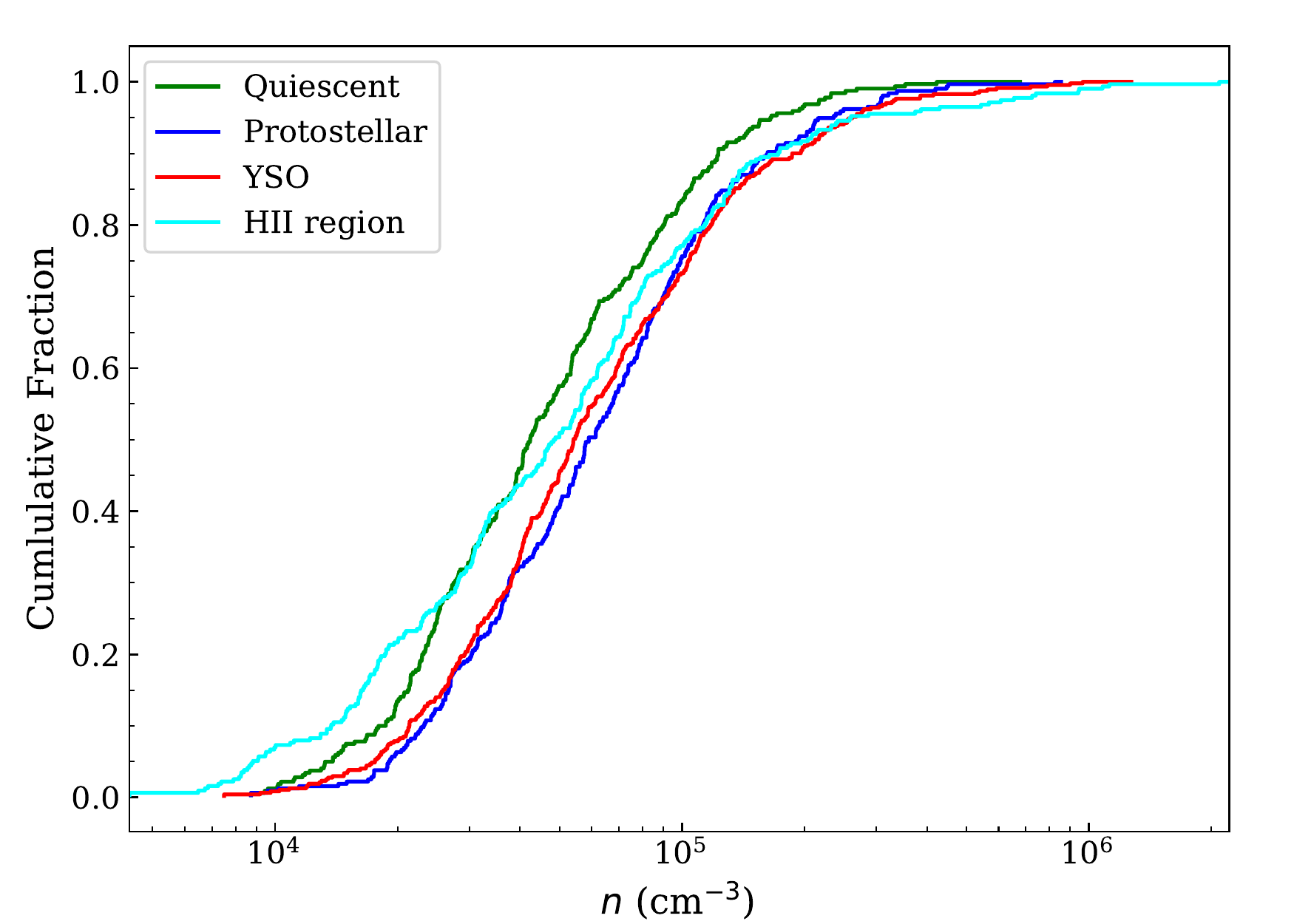}
   \includegraphics[width=.45\textwidth, trim= 0 0 0 0]{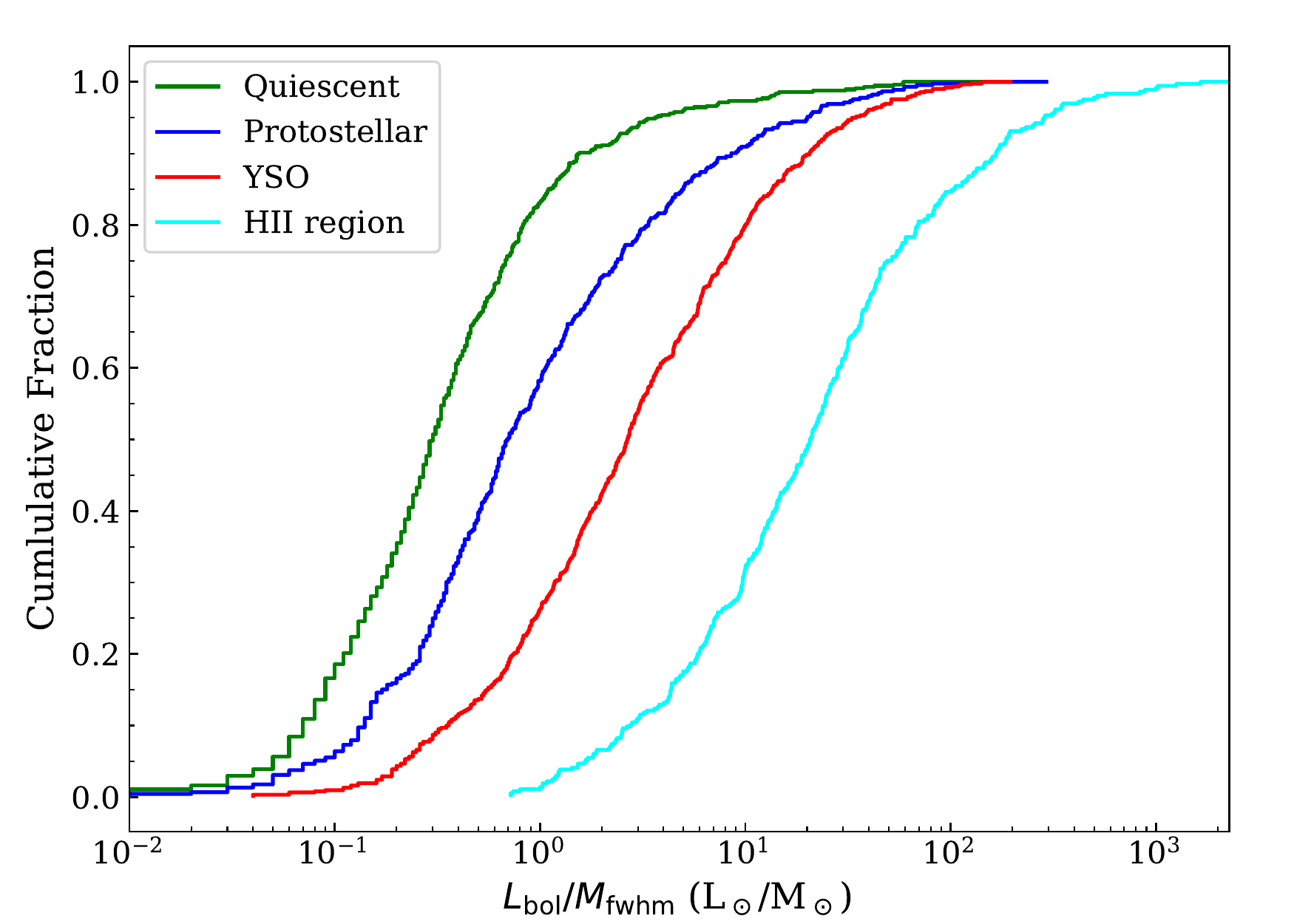}

\caption{Cumulative distribution functions for the main evolutionary categories. The left panels show the distributions  for the FWHM radius, mass, and volume densities, which are similar for all of the subsamples. The right panels  show the dust temperature, bolometric luminosity, and the luminosity-to-mass ratio. These latter values are all  good evolutionary diagnostics and show very clear differences between the different evolutionary subsamples. For all of these plots we have used a distance-limited sample of 2-4\,kpc to avoid any possible distance bias. }

\label{fig:properties_cdf}
\end{figure*}

In the left panels of Fig.\,\ref{fig:properties_cdf}, we show the cumulative distribution functions for the sizes, masses, and volume densities for all four evolutionary stages. Inspection of the sizes reveals a significant difference between the quiescent clumps and the other three evolutionary stages (KS-test returns a $p$-value $\ll$0.0013). The sizes of the other three stages are all similar. Turning our attention to the clump masses, we notice a trend for decreasing mass in the centres of the clumps as the embedded objects evolve towards the main sequence (see middle left panel of Fig.\,\ref{fig:properties_cdf}; the $p$-values for the quiescent and protostellar clumps is 0.0006 and for the protostellar and YSOs it is 0.0002, confirming the observed difference is statistically significant). Looking at the volume density distribution (lower left panel of Fig.\,\ref{fig:properties_cdf}) we see that the protostellar and YSO distributions are indistinguishable from each other. The quiescent clumps have a similar-shaped curve to that of the YSOs and protostars but shifted to lower densities ($p$-value $\ll 0.0013$). The \hii\ region distribution is significantly different from the other three stages, having a similar shape as the YSOs and protostars at high volume densities but a shallower slope, and extends to include some of the lowest volume densities. The differences observed for some of these parameters could be attributed to the evolution of the embedded object. We will see in the next section, however, that these differences are in fact the result of observational biases.

In the right panels of Fig.\,\ref{fig:properties_cdf}, we show the cumulative distribution functions for temperature, luminosity, and luminosity-to-mass ratio. All of these parameters increase as a function of evolution, as expected.  The increase in these parameters has been reported in previous studies (\citealt{urquhart2018, elia2017, konig2017}) and so these results are not new but demonstrate that we have produced a large and reliable set of evolutionary subsamples of clumps.

\section{Discussion}
\label{sect:discussion}

\subsection{Physical correlations}
\label{sect:physical_correlations}

   \begin{figure}
    \centering
  
    \includegraphics[width=.45\textwidth, trim= 0 0 0 0]{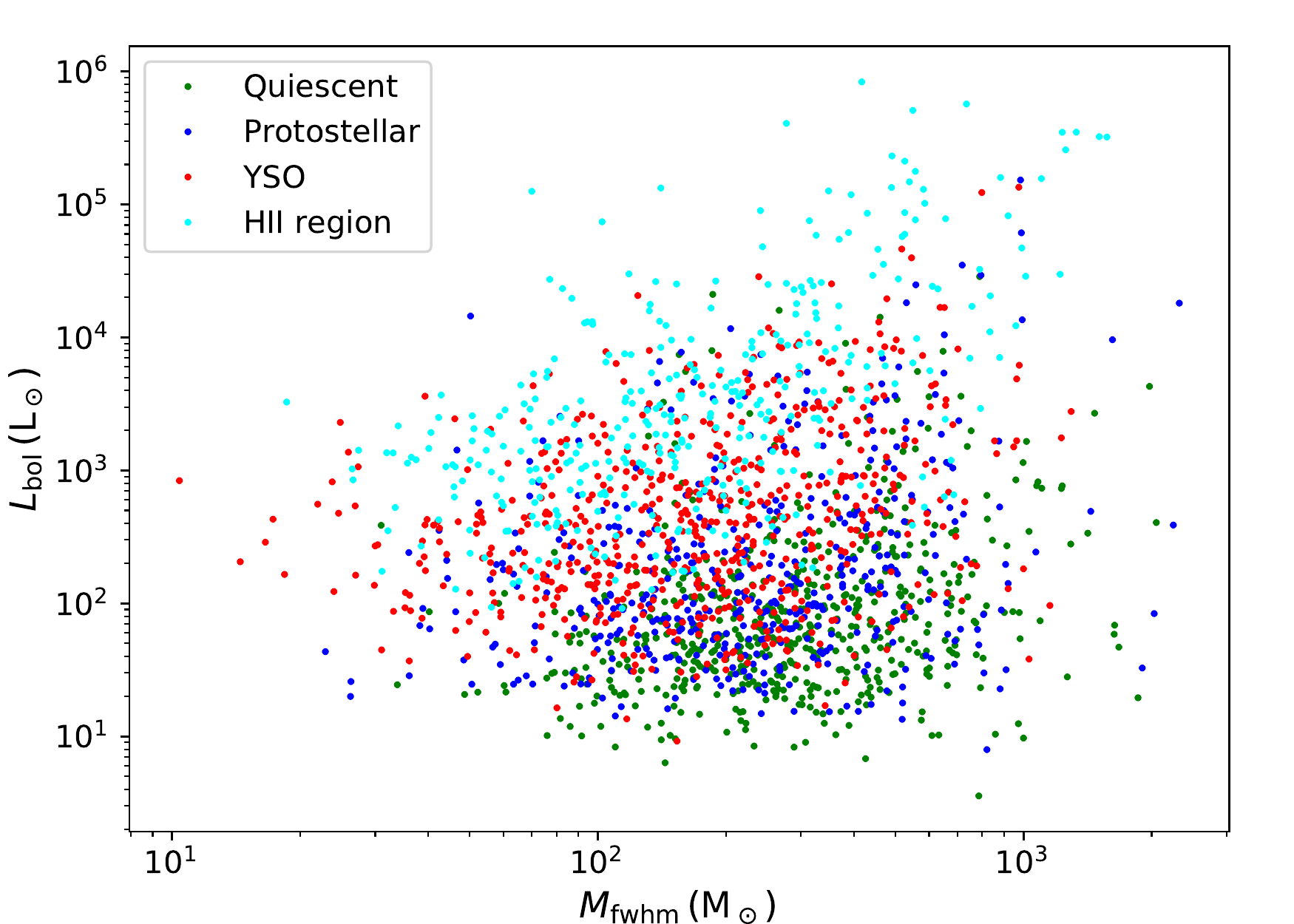}
   \includegraphics[width=.45\textwidth, trim= 0 0 0 0]{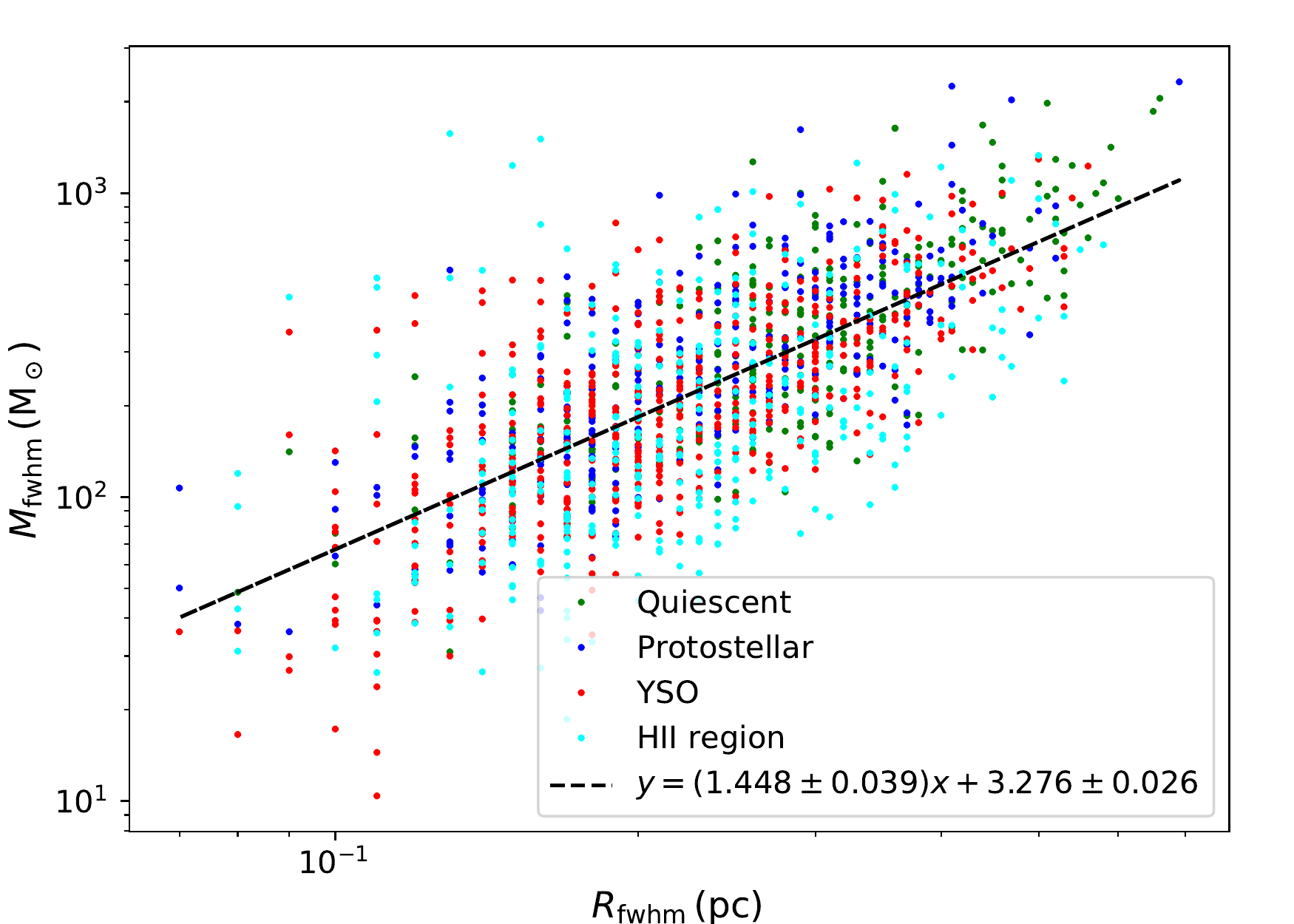}\\

\caption{Luminosity-mass and mass-radius relationships are shown as a function of the evolutionary source type for all clumps in the distance limited sample.  }

\label{fig:lum_mass_dist}
\end{figure}

We illustrate the luminosity-mass and mass-radius distributions by evolutionary source type (as identified in Sect.\,\ref{sect:evolutionary_sequence}) in Figure\,\ref{fig:lum_mass_dist}.
The linear least-squares fit to the luminosity-mass distribution of a distance-limited sample (2-4\,kpc) of the \hii\ regions (shown in Fig.\,\ref{fig:lum_mass_dist_HII}) gives a slope of $1.224\pm0.0916$ and intercept of 0.777$\pm$0.206 compared with the value of $1.314\pm0.019$ determined in \citet{urquhart2018}. The change in the way the mass is determined has resulted in a decrease in the steepness of the slope, but, the new slope is consistent with the older one within 3$\sigma$.
The slope fit to the mass-radius distribution of the full star-forming sample shown in Figure\,\ref{fig:lum_mass_dist} is $1.448\pm0.039$, which is again slightly shallower than the value of $1.647\pm0.012$ determined in \citet{urquhart2018}. 

   \begin{figure}
    \centering
  \includegraphics[width=.45\textwidth, trim= 0 0 0 0]{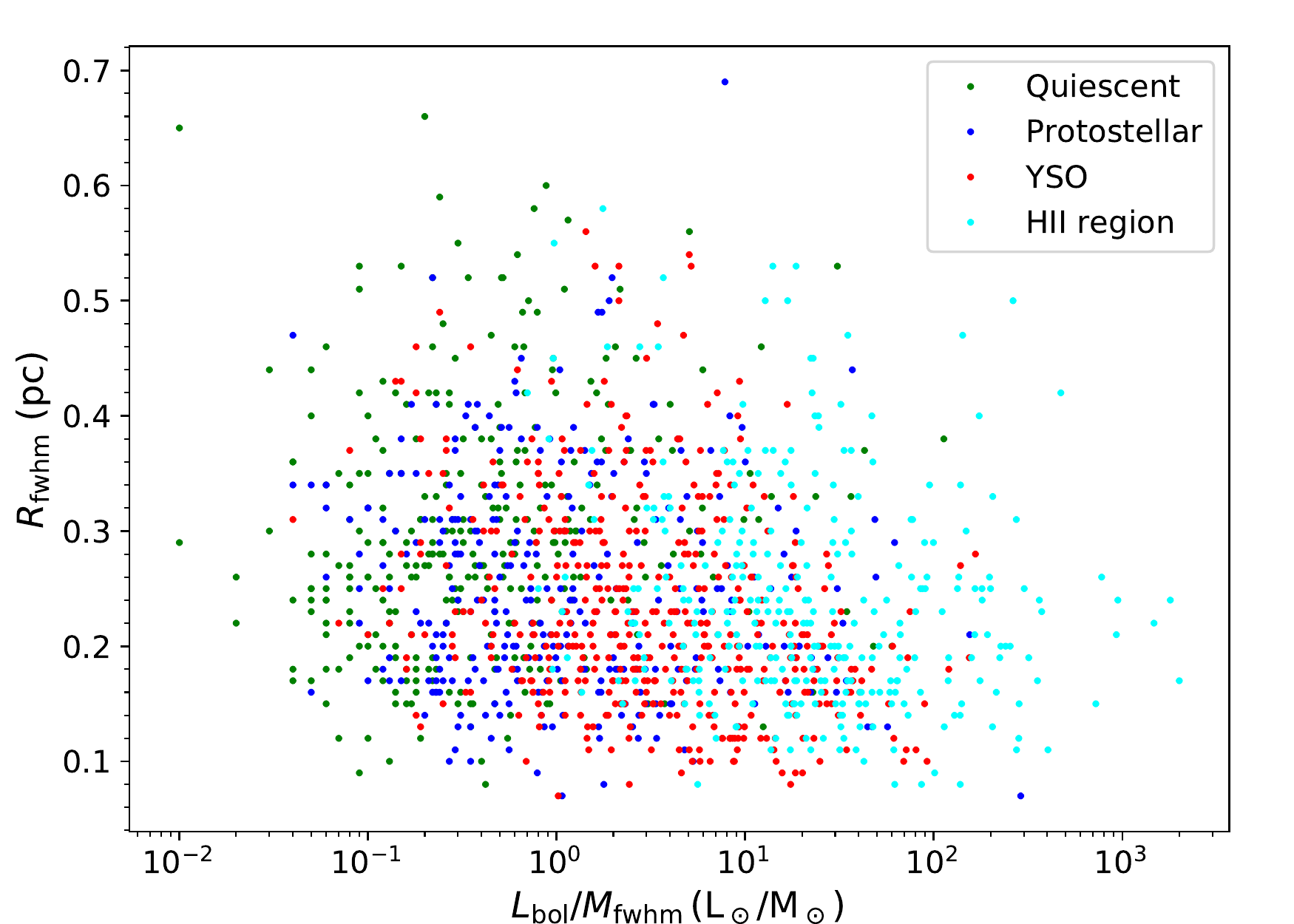}
    \includegraphics[width=.45\textwidth, trim= 0 0 0 0]{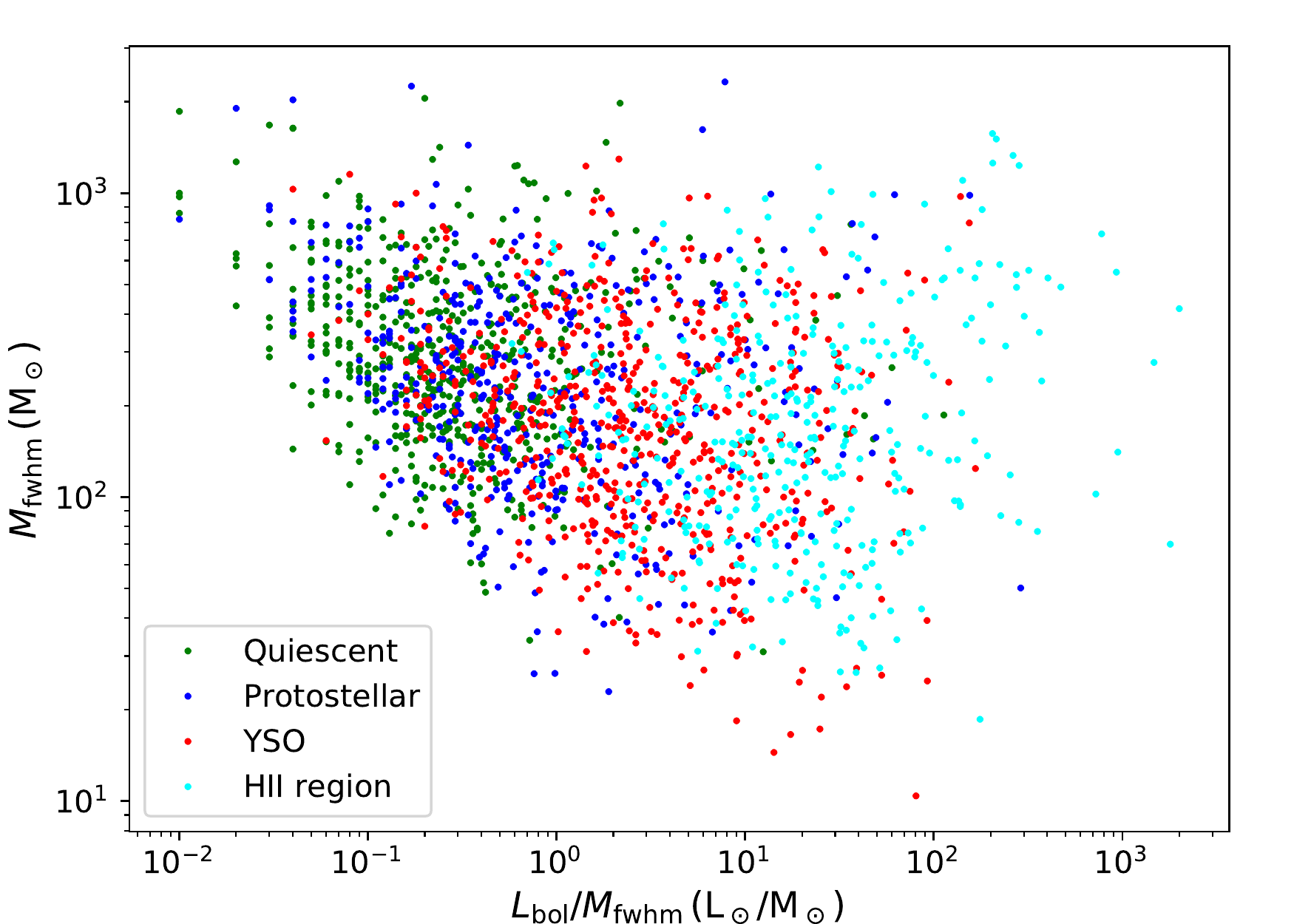}
    \includegraphics[width=.45\textwidth, trim= 0 0 0 0]{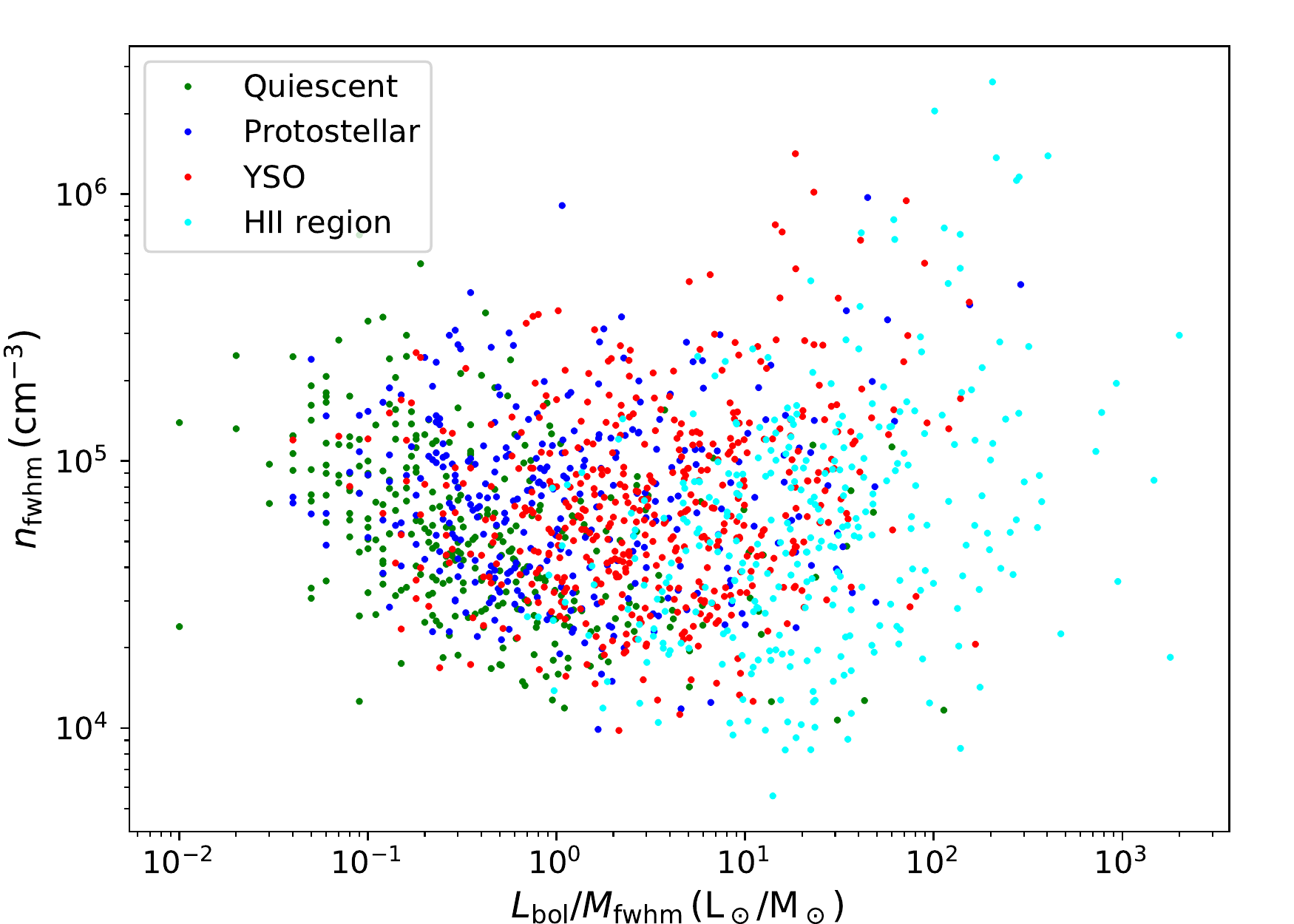}
   
\caption{The radius, mass and volume density for the the evolutionary source types as a function of the luminosity-to-mass ratio. These plots include all clumps in the distance-limited sample.  }

\label{fig:mass_vol_lm}
\end{figure}
    
We show the radius, mass, and volume density distributions as functions of the luminosity-to-mass (\lm) ratio in Figure\,\ref{fig:mass_vol_lm}. The distributions of the mass and volume density are similar, with the evolutionary type changing with increasing \lm-ratio as we would expect.  In the previous section, we noted some significant trends in the cumulative distribution functions of these parameters but cautioned that these are likely to be due to observational bias rather than the result of evolution of an embedded object. There is no obvious correlation between size and evolution (upper panel; the Spearman rank correlation coefficient is $r= -0.087$). A cursory glance at the mass and density distribution, however, suggests that these parameters are correlated with evolution, particularly in the mass-\lm-ratio plot. This trend is largely due to the presence of a significant deficit of clumps towards the lower-left region of the parameter space (middle and lower panels of  Figure\,\ref{fig:mass_vol_lm}). ATLASGAL is a flux-limited survey. Since the clump mass is a function of flux and dust temperature, it has a poorer mass sensitivity for colder objects leading to the deficits in the Fig.\,\ref{fig:mass_vol_lm} and resulting in the perceived change in mass as a function of evolution noted in the middle and lower-left panels of Fig.\,\ref{fig:properties_cdf}. One final interesting feature seen in the volume density distribution plot (lower panel of Fig.\,\ref{fig:mass_vol_lm}) is a small but noticeable increase in the volume density above a \lm-ratio of 10\,\lsun/\msun. This is possibly related to the faster free-fall collapse times found for higher-density clumps, resulting in lifetimes so short that they are very unlikely to be observed in the earlier evolutionary stages. These high-density and rapidly evolving clumps are only seen towards the end of their evolution because it takes them a while to disperse their natal material once the \hii\ region has formed.

If we limit the analysis to a mass and density range in which we are complete (i.e., \mclump\ $> 200$\,\msun\ and $n_{\rm fwhm}$ between $3\times 10^4$\,cm$^{-3}$ and $4\times 10^5$\,cm$^{-3}$), then the differences between the different evolutionary stages disappear (the Spearman rank correlation coefficients are $r= -0.03$ and $r= -0.02$ for the $M_{\rm fwhm}$-\lm-ratio and $n_{\rm fwhm}$-\lm-ratio respectively for the ranges mentioned). There is, therefore, no obvious correlation between radius, mass or volume density with evolution (i.e., the overall distributions are flat).  The lack of any significant change in the clump mass and density as a function of the \lm-ratio further supports these findings (see also \citealt{urquhart2018} and \citealt{billington2019_meth}). This result is somewhat surprising, given that significant global infall motions in clumps  are often reported in the literature (e.g., 0.3- 16\x10$^{-3}$\,\msun\,yr$^{-1}$;  \citealt{wyrowski2016}), which would lead to a significant increase in the clump mass if these rates were to continue over the star formation  time frame (i.e., several 10$^5$\,yr). 

Recent analysis by \citet{jackson2019} of a sample of $\sim$1000 MALT90 clumps found that, while high-mass stars are undergoing gravitational collapse, the blue asymmetry in the line emission measured for quiescent, protostellar and compact \hii\ regions was significantly larger than for more extended \hii\ regions and PDRs. These results suggest that global collapse is underway even in the earliest quiescent stage but decreases in the later stages. This hypothesis is supported by a more recent set of observations  of the $J=3-2$ and $J=4-3$ rotational transitions of the HCO$^+$, HNC, HCN and N$_2$H$^+$ towards a sample of ATLASGAL clumps that found the incidence of infall motions decreases as the evolutionary stage of the clumps increases (Neupane et al. 2022, in prep.), presumably due to the increasing radiative and mechanical feedback from the forming proto-cluster. 
In fact, the gas motions of 3 of the 9 sources investigated by \citet{wyrowski2016} were found to be dominated by strong outflowing/expanding motions. Two of these were classified as \uchii\ regions and one as a MYSO and so these can all be considered to be relatively evolved. A decreasing infall rate may significantly limit any increase in the clump masses and would be consistent with our findings. The results reported by \citet{jackson2019}, however, are indicative rather than definitive and, while the results obtained from the higher excitation transitions reported by Neupane et al. (2021, in prep.) show a clear trend, they are drawn from a relatively small number of sources and further work is required confirm this hypothesis.

The process of turning the gas and dust into stars will also lead to a decrease in the clump mass over time, however, some of the gas used to create stars is expected to be replaced by infalling material onto the clump from the parental GMC. Star formation efficiencies ($\varepsilon$) are expected to range from 5-30\,per\,cent (e.g. \citealt{lada2003, rugel2019}). If we assume an efficiency to be $\sim$20\,per\,cent during the star formation process and compare this to the uncertainties in clump masses (which is a factor of a few) we find that we are unlikely to notice this change, particularly since any decrease in mass is likely to be counteracted by an increase in the mass due to infall. Although there is no evidence of any significant change in the clump mass during the star formation process it does not preclude changes in the internal structure on smaller scales. In fact, we expect gas to continue to contract to form higher-density substructures and to feed accretion onto forming protostellar objects (e.g., \citealt{csengeri2017_spark,csengeri2018_spark}). Rather, it is just that these motions are not observed on the scales ATLASGAL is probing (i.e., $R_{\rm fwhm} \approx0.5-1$\,pc). Indeed, a recent deep study of a $2\degr \times 1\degr$ field  by \citet{rigby2021}  reported a small but significant increase in the average clump mass from the earliest to the middle stages, before decreasing towards the later stages. Their observations have a modest improvement in resolution compared to ATLASGAL (12\arcsec, corresponding to 0.2-0.4\,pc), which might explain the difference in the results. The way the masses are measured, frequency of the observations and coverage, however, are all quite different making a direct comparison difficult.

\subsection{Linking the \lm-ratio to a star formation time scale}
\label{sect:timescale}

The right panels of Fig.\,\ref{fig:properties_cdf} show the cumulative distribution functions for temperature, luminosity, and luminosity-to-mass ratio for the four evolutionary subsamples. All of these distributions look relatively smooth as one would expect if the properties of these subsamples are normally distributed. If, however, these observationally classified subsamples do represent distinct stages in the star formation process, we might expect to see points of inflection and jumps in the cumulative distribution function at the transition points of a clump's evolution. 

\begin{figure}
    \centering
    \includegraphics[width=.45\textwidth, trim= 0 0 0 0]{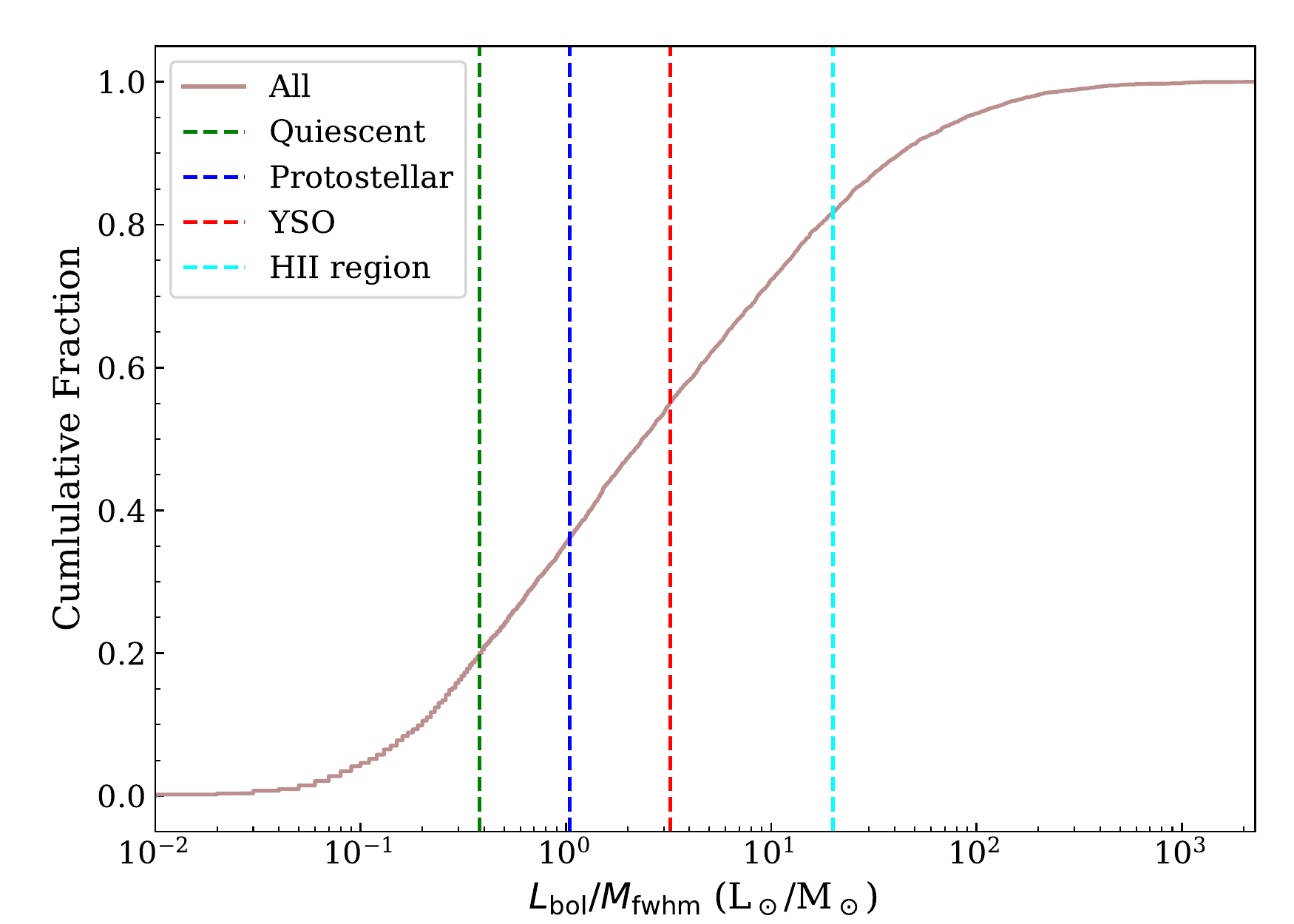}

\caption{Cumulative distribution function for the \lm-ratio for the combined star formation sample (rose curve). The vertical lines indicate the mean of the lognormal distribution of the four subsamples identified that constitute the star formation sample. }

\label{fig:luminosity_lm_ratio_all}
\end{figure}

We show the cumulative distribution for the combined evolutionary sample in Figure\,\ref{fig:luminosity_lm_ratio_all}. In this plot and those that follow we use the full evolutionary sample as \lm-ratio is a distance-independent parameter and there is no significant difference between the full sample and the distance-limited sample. This plot shows that the central part of the curve is still smooth and devoid of any features that are coincident with the mean values of \lm-ratios for the four evolutionary stages. This lack of abrupt changes suggests that star formation is a rather smooth and continuous process. Also the stages are useful in identifying groups of protostellar objects with similar properties and/or ages but do not themselves represent fundamentally different stages or changes in the physical mechanisms involved. The curve flattens out at both ends but the source numbers are too low in these regions of the parameter space to draw any conclusions from that behaviour. We can speculate as to their nature, however. The steepening of the slope at the low-\lm-ratio end of the curve is possibly related to the onset of star formation and the fact this steepening is seen below the mean of the quiescent stage indicates that star formation is ongoing in many of these apparently `starless' clumps. This hypothesis is strongly supported by the association of many of the quiescent clumps with outflows as discussed in the following subsection. On the other hand, the turnover of the slope at the high-\lm-ratio end starts just after the mean of the \hii\ region stage, which is possibly related to the increasingly disruptive influence of high Lyman photon flux and expanding ionization front.

\begin{figure}
    \centering
    \includegraphics[width=.45\textwidth, trim= 0 0 0 0]{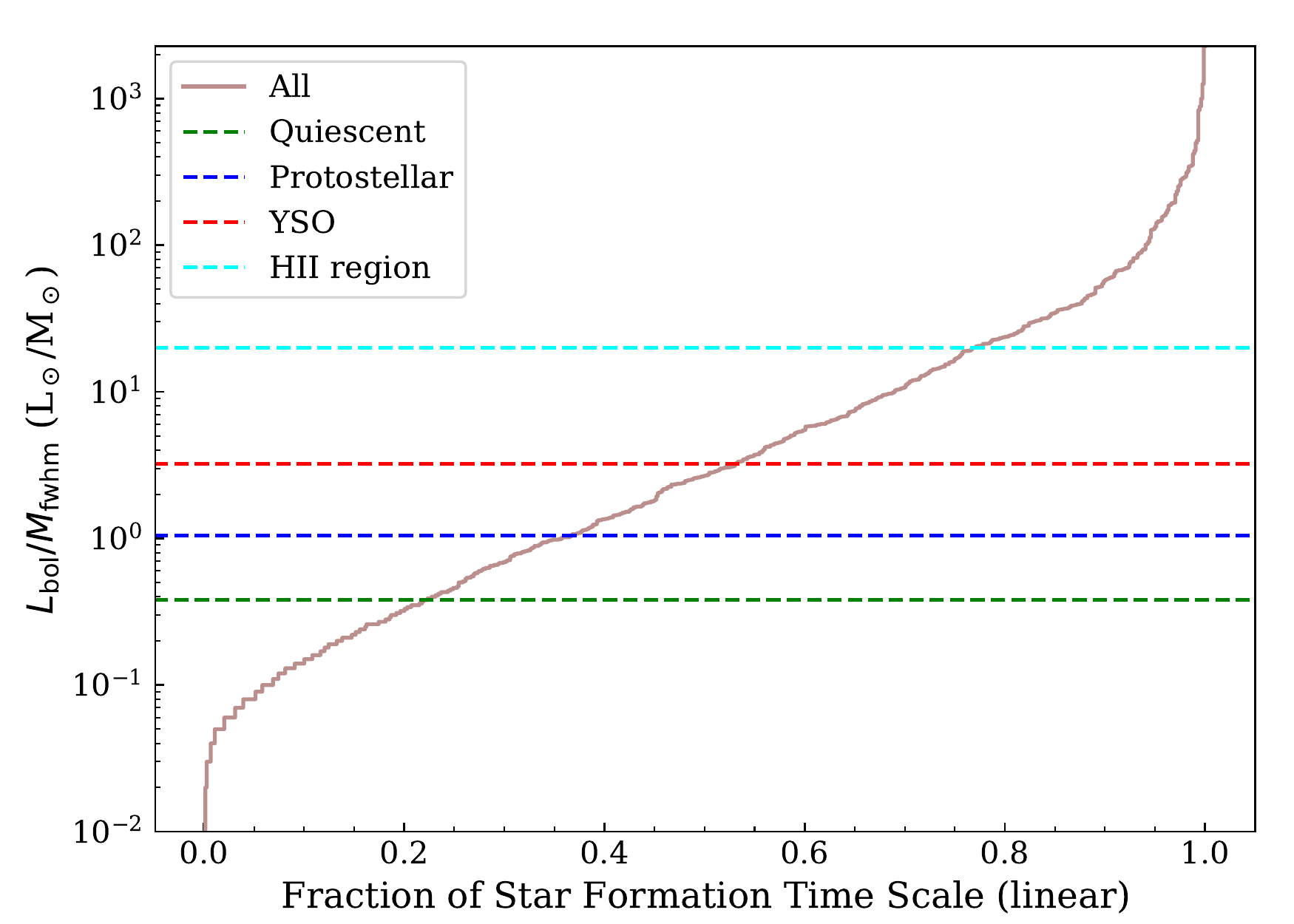}
      \includegraphics[width=.45\textwidth, trim= 0 0 0 0]{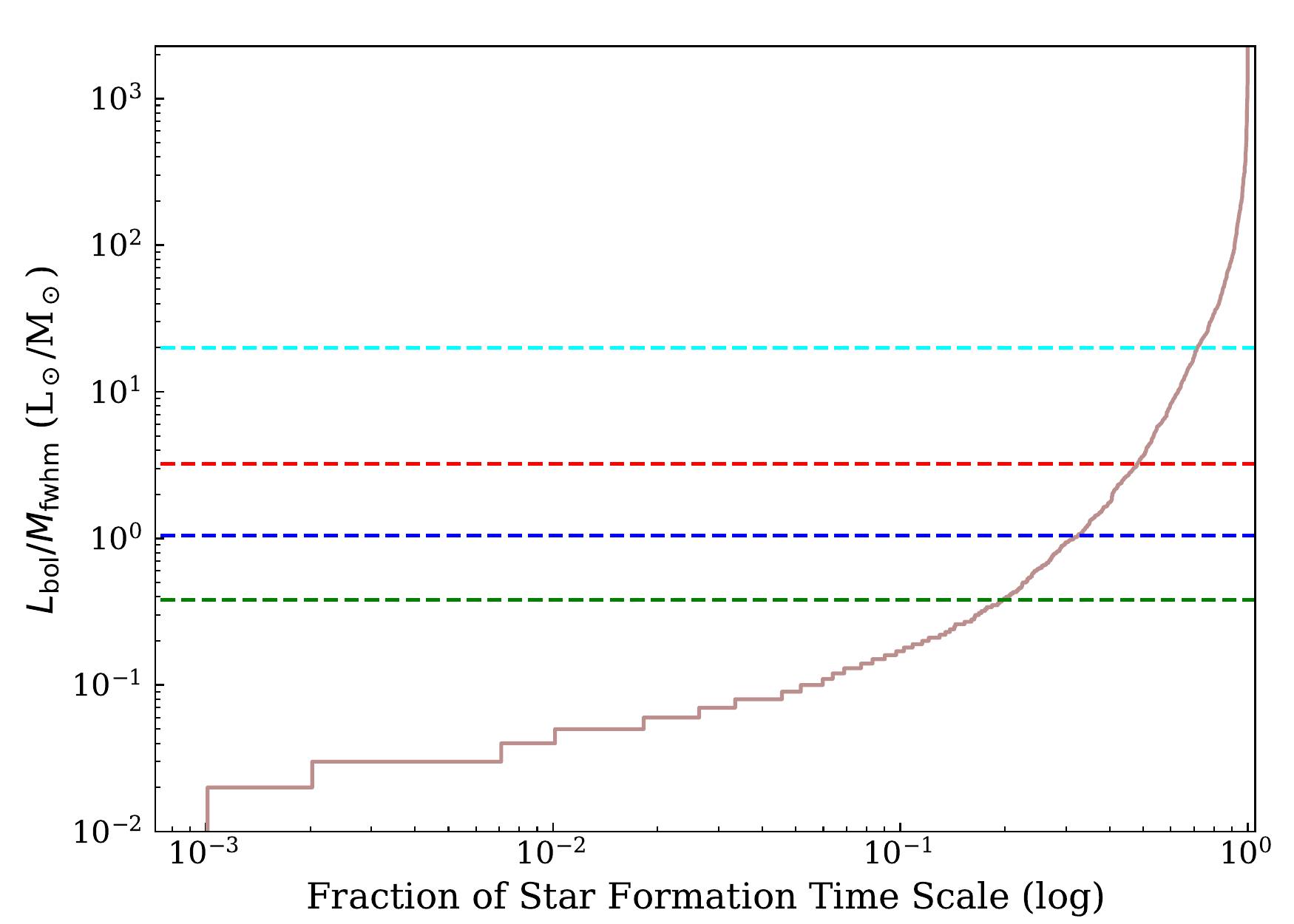}

\caption{The \lm-ratio as a function of the fractional free-fall collapse times of the clumps. The two plots show the same data but in the upper panel the $x$-axis is linear and in the lower panel it is plotted on a log scale. The horizontal lines indicate the mean of the lognormal distribution of the four evolutionary types.} 

\label{fig:lm_vs_evolution_timescale}
\end{figure}

We can take our analysis of the \lm-ratio a step further by linking it to  evolutionary time scales. If we assume that the number of sources in each \lm\ bin is proportional to the fraction of the star formation time scale that a protostar spends in each \lm\ range, then the y-axis of the cumulative distribution function is also proportional to the star's formation time. In Figure\,\ref{fig:lm_vs_evolution_timescale} we have re-plotted Fig.\,\ref{fig:luminosity_lm_ratio_all} but have switched the axes to have the independent variable on the x-axis and have changed the label to accordingly. These plots show how the \lm-ratio changes during the star formation process. Looking in a little more detail at the curve presented in the lower panel of Fig.\,\ref{fig:lm_vs_evolution_timescale} we see that nearly all clumps fall on a smooth curve of increasing \lm-ratio and so are therefore increasing their luminosity very slowly in the very early stages. This also reveals that the rate of increase in the \lm-ratio accelerates as the embedded object evolves through the protostellar and YSO stages before turning into what appears to be a runaway process when it reaches the \hii\ region stage. This is to be expected, as rapidly expanding ionised bubbles begin to disrupt and dissipate their natal environment.

In the early stages of a protostar's evolution, the majority of the luminosity is generated by accretion rather than fusion and so the acceleration of the \lm-ratio with evolution is therefore related to an increase in the accretion rate over time for the earliest stages. High-mass stars reach the main sequence while still deeply embedded and can continue to accrete material and so in the later stages an increasing proportion of their luminosity will be generated by fusion, which is perhaps why we see a shape increase in the \lm-ratio for the most evolved sources. The steady increase can be seen in the upper panel of Fig.\,\ref{fig:lm_vs_evolution_timescale} lends support for high-mass star formation models where the rate of accretion also increases over time, which were previously found to be consistent with the results reported by \citet{davies2011} from the modelling of the RMS survey (\citealt{lumsden2013}).  

In Section\,\ref{sect:physical_properties}, we used the mean volume densities to estimate the free-fall time scales for the clumps. The calculation of the free-fall time by definition does not take account of internal support or feedback, both of which will slow down the collapse process and result in longer collapse times. The free-fall time scales, however,  were a few times 10$^5$\,yr, i.e., similar to the \hii\ region formation time scales derived empirically by \citet{mottram2011b} and from a numerical model by \citet{davies2011}. This similarity suggests that, although the free-fall times are a lower limit, they are perhaps only a few times shorter than the star formation time scale and can therefore be useful to gain some insight to how the star formation process proceeds.

\begin{figure}
    \centering
    \includegraphics[width=.45\textwidth, trim= 0 0 0 0]{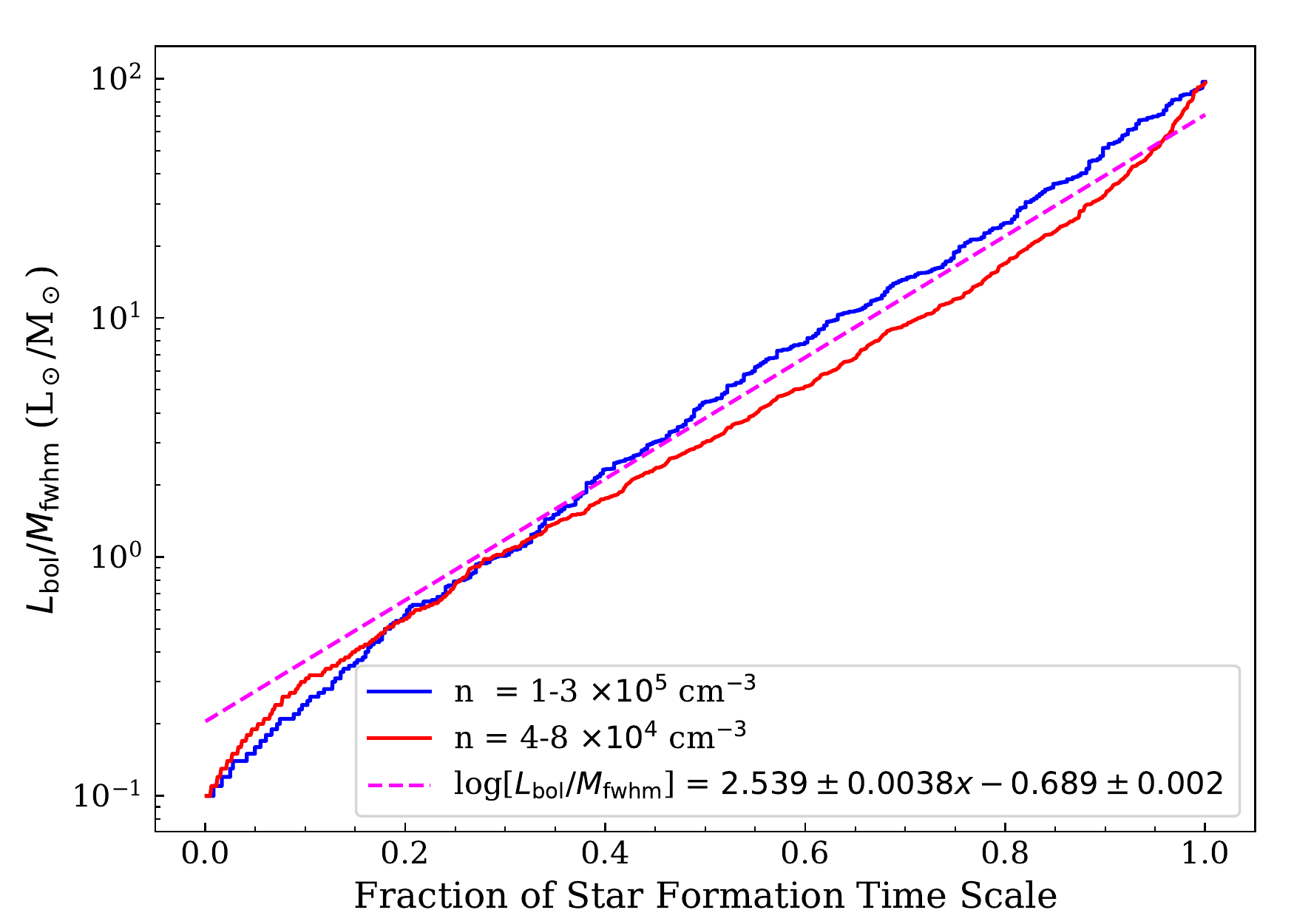}
      \includegraphics[width=.45\textwidth, trim= 0 0 0 0]{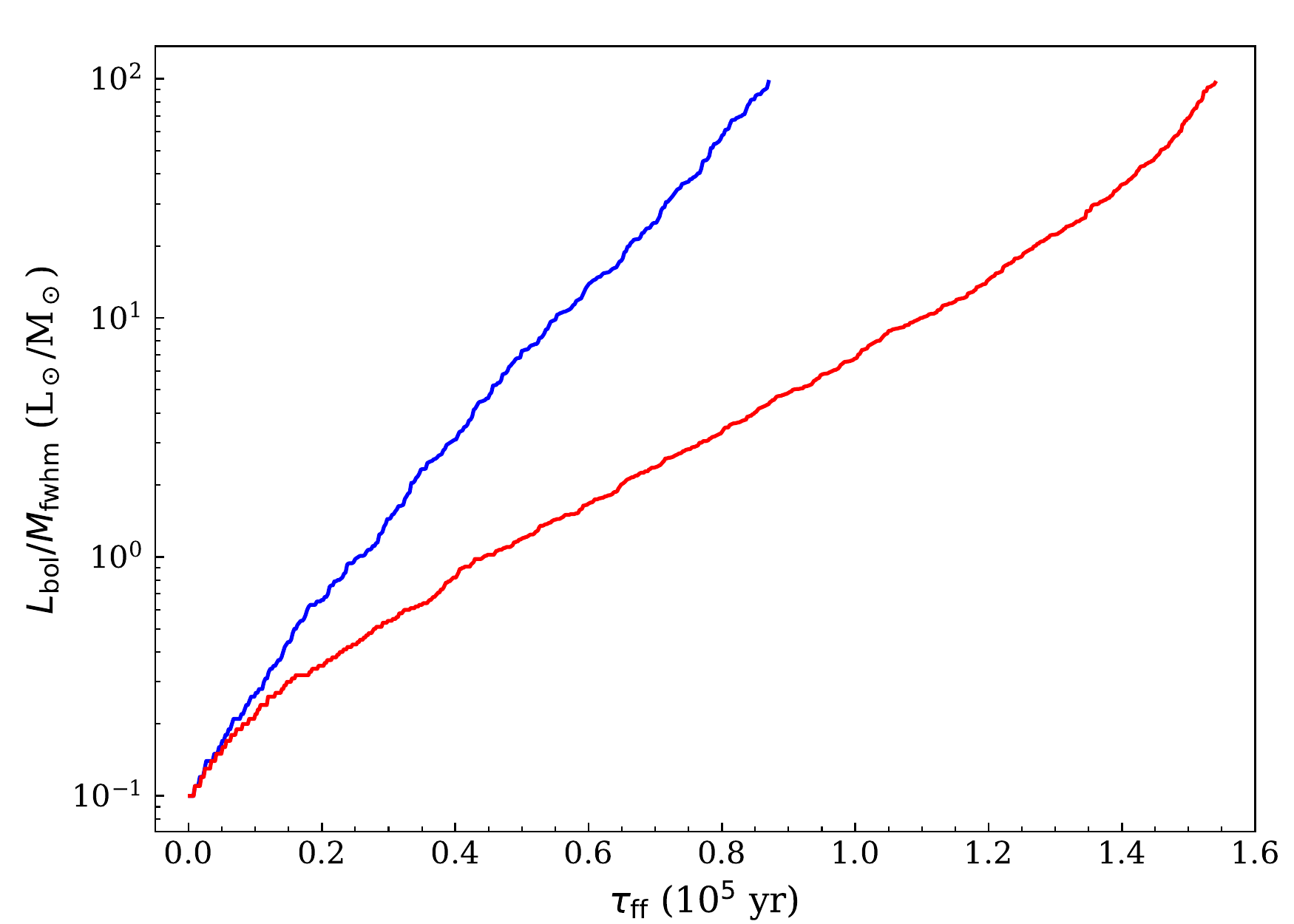}

\caption{The \lm-ratio as a function of the star formation time scale (upper panel) and as a function of the free-fall collapse times of the clumps (lower panel) for two volume density ranges. We have restricted the \lm\ range to be between 0.1 and 100 \lsun/\msun\ to concentrate the analysis on the region of parameter space for which we have reliable statistics. A linear least-squares fit to the \lm-ratio track of all clumps with volume densities between 1-50$\times 10^5$\,cm$^{-3}$ is shown in the upper panel.}

\label{fig:lm_vs_ff}
\end{figure}

In Figure\,\ref{fig:lm_vs_ff}, we have plotted the \lm\ curve for two small ranges in volume density as a function of the star formation time scale (upper panel) and as a function of the corresponding free-fall time calculated in Sect.\,\ref{sect:physical_properties} (lower panel). A linear-least square fit to the log of the L/M-ratio is shown in the upper panel and gives a power-law relation between the L/M-ratio and the star formation \underline{time scale} of  10$^{2.5\times \tau_{\rm ff} - 0.7}$.
Comparing the tracks for the two different density ranges reveals them to be very similar and a KS-test returns a $p$-value of 0.03 confirming that the tracks are not significantly different. From this similarity, we conclude that the process by which stars accrete material is not influenced by the mean volume density of the clumps. Hence, the star formation time scale and the relative time in the different stages both scale with clump density and so with free-fall time.  The \underline{process} is therefore similar and scaleable, it just goes faster in denser clumps. The lower panel of Fig.\,\ref{fig:lm_vs_ff} reveals how the \lm-ratio changes as a function of free-fall time for different density ranges, with the denser clumps forming stars much more quickly, as expected. 


\subsection{Correlation with other samples}
\label{sect:other_samples}

We have cross-matched the ATLASGAL sample with four other star formation tracers to test the validity of the evolutionary sequence identified in this paper. We have used the catalogue of $\sim$300 extended green objects (EGOs; \citealt{cyganowski2008}), catalogues of molecular outflows produced by \citet{maud2015, de_villiers2014,yang2018,yang2021}, the methanol multibeam (MMB) survey (\citealt{green2009_full}) and  water masers identified by HOPS (\citealt{walsh2011,walsh2014}). EGOs are identified by their enhanced emission in the 4.5-\mum\ band that contains the rotationally excited H$_2$ ($v = 0-0$, S(9, 10, 11)) and CO ($v = 1-0$) band-head lines, which are indicative of outflow activity and active accretion (\citealt{cyganowski2008}). Molecular outflows are intimately associated with accretion disks and are therefore a strong indication that star formation is taking place in a clump. Class II Methanol masers (\citealt{menten1991}) are thermally pumped and require the presence of high densities and a strong mid-infrared radiation source and so are thought to be associated exclusively with the closest environment of  high-mass protostellar objects (\citealt{minier2003,breen2013, billington2019_meth}). Water masers are collisionally pumped and, when encountered in star formation environments, are thought to be excited in the cavities of molecular outflows.

\begin{figure}
    \centering
 
        \includegraphics[width=.49\textwidth, trim= 0 100 0 0]{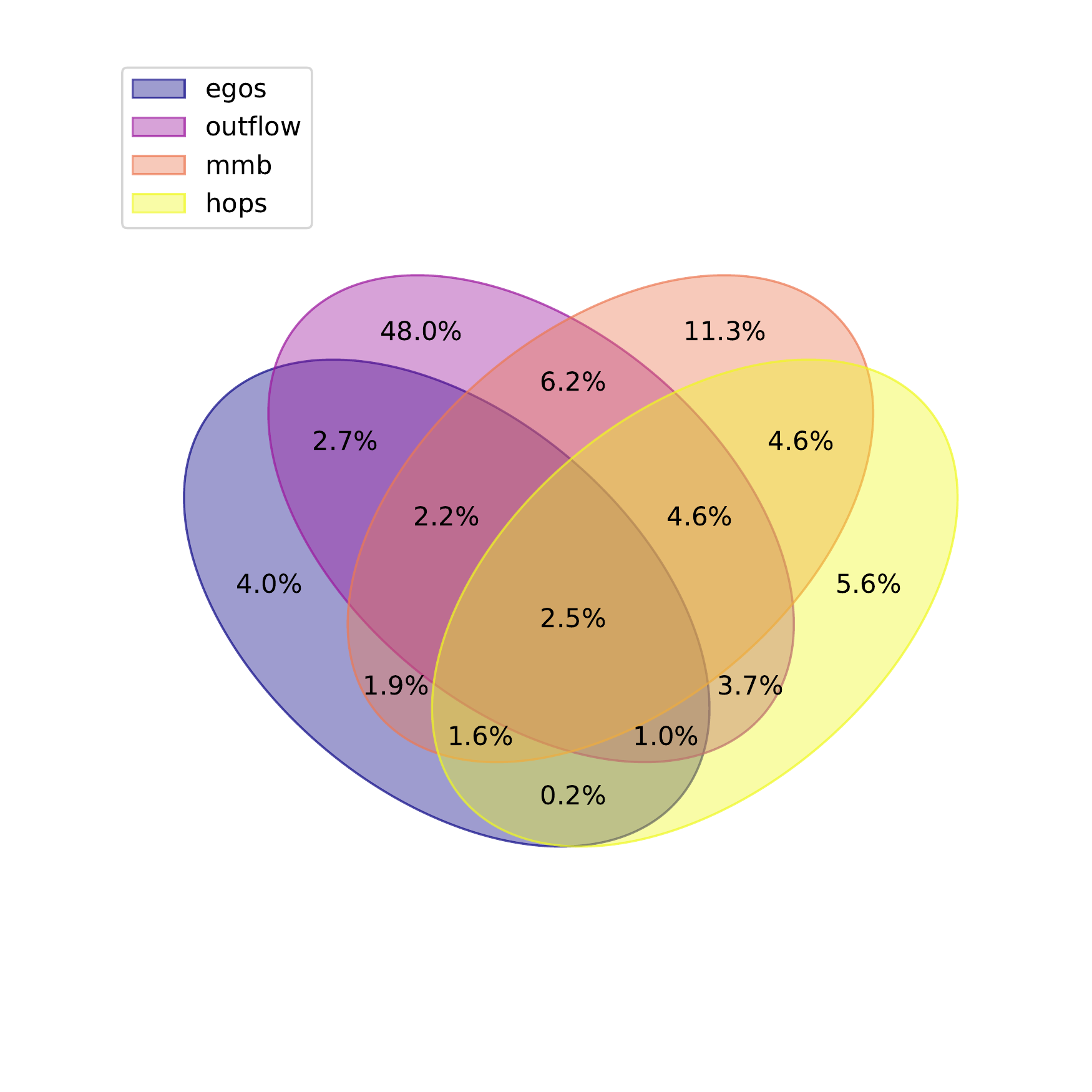}

\caption{Venn diagram showing the correlation between the four different star formation tracers associated with ATLASGAL clumps.  }

\label{fig:venn_assoc}
\end{figure}

We give the statistics of these matches as a function of the evolutionary stage in Table\,\ref{tbl:assoc_statistics}. For each tracer, we give the total number of associations, the overall percentage, and the percentage for the region where all of the surveys overlap. It is the latter of these that is most useful when comparing trends in the association rates as a function of evolution. We show a Venn diagram of the correlation between the different tracers in Figure\,\ref{fig:venn_assoc}. There are no strong correlations between the different tracers, which would suggest they are somewhat random.

\setlength{\tabcolsep}{6pt}

\begin{table*}

\begin{center}
\caption{Summary of associations with other star formation tracers by evolutionary stage. For each tracer we give the total number of associations, the overall percentage and the percentage for the region where all of the surveys overlap (i.e. 300\degr $ < \ell< 350\degr$ and $|b| < 0.5\degr$).}
\label{tbl:assoc_statistics}
\begin{minipage}{\linewidth}
\small
\begin{tabular}{lccccccccccccc}
\hline \hline
  \multicolumn{1}{l}{Classification}&  
  \multicolumn{1}{c}{Number}&
  \multicolumn{3}{c}{EGOs}   &
  \multicolumn{3}{c}{H$_2$O masers}   &
  \multicolumn{3}{c}{Methanol masers}   &	
  \multicolumn{3}{c}{Outflows} \\

    \multicolumn{1}{c}{type }&  
      \multicolumn{1}{c}{}&  
\multicolumn{1}{c}{(\#)} &
    \multicolumn{1}{c}{(\%)}&
    \multicolumn{1}{c}{(\%)}&
   \multicolumn{1}{c}{(\#)} &
    \multicolumn{1}{c}{(\%)}&
    \multicolumn{1}{c}{(\%)}&
    \multicolumn{1}{c}{(\#)} &
    \multicolumn{1}{c}{(\%)}&
    \multicolumn{1}{c}{(\%)}&
    \multicolumn{1}{c}{(\#)} &
    \multicolumn{1}{c}{(\%)}&
    \multicolumn{1}{c}{(\%)} \\
\hline

Quiescent	&	1206	&	1	&	0.1	&	0	&	1	&	0.1	& 0.3		&	2	&	0.2	&	0.0	&	106	&	8.8	& 15.7		\\
Protostellar	&	941	&	46	&	4.9	&	7.4	&	37	&	3.9	&	5.9	&	97	&	10.3	&	11.5	&	120	&	12.8	&	19.7	\\
YSO	&	1510	&	120	&	7.9	&	8.9	&	90	&	6.0	& 10.9		&	231	&	15.3	&	16.8	&	257	&	17.0	& 23.8		\\
HII	&	1222	&	51	&	4.2	&	5.6	&	116	&	9.5	&	14.8	&	242	&	19.8	& 19.2		&	359	&	29.4	& 41.6		\\
\hline
Total &	8417	&	257	&	3.1	&		&	317	&	3.8	&		&	723	&	8.6	&		&	1458	&	17.3	&		\\
\hline\\
\end{tabular}\\

\end{minipage}

\end{center}
\end{table*}
\setlength{\tabcolsep}{6pt}

\begin{figure}
    \centering
 
      \includegraphics[width=.45\textwidth, trim= 0 0 0 0]{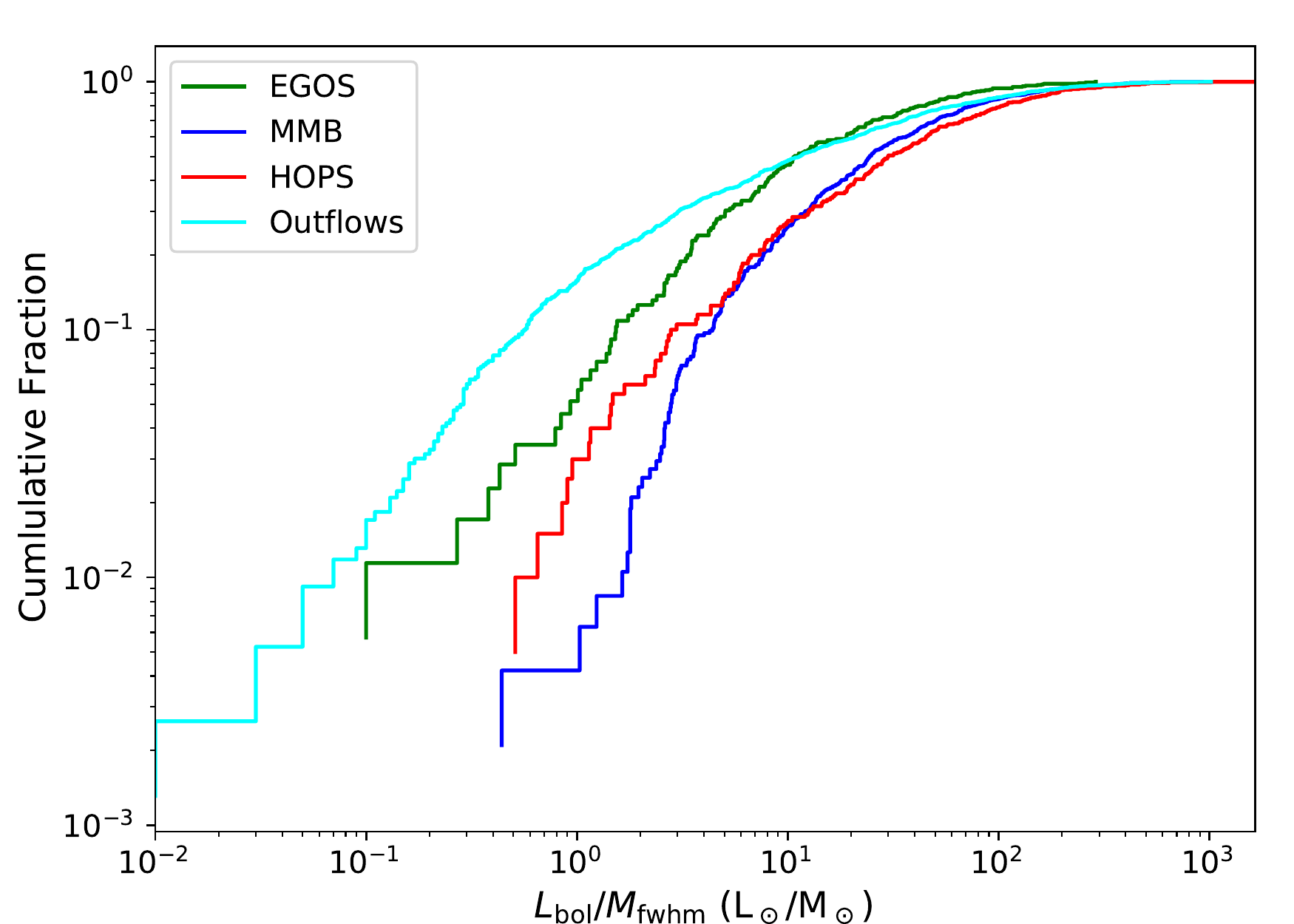}

\caption{Rate of change in the \lm-ratio as a function of the star formation tracers. See Sect.\,\ref{sect:other_samples} for details of the surveys used and how the associations have been made. }

\label{fig:lm_vs_tracers}
\end{figure}

It is clear from Table\,\ref{tbl:assoc_statistics} that the association rates increase as a function of evolution, and this trend holds even up to the \hii\ region stage for all but the EGOs. For these, the association rate peaks at the YSO stage and then begins to decrease. It is also interesting to note that the association rates for EGOs, water and methanol masers with clumps classified as quiescent are effectively zero, which is consistent with their classification as starless (as suggested by the lack of a 70-\mum\ point source).  We do, however, find that a significant fraction are associated with molecular outflows, which is a strong indication that star formation is already underway in them. Figure\,\ref{fig:lm_vs_tracers} shows the cumulative distribution functions for the \lm-ratio of clumps associated with each of the four star formation tracers. This plot reveals that molecular outflows are the earliest signpost for star formation, followed by EGOs, and then by water and methanol masers (for more detailed analysis of methanol and water masers and star formation evolutionary sequences, see \citealt{breen2018,billington2020_timescale, ladeyschikov2020}). 

\begin{figure}
    \centering
 
      \includegraphics[width=.45\textwidth, trim= 0 0 0 0]{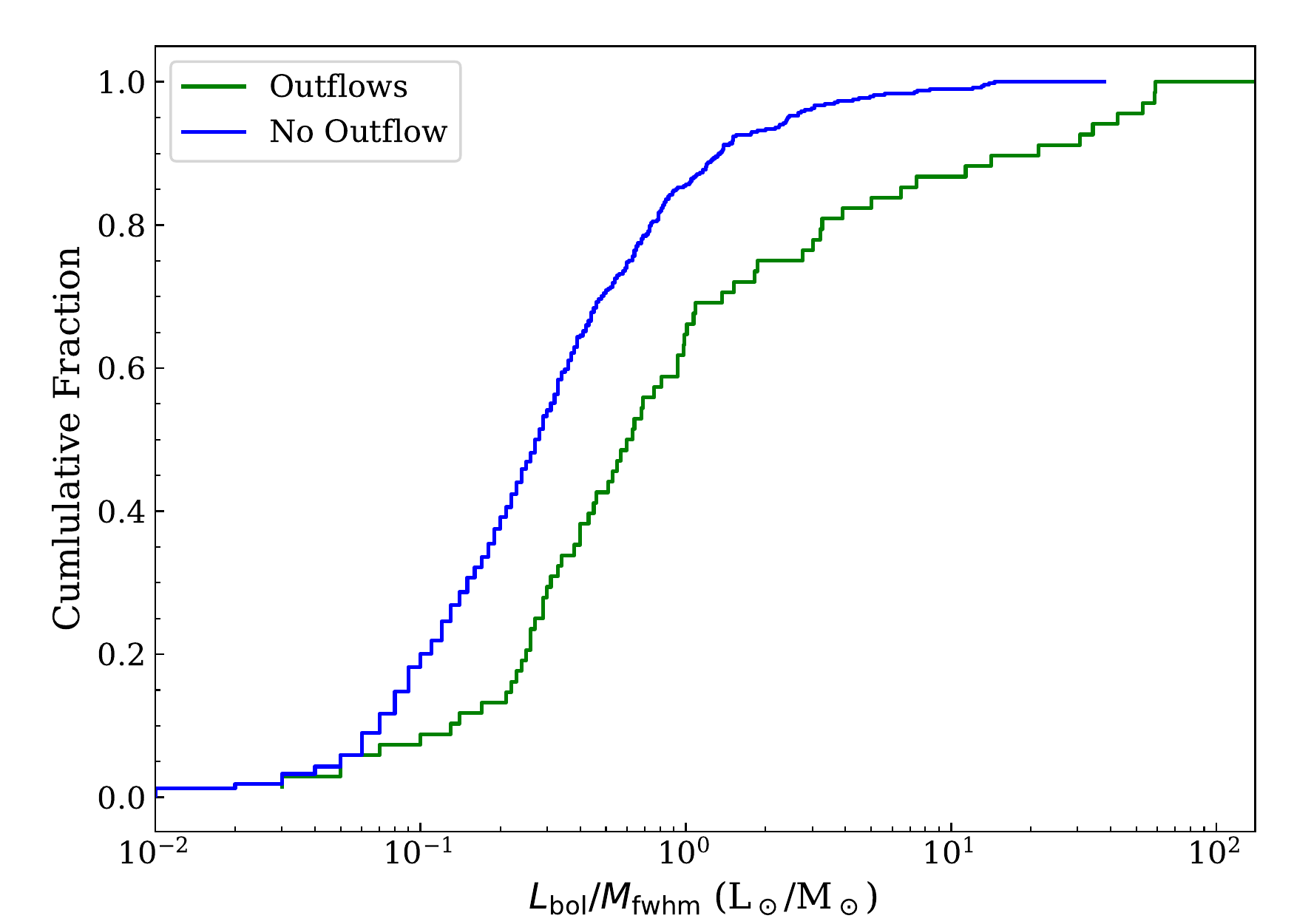}

\caption{The \lm-ratio for quiescent clumps with and without outflow associations. The mean values for  the \lm-ratio  of clumps with and without outflows are 7.8\,\lsun/\msun\ and 0.8\,\lsun/\msun\ respectively.}

\label{fig:quiescent_clump_lm_ratio}
\end{figure}

The number of clumps associated with molecular outflows is likely a lower limit, as the sensitivity of the available CO data is often not sufficient to detect high-velocity wings in the spectra. If one includes only the clumps towards which good CO data are available in the statistical analysis, the association rates approximately three times larger (\citealt{yang2018}). This result suggests that star formation is already under way in perhaps half of the quiescent clumps. This behaviour is consistent with the observed increases in the \lm-ratio seen in Fig.\,\ref{fig:luminosity_lm_ratio_all} prior to the appearance of a protostellar object and explains why there is no sharp jump in the cumulative distribution curve between the quiescent and protostellar stages. Comparing the \lm-ratio for quiescent clumps with and without an associated outflow (see Fig.\,\ref{fig:quiescent_clump_lm_ratio}) reveals that those associated with outflows have significantly larger ratios ($p$-value $\ll$ 0.0013) and are therefore more evolved.

We therefore conclude that the presence of molecular outflows is the earliest signpost for star formation -- even preceding maser activity -- and that star formation is already underway in a significant number of quiescent clumps identified in the ATLASGAL catalogue.

\section{Summary and conclusions}
\label{sect:summary}

The ATLASGAL compact source catalogue consists of approximately 10\,000 clumps located across the inner Galactic plane. In this paper, we use new molecular-line information to determine reliable velocities and distances to many hundreds of clumps located outside of the Galactic centre ($3\degr<|\ell|<60\degr$). We used these new distances to refine the membership of associations of clumps determined using a friends-of-friends analysis. In total, 877 associations with two or more clump members have been identified.  We also use the distances to determine a complete set of unbiased physical properties for the clumps (masses, luminosities, radii, volume densities, and free-fall times). Finally, we have performed a visual inspection of multi-wavelength images (8--870\,\mum, including Spitzer, WISE, and HiGAL bands) of all sources located outside the Galactic centre to identify robust evolutionary samples (quiescent, protostellar, YSO, and \hii\ region) and compared their physical properties to identify trends in the data that provide insight into the star formation process.\\

\noindent The main results of this analysis are:

\begin{itemize}
  
    \item We have allocated velocities to 8899 clumps of the 9817 clumps located in the main ATLASGAL survey region (i.e., $300\degr < \ell < 60\degr$ and $|b| < 1.5$\degr), corresponding to $\sim$90\,per\,cent of the whole sample. Excluding clumps within 3\degr\ of the Galactic centre, where kinematic distances are unreliable, results in a sample of 8417 clumps. We have determined distances to 8130 of these ($\sim$97\,per\,cent). \\
    
     \item The visual classification has resulted in the identification of 5007 clumps in one of  four evolutionary stages (1218 quiescent, 1010 protostellar, 1543 YSO and 1236 \hii\ region). The fact that we have identified roughly equal numbers in each of the four evolutionary stages suggests that the statistical lifetimes of each stage are comparable. These are considered to be a high-reliability sample of clumps with well-determined physical properties and visual classification. \\
     
     \item A comparison of the physical properties reveals that the masses, sizes, and densities are similar for all four evolutionary stages, suggesting that the mass and structure of the clumps do not change significantly during the star-formation process (typically a few $\times 10^5$\,years). Conversely, we find significant and systematic increases in the temperature, luminosity, and luminosity-to-mass ratio that are consistent with the evolutionary stages identified.\\
     
     \item We have investigated the correlations between the luminosity, mass and size and find that the luminosity-mass and mass-radius relations are broadly consistent with previously reported values. These relations are therefore fairly robust and are relatively insensitive to the method used to determine the parameter values.\\
     
     \item We find that the cumulative distribution function of the \lm-ratio is very smooth and reveals no significant changes in the slope or jumps at the mean values determined for the four evolutionary stages or in the intervening regions where one would expect the embedded objects transition between one stage and another. This behaviour leads us to conclude that star formation is a rather monotonic and continuous process and the observational stages, while useful in identifying groups of protostellar objects with similar properties and/or ages, do not in themselves represent fundamentally different stages or changes in the physical mechanisms involved on clump scales.\\
     
     \item If we assume that the number of clumps in each \lm-ratio interval reflects the fraction of the statistical lifetime spent in each interval, we are able to equate the cumulative distribution function to a statistical lifetime. This step allows us to investigate how the \lm-ratio changes with time, and doing so we find that the rate of change of \lm-ratio continues to accelerate as the clump evolves, turning into a runaway process towards the end. This behaviour supports star-formation models in which the accretion rate increases over time.\\
     
     \item Cross-matching the evolutionary sample with four other star-formation samples (outflows, EGOs, and methanol and water masers) we find an overall association rate of $\sim$15-20\,per\,cent. There is a clear correlation between association rate and evolutionary stage, with the former increasing as a function of evolution. The association rate for EGOs and masers is effectively zero for the quiescent stage, as one would expect for starless clumps. A significant fraction of these sources are associated with molecular outflows, however, indicating that star formation is already underway in many such clumps. Molecular outflows are therefore the earliest manifestations of star formation, appearing before the protostar becomes visible in the far-infrared or maser emission is detected.\\
 
 \end{itemize}

\section*{Acknowledgements}

We thank the referee for their comments and suggestions that have helped improve the clarity of this work. SL acknowledges financial support from INAF through the project INAF-PRIN 2019 ONSET (project number 1.05.01.85.05). DC acknowledges support by the German \emph{Deut\-sche For\-schungs\-ge\-mein\-schaft, DFG\/} project number SFB956A. This research made use of Astropy,\footnote{http://www.astropy.org} a community-developed core Python package for Astronomy \citep{astropy2013, astropy2018}; matplotlib \citep{matplotlib2007}; numpy and scipy \citep{scipy2020}. This  research  has  made  use of  the  SIMBAD  database  and  the  VizieR  catalogue,  operated  at  CDS,  Stras-bourg, France. This document was prepared using the Overleaf web application, which can be found at www.overleaf.com.

\section*{Data Availability}

The data presented in this article is available from a dedicated website:
\url{https://atlasgal.mpifr-bonn.mpg.de/}


\bibliographystyle{mnras}
\bibliography{urquhart_2019,wells}

\label{lastpage}

\end{document}